\documentclass[12pt]{article}
\usepackage{a4wide,latexsym,graphicx,epsfig,psfrag,here}
\newcommand{\g}{\gamma}

\newcommand{\eps}{\epsilon}
\newcommand{\epsp}{\epsilon^\prime}

\newcommand{\bea}{\begin{eqnarray}}
\newcommand{\eea}{\end{eqnarray}}
\newcommand{\beq}{\begin{equation}}
\newcommand{\eeq}{\end{equation}}
\newcommand{\nn}{\nonumber}
\newcommand{\nl}{\nonumber\\}
\newcommand{\fr}{\frac}
\newcommand{\dg}{\dagger}
\newcommand{\hl}{\hline}

\newcommand{\real}{{\rm Re}}
\newcommand{\imag}{{\rm Im}}

\newcommand{\PL}[3]{{Phys. Lett.} {\bf#1} {(#2)} {#3}} 
\newcommand{\PRL}[3]{{Phys. Rev. Lett.}  {\bf#1} {(#2)} {#3}} 
\newcommand{\PR}[3]{{Phys. Rev.} {\bf#1} {(#2)} {#3}} 
\newcommand{\NP}[3]{{Nucl. Phys.} {\bf#1} {(#2)} {#3}} 
\newcommand{\EPJ}[3]{{Eur. Phys. J.} {\bf#1} {(#2)} {#3}} 
\newcommand{\ZP}[3]{{Z. Phys.} {\bf#1} {(#2)} {#3}}
\newcommand{\JHEP}[3]{{JHEP} {\bf#1} {(#2)} {#3}}  
\newcommand{\hepph}[1]{{\tt hep-ph/#1}}

\newcommand{\cO}{{\cal O}}
\newcommand{\cL}{{\cal L}}

\newcommand{\cA}{{\cal A}}

\newcommand{\ba}{\begin{array}{c}}
\newcommand{\bat}{\begin{array}{cc}}
\newcommand{\ea}{\end{array}}
\def\eqn#1{(\ref{#1})}

\newcommand{\no}{\nonumber}
\newcommand{\pe}{$\pi^0-\eta$}  
\def\slashchar#1{\setbox0=\hbox{$#1$}\dimen0=\wd0%
\setbox1=\hbox{/}\dimen1=\wd1%
\ifdim\dimen0>\dimen1%                     
\rlap{\hbox to
\dimen0{\hfil/\hfil}}#1\else                                     
\rlap{\hbox to \dimen1{\hfil$#1$\hfil}}/\fi}
\renewcommand{\arraystretch}{1.5}
\begin{document}
\parskip=3pt plus 1pt

\begin{titlepage}
%\vskip 1cm
\begin{flushright} 
{IFIC/03-39 \\ UWThPh-2003-17\\MAP-292}
\end{flushright}
\vskip 2cm

\begin{center} 
{\LARGE \bf 
Isospin Breaking in $K \rightarrow \pi \pi$ Decays$^*$ 
} 
\\[40pt] 
V. Cirigliano$^{1,2}$, G. Ecker$^{3}$, H. Neufeld$^{3}$ 
and A. Pich$^{1}$ 
 
\vspace{1cm}
${}^{1)}$ Departament de F\'{\i}sica Te\`orica, IFIC, CSIC --- 
Universitat de Val\`encia \\ 
Edifici d'Instituts de Paterna, Apt. Correus 22085, E-46071 
Val\`encia, Spain \\[10pt]

${}^{2)}$  Department of Physics, California Institute of Technology \\ 
Pasadena, California 91125, USA  \\[10pt]  

${}^{3)}$ Institut f\"ur Theoretische Physik, Universit\"at 
Wien\\ Boltzmanngasse 5, A-1090 Vienna, Austria \\[10pt] 
\end{center} 
 
\vfill 

\begin{abstract}
We perform a complete analysis of isospin breaking in $K \to 2 \pi$
amplitudes in chiral perturbation theory, including both strong
isospin violation ($m_u \neq m_d$) and electromagnetic corrections to
next-to-leading order in the low-energy expansion.  The unknown chiral 
couplings are estimated at leading order in the $1/N_c$ expansion.
We study the impact of isospin breaking on CP conserving amplitudes
and rescattering phases.  In particular, we extract the effective
couplings $g_8$ and $g_{27}$ from a fit to $K \rightarrow \pi \pi$
branching ratios, finding small deviations from the isospin-limit
case. The ratio $\real A_0 / \real A_2$ measuring the $\Delta I= 1/2$ 
enhancement is found to decrease from $22.2 \pm 0.1$ in the isospin
limit to $20.3 \pm 0.5$ in the presence of isospin breaking. 
We also analyse the effect of isospin violation on the CP violation
parameter $\epsp$, finding a destructive interference between three
different sources of isospin violation.  Within the uncertainties of
large-$N_c$ estimates for the low-energy constants, the isospin
violating correction for $\epsp$ is below 15 $\%$.
\end{abstract}

\vfill
 
\noindent 
*~Work supported in part by HPRN-CT2002-00311 (EURIDICE) and by
Acciones Integradas, Project No. 19/2003 (Austria),
HU2002-0044 (MCYT, Spain).
\end{titlepage} 
\newpage

\tableofcontents

\section{Introduction}
\label{sec:intro}
\renewcommand{\theequation}{\arabic{section}.\arabic{equation}}
\setcounter{equation}{0}
A systematic treatment of isospin violation in nonleptonic weak
interactions is needed for many phenomenological applications. The
generically small effects induced by electromagnetic corrections and 
by the quark mass difference $m_u - m_d$ are enhanced in subdominant 
amplitudes with $\Delta I > 1/2$ because of the $\Delta I=1/2$ rule. 
For one, a quantitative understanding of the $\Delta I=1/2$ rule 
itself is only possible with isospin violating effects included. 
Another area of application is CP violation in the $K^0-\overline{K^0}$ 
system where isospin breaking is crucial for a precision calculation 
of $\epsp/\eps$. 

Isospin violation in $K \to 2 \pi$ decays has already been addressed 
in recent works \cite{Cirigliano:fv,gv99,Cirigliano:1999ie,Cirigliano:1999hj,
Cirigliano:2000zw,gv00,Wolfe:1999br,Wolfe:2000rf}.    
In this paper, we reanalyse the $K \to \pi\pi$ decay amplitudes
to perform a comprehensive study of all isospin violating 
effects to next-to-leading order in the low-energy expansion of the 
standard model. More precisely, we shall work to first order in $\alpha$
and in $m_u - m_d$ throughout, but to next-to-leading order in the chiral 
expansion. In view of the observed octet dominance of the nonleptonic weak 
interactions, we therefore calculate to 
$\cO(G_8 p^4,G_8 (m_u-m_d) p^2,e^2 G_8 p^2)$ and to $\cO(G_{27} p^4)$
for octet and 27-plet amplitudes, respectively. 

At this order, many 
a priori unknown low-energy constants (LECs) appear. With few
exceptions to be discussed in Sec.~\ref{sec:LECs}, we adopt 
leading large-$N_c$ estimates for the LECs. The advantage is
that we employ a systematic approximation scheme with solid theoretical 
foundation that can in principle be carried through beyond leading
order. On the other hand, the importance of subleading large-$N_c$
effects is at present not known in general. We shall estimate the 
uncertainties by varying 
the two scales entering those estimates: the renormalization scale for
evaluating Wilson coefficients (short-distance scale) and the chiral scale
(long-distance scale) at which the large-$N_c$ results are supposed to apply.

In performing electromagnetic corrections, a careful analysis of radiative
events is necessary as emphasized in Ref.~\cite{Cirigliano:2000zw}. We shall 
perform such an analysis for the new KLOE measurement \cite{KLOE} of the ratio
$\Gamma(K_S\to \pi^+\pi^-[\gamma])/\Gamma(K_S\to \pi^0\pi^0)$ with a fully 
inclusive $\pi^+\pi^-[\gamma]$ final state. The KLOE result 
influences  the phase difference $\chi_0 - \chi_2$ of the two
isospin amplitudes strongly. Together with this phase
difference, the effective weak octet and 27-plet couplings $G_8$, $G_{27}$
will be the primary output of our analysis. With that output, several 
phenomenological issues can be addressed such as the relation of the phases
$\chi_0, \chi_2$ to the s-wave pion-pion scattering phase shifts or the impact
of isospin breaking on $\epsp/\eps$. 

The content of the paper is as follows. In the subsequent section, we
introduce the decay amplitudes and the relevant effective
chiral Lagrangians. The amplitudes at leading order in the low-energy
expansion are presented in Sec.~\ref{sec:chptLO}. The amplitudes at 
next-to-leading order are investigated in Sec.~\ref{sec:chptNLO},
distinguishing between \pe~mixing and all other contributions arising
at that order. The amplitudes are divided into various parts depending
on the source of isospin violation. The local amplitudes of
next-to-leading order are explicitly given here. Sec.~\ref{sec:LECs}
analyses the LECs at leading order in $1/N_c$. To determine weak
and electroweak LECs for $N_c \to \infty$, one needs input for the
strong [up to $\cO(p^6)$] and electromagnetic couplings [up to 
$\cO(e^2 p^2)$], in addition to the relevant Wilson coefficients. 
We discuss to which extent the necessary information
is available. The numerical calculations of the various amplitudes to
next-to-leading order in chiral perturbation theory (CHPT) are presented 
in Sec.~\ref{sec:num}.  
Dispersive and absorptive components of the loop amplitudes are given
together with CP-even and CP-odd parts of the local amplitudes. Those
amplitudes are then used in Sec.~\ref{sec:pheno1} to extract the
lowest-order nonleptonic couplings $G_8, G_{27}$ and the phase
difference $\chi_0 - \chi_2$ from $K \to \pi\pi$ data. We compare
those quantities at lowest and next-to-leading order, the latter with
and without isospin violation included. With this information, we then
analyse the relation of the phase difference $\chi_0 - \chi_2$ to the
corresponding difference of $\pi\pi$ phase shifts. In
Sec.~\ref{sec:pheno2} we discuss isospin violating contributions to
the parameter $\epsp$ of direct CP violation in $K^0 \to \pi\pi$
decays. Sec.~\ref{sec:concl} contains our conclusions. Various
technical aspects are treated in several appendices: 
next-to-leading-order effective chiral Lagrangians; explicit loop 
amplitudes; an alternative convention for LECs of lowest order;
details for the analysis of the phase difference.

\section{Nonleptonic weak interactions in CHPT }
\label{sec:general}
\renewcommand{\theequation}{\arabic{section}.\arabic{equation}}
\setcounter{equation}{0}
In this section, we define our notation for the $K\to \pi\pi$ amplitudes and
we introduce the relevant effective chiral Lagrangians.

\subsection{Decay amplitudes}
Using the isospin decomposition of two-pion final states, we write  
the $K \rightarrow \pi \pi$ amplitudes in the charge basis in terms 
of three amplitudes\footnote{We shall use the
invariant amplitudes $\cA_{n}$ defined as follows: 
$$
\langle (\pi \pi)_n | T \bigg( e^{\displaystyle 
i \int  dx\, {\cal L} (x) } \bigg) 
| K \rangle\, =\, i \, (2 \pi)^4 \, \delta^{(4)} (P_f - P_i) \, 
\times \left( -i \, \cA_{n} \right) \ . 
$$
}
$\cA_{\Delta I}$ that are generated by the $\Delta I$ component of 
the electroweak
effective Hamiltonian in the limit of isospin conservation:
\begin{eqnarray}   \cA_{+-} &=& \cA_{1/2} +
{1 \over \sqrt{2}} \left( \cA_{3/2} + \cA_{5/2} \right) \nn \\
\cA_{00} &=& \cA_{1/2} - \sqrt{2} \left( \cA_{3/2} + \cA_{5/2} \right)
\label{eq:intro1} \\ \cA_{+0} &=& {3 \over 2} \, \left( \cA_{3/2} - 
{2 \over 3} \cA_{5/2} \right) ~.\nn 
\end{eqnarray}
In the standard model, the $\Delta I = 5/2$ piece is absent in the isospin 
limit, thus reducing the number of
independent amplitudes to two.  Each amplitude $\cA_n$ has a
dispersive (${\cal D}isp \, \cA_n$) and an absorptive 
(${\cal A}bs \, \cA_n$) component. 
In order to carry out 
phenomenological applications and to keep the notation as close as 
possible to the standard analysis in the isospin limit, we  write 
\begin{eqnarray}  
A_0 \,e^{i \chi_0 } &=&   \cA_{1/2}
\nn \\
 A_2 \,e^{i \chi_2 }&=& \cA_{3/2} + \cA_{5/2}
\label{eq:intro2}   \\
A_2^{+} \,e^{i \chi_2^{+} } &=& \cA_{3/2} - 2/3 \,  \cA_{5/2}~, 
\nn 
\end{eqnarray}  
where we explicitly separate out the phases $\chi_I$. In the
limit of CP conservation, the amplitudes $A_{0}, A_{2}$ and  $A_{2}^+$
are real and positive.   
In the isospin limit, the $A_I$ are the standard isospin amplitudes 
and the phases $\chi_I$ are identified with the 
s-wave $\pi \pi$ scattering phase shifts $\delta_I (\sqrt{s} =
M_K)$. 

For the phenomenological analysis (see Secs.~\ref{sec:pheno1} and 
\ref{sec:pheno2}), we therefore adopt 
the following parametrization of $K \rightarrow \pi \pi$ 
amplitudes: 
\begin{eqnarray}
{\cal A}_{+-} &=& 
A_{0} \, e^{i \chi_0}  + { 1 \over \sqrt{2}} \,   A_{2}\,  e^{i\chi_2 }
 \nn \\
{\cal A}_{00} &=& 
A_{0} \, e^{i \chi_0}  - \sqrt{2} \,   A_{2}\,  e^{i\chi_2 }
\label{eq:intro3}
\\ 
{\cal A}_{+0} &=& {3 \over 2} \,  
A_{2}^{+} \,  e^{i\chi_2^{+}}  ~. 
\nonumber
\end{eqnarray}
This parametrization holds for the infrared-finite amplitudes where the
Coulomb and infrared parts (defined in Sec.~\ref{sec:chptNLO}) 
have been removed from $\cA_{+-}$.  

In the absence of electromagnetic interactions $\cA_{5/2}=0$ and 
therefore $A_2 = A_2^+$.  To set the stage, we extract the isospin
amplitudes $A_0, A_2$ and the phase difference $\chi_0 - \chi_2$ from
a fit to the three $ K \rightarrow \pi \pi $ branching ratios
\cite{KLOE,PDG02}:
\begin{eqnarray} 
A_0 &=& (2.715 \pm 0.005) \cdot 10^{-7} \mbox{ GeV} \nl
A_2 &=& (1.225 \pm 0.004) \cdot 10^{-8} \mbox{ GeV} \\
\chi_0 - \chi_2 &=& (48.6 \pm 2.6)^{\circ} ~. \nn
\end{eqnarray}       
These values hold in the isospin limit except that the physical pion
masses have been used for phase space. The substantial reduction in
the phase difference $\chi_0 - \chi_2$ (from about $58^{\circ}$
during the past 25 years \cite{PDG02}) is
entirely due to the new KLOE measurement of the ratio
$\Gamma(K_S\to \pi^+\pi^-(\gamma))/\Gamma(K_S\to \pi^0\pi^0)$ 
\cite{KLOE}.

\subsection{Effective chiral Lagrangians}
\label{sec:ECL}
In the presence of isospin violation, the physics of $K \to \pi\pi$
decays involves an interplay of the nonleptonic weak, the strong and
the electromagnetic interactions. Consequently, a number of
effective Lagrangians are needed to describe those transitions. We use
the well-known Lagrangian for strong interactions to
$\cO(p^6)$ \cite{gl85a,fs96,bce99}, the nonleptonic weak Lagrangian to 
$\cO(G_F p^4)$ \cite{cronin67,kmw90,ekw93,bpp98}, the electromagnetic
Lagrangian to $\cO(e^2 p^2)$ \cite{egpr89,urech95} and, finally, the
electroweak Lagrangian to $\cO(e^2 G_8 p^2)$ \cite{bw84,grw86,eimnp00}. 

Only the leading-order (LO) Lagrangians are written down explicitly
here. The relevant
parts of the next-to-leading-order (NLO) Lagrangians can be found in 
App.~\ref{app:NLOLag} along with further details.\\[.3cm] 
\noindent
{\bf Strong Lagrangian:}
\begin{eqnarray}
\cL_{\rm strong}  &=& \frac{F^2}{4} \,\langle D_\mu U D^\mu U^\dagger + 
\chi U^\dagger + \chi^\dagger  U \rangle \nn\\
& + & \sum_i\; L_i\, O^{p^4}_i + \sum_i\; X_i\ F^{-2}\, O^{p^6}_i 
~. \label{eq:Lstrong}
\end{eqnarray}
$F$ is the pion decay constant in the chiral limit, the $SU(3)$ matrix
field $U$ contains the pseudoscalar fields and the scalar field $\chi$
accounts for explicit chiral symmetry breaking through the quark
masses $m_u, m_d, m_s$. The relevant operators  $O^{p^4}_i$ are listed
in App.~\ref{app:NLOLag}. The LECs $X_i$ of $O(p^6)$ will only enter
through the large-$N_c$ estimates of the electroweak couplings in 
Sec.~\ref{sec:LECs}. $\langle A \rangle$ denotes the $SU(3)$
flavour trace of $A$.\\[.3cm] 
\noindent
{\bf Nonleptonic weak Lagrangian:}
\begin{eqnarray} 
\cL_{\rm weak} &=&  G_8  F^4  \,\langle\lambda D^\mu U^\dagger
 D_\mu U \rangle + G_{27} F^4 \left( L_{\mu 23} L^\mu_{11} + 
{2\over 3} L_{\mu 21} L^\mu_{13}\right)\nn\\
& + &    \sum_i\;  G_8 N_i F^2\, O^8_i +
  \sum_i\; G_{27} D_i F^2\, O^{27}_i \ + \ {\rm h.c.} \label{eq:Lweak}
\end{eqnarray}
The matrix $L_{\mu}=i U^\dagger D_\mu U$  represents the octet of
$V-A$ currents to lowest order in derivatives;
$\lambda = (\lambda_6 - i \lambda_7)/2$ projects onto the
$\bar s\to \bar d$ transition. 
%[$\lambda_{ij} =\delta_{i3}\delta_{j2}$]. 
Instead of $G_8$, $G_{27}$ we will also use the dimensionless couplings 
$g_8$, $g_{27}$ defined as
\begin{equation} 
G_{8,27} = -{G_F \over \sqrt{2}}\,  V_{ud}^{\phantom{*}} V_{us}^*  
\  g_{8,27}~.
\end{equation}
One of the main tasks of this investigation will be the determination 
of $g_8$, $g_{27}$ in the presence of isospin violation to NLO.  
The LECs $N_i$, $D_i$ of $\cO(G_F p^4)$ are
dimensionless. The monomials $O^8_i$, $O^{27}_i$ relevant for $K \to 2
\pi$ transitions can be found in App.~\ref{app:NLOLag}. \\[.3cm]  
\noindent
{\bf Electromagnetic Lagrangian:}
\begin{eqnarray} 
\cL_{\rm elm} &=&  e^2 Z  F^4  \,\langle Q  U^\dagger Q U\rangle 
\nn \\
&+& e^2 \sum_i\; K _i F^2\, O^{e^2 p^2}_i ~. \label{eq:Lelm}
\end{eqnarray}
The quark charge matrix is given by 
$Q= {\rm diag}(2/3,-1/3,-1/3)$. The
lowest-order LEC can be determined from the pion mass difference to
be $Z\simeq 0.8$. The NLO LECs $K_i$ are 
dimensionless and the relevant monomials $O^{e^2 p^2}_i$ can again be
found in App.~\ref{app:NLOLag}.\\[.3cm]
\noindent 
{\bf Electroweak Lagrangian:}
\begin{eqnarray} 
\cL_{\rm EW} &=& e^2 G_8  g_{\rm ewk} F^6 \,\langle\lambda U^\dagger Q 
U\rangle \nn \\
& + &  e^2    \sum_i\; G_8 Z_i F^4\, O^{EW}_i \ + \ {\rm h.c.}  
\label{eq:Lelweak} 
\end{eqnarray} 
The value of the LO coupling $g_{\rm ewk}$ is discussed in
Sec.~\ref{sec:LECs}. 
The LECs $Z_i$ are dimensionless and the associated monomials 
$O^{EW}_i$ are
collected in App.~\ref{app:NLOLag}. We do not include isospin
violating corrections for 27-plet amplitudes. 

The low-energy couplings $L_i,N_i$, $D_i$, $K_i$, $Z_i$ are in general
divergent.  They absorb the divergences appearing in the one-loop
graphs via the renormalization 
\begin{eqnarray} 
L_i &=& L_i^r (\nu_\chi) + \Gamma_i \,\Lambda (\nu_\chi) \nl 
N_i &=& N_i^r (\nu_\chi) + n_i \,\Lambda (\nu_\chi) \nl 
D_i &=& D_i^r (\nu_\chi) + d_i \,\Lambda (\nu_\chi) \label{eq:beta} \\ 
K_i &=& K_i^r (\nu_\chi) + \kappa_i \,\Lambda (\nu_\chi) \nl 
Z_i &=& Z_i^r (\nu_\chi) + z_i \,\Lambda (\nu_\chi) \ , \nn  
\end{eqnarray}
where $\nu_\chi$ is the chiral renormalization scale and the 
divergence is included in the factor
\beq
\Lambda (\nu_\chi) = \frac{\nu_\chi^{d-4}}{(4 \pi)^2} \, \left\{ 
\frac{1}{d-4} - \frac{1}{2} 
\bigg[ \log (4 \pi) + \Gamma ' (1) + 1 \bigg] 
\right\} \ .  
\label{eq:div}
\eeq
The divergent parts of the couplings are all known 
\cite{gl85a,kmw90,ekw93,urech95,eimnp00} and they allow for a
nontrivial check of the loop calculation. On the other hand,
many of the renormalized LECs contributing to the decay amplitudes 
are not known. Our strategy will be to use 
LO large-$N_c$ estimates. A comprehensive discussion of 
all relevant LECs will be presented in Sec.~\ref{sec:LECs}.

\section{Amplitudes at leading order [$\cO (G_F p^2, e^2 G_8 p^0)$]}
\label{sec:chptLO}
\renewcommand{\theequation}{\arabic{section}.\arabic{equation}}
\setcounter{equation}{0}
With the most general effective chiral Lagrangian of the previous
section, we can now proceed with the construction of physical amplitudes. 
At LO in the low-energy expansion, the procedure is
straightforward: chiral power counting tells us
that the amplitudes are obtained by summing all tree-level Feynman
diagrams with one insertion from either $\cL_{\rm weak}$ of $\cO(G_F
p^2)$ or $\cL_{\rm EW}$ of $\cO(e^2 G_8 p^0)$, at most one insertion
of $\cL_{\rm elm}$ of $\cO(e^2 p^0)$
and any number of insertions from the $\cO(p^2)$ part of the strong
Lagrangian (\ref{eq:Lstrong}).

In addition to contributions proportional to the electroweak coupling
$g_{\rm ewk}$, isospin breaking occurs also in the pseudoscalar
mass matrix, generating in particular non-diagonal terms in the fields 
$(\pi_3, \pi_8)$ ($\pi^0 - \eta$ mixing).  Upon diagonalizing the 
tree-level mass matrix one obtains the relation between the
LO mass eigenfields $(\pi^0,\eta)$ and the original 
fields $(\pi_3, \pi_8)$ (to first order in $m_u - m_d$): 
\beq 
\left( \begin{array}{c} \pi_3 \\ \pi_8  \end{array} \right) = 
\left( \begin{array}{cc} 1 &  - \varepsilon^{(2)} \\ 
\varepsilon^{(2)}  & 1  \end{array} \right) \, 
\left( \begin{array}{c} \pi^0 \\ \eta  \end{array} \right)_{\rm LO} \ ,  
\label{eq:LOmixing}
\eeq
with the tree-level $\pi^0-\eta$ mixing angle $\varepsilon^{(2)}$  given by
\beq
\varepsilon^{(2)} = \frac{\sqrt{3}}{4} \; 
\frac{m_d - m_u}{m_s - \widehat{m}}  ,
\label{epsilon}
\eeq 
where $\widehat{m}$ stands for the mean value of the
light quark masses,
\beq
\widehat{m} = \frac{1}{2} (m_u + m_d) \  .
\eeq
The physical amplitudes are then obtained by considering the relevant 
Feynman graphs with insertions from the LO effective Lagrangian 
expressed in terms of the LO mass eigenfields.  

Apart from $\pi^0 - \eta$ mixing, isospin breaking manifests itself
also in the mass differences between charged and neutral mesons, due to 
both the light quark mass difference and electromagnetic contributions.  
We choose to express all masses in terms of those of the neutral
kaon and pion (denoted from now on as $M_K$ and $M_\pi$, respectively).
In terms of quark masses and LO couplings ($B_0$ is related 
to the quark condensate in the chiral limit by $\langle 0 | 
\overline{q} q | 0 \rangle = -F^2 B_0$), the pseudoscalar meson masses 
read:
\bea
 M^2_{\pi} &=& 2 B_0 \,\widehat{m}   \nl
 M^2_{\pi^\pm} &=& M_\pi^2 + 2\, e^2 Z F^2   \nl
 M^2_{K} &=& B_0 \left( m_s + m_d \right)  \label{eq:treemass} \\
 M^2_{K^\pm} &=& M_K^2 
-  \frac{4 \,\varepsilon^{(2)}}{\sqrt{3}}\,  B_0 (m_s - \widehat{m})  
+ 2\, e^2 Z F^2    \nl
 M^2_\eta &=& \frac{1}{3} \, \left( 4 M_K^2  - M_\pi^2 \right) 
-  \frac{8 \,\varepsilon^{(2)}}{3 \sqrt{3}}\,  B_0 (m_s - \widehat{m})
~.\nn
\eea

We are now in the position to write down the three independent 
amplitudes relevant for $K \rightarrow \pi \pi$ decays. 
In the physical ``charge'' basis the LO amplitudes are 
\bea
\cA_{+-} & = & 
{2 \over 3}  \sqrt{2}\, G_{27} F \left( M_K^2 - M_\pi^2 \right)  + 
\sqrt{2}\, G_8 F \Bigg[ M_K^2 - M_\pi^2 - e^2 F^2 \left( g_{\rm ewk} + 
2 Z \right) 
\Bigg]   \nn \\
\cA_{00} & = & 
- \sqrt{2}\, G_{27} F \left( M_K^2 - M_\pi^2 \right) + 
\sqrt{2}\, G_8 F \left( M_K^2 - M_\pi^2 \right) \left(1 - {2 \over \sqrt{3}} 
\,\varepsilon^{(2)} \right)
\label{eq:LOamp1}  
\\ 
\cA_{+0} & = & {5 \over 3}\,  G_{27} F \left( M_K^2 - M_\pi^2 \right) + 
 G_8 F  \left[  \left( M_K^2 - M_\pi^2 \right) {2 \over \sqrt{3}}
  \,\varepsilon^{(2)}   - 
e^2 F^2 \left( g_{\rm ewk} + 2 Z  \right) \right] ~.  \nn
\eea
We recall that we do not include isospin violation for the 27-plet
amplitudes. In the isospin basis, more convenient for phenomenological
applications, the LO amplitudes are given by (see Eq.~(\ref{eq:intro1}) 
for the relation between the two bases)
\bea
\cA_{1/2} & = & 
{\sqrt{2} \over 9}\, G_{27} F \left( M_K^2 - M_\pi^2 \right)  \nn \\
 & + &  
\sqrt{2}\, G_8 F \left[  \left( M_K^2 - M_\pi^2 \right) 
 \left(1 - {2 \over 3 \sqrt{3}} \,\varepsilon^{(2)} \right)  
- {2 \over 3}\, e^2 F^2 \left( g_{\rm ewk} + 2 Z \right)  \right]
\label{eq:LOamp2}  
\\
\cA_{3/2} & = & 
{10 \over 9}\,  G_{27} F \left( M_K^2 - M_\pi^2 \right) + 
G_8 F \left[  \left( M_K^2 - M_\pi^2 \right) {4 \over 3  \sqrt{3}}\, 
\varepsilon^{(2)}  -  {2 \over 3}\, e^2 F^2 \left( g_{\rm ewk} + 2 Z \right)
\right] \nn   \\ 
\cA_{5/2} & = & 0 ~. \nn 
\eea
The parameter $F$ can be identified with the pion
decay constant $F_\pi$ at this order.  The effect
of strong isospin breaking (proportional to $\varepsilon^{(2)}$) is
entirely due to $\pi^0 - \eta$ mixing at LO. 
Electromagnetic interactions contribute through mass splitting (terms 
proportional to $Z$) and insertions of $g_{\rm ewk}$. 
As a consequence of imposing CPS symmetry \cite{Bernard}
on the effective Lagrangian, 
electromagnetic corrections to the octet weak Hamiltonian do not
generate a $\Delta I= 5/2$ amplitude at LO in the quark mass 
expansion.

\section{Amplitudes at next-to-leading order [$ \cO (G_F p^4,e^2 G_8 p^2)$]}
\label{sec:chptNLO}
\renewcommand{\theequation}{\arabic{section}.\arabic{equation}}
\setcounter{equation}{0}
Let us now outline the construction of NLO amplitudes.
As always, chiral power counting is the guiding principle: it tells us
that both one-loop and tree-level diagrams now contribute.  In the
one-loop diagrams, one has to consider one insertion from either 
$\cL_{\rm weak}$ of $\cO(G_F p^2)$ or $\cL_{\rm EW}$ of $\cO(e^2 G_8 
p^0)$, at most one insertion of $\cL_{\rm elm}$ of $\cO(e^2 p^0)$
and any number of insertions from the $\cO(p^2)$ part of the strong
Lagrangian, with the LO effective Lagrangians expressed in terms of the LO
mass eigenfields. For the tree-level diagrams, one has to apply 
one insertion from the NLO effective Lagrangian and any number 
of insertions from the LO Lagrangian. The strangeness changing vertex 
can come from either the LO or NLO effective Lagrangians.
This straightforward prescription leads to a large 
number of explicit diagrams for each mode, due to several 
topologies and several possibilities to insert isospin breaking 
vertices from the LO effective Lagrangian (in the weak vertex, 
in the strong vertex, in the internal propagators, in the external legs). 
We begin with the well-defined class of NLO corrections to the
pseudoscalar meson propagators, focusing afterwards on the other
corrections.  

\subsection{$\pi^0 - \eta$ mixing at NLO}
As in the LO case, it is convenient to first analyse isospin breaking in the 
two-point functions (inverse propagators) and to define 
renormalized fields in which the propagator has a diagonal form with
unit residues at the poles\footnote{This way, no further wave function 
renormalization effect has to be included.}.
At NLO two main new features arise: 
\begin{itemize}
\item Not only the $(\pi_3, \pi_8)$ mass matrix acquires off-diagonal 
matrix elements, but also the purely kinetic part of the propagator does so. 
\item Electromagnetic interactions contribute to this phenomenon, in 
addition to the up-down mass splitting.  
\end{itemize}
Results on NLO mixing effects induced by quark mass splitting already
appear in Refs.~\cite{gl85a,Ecker:1999kr,Wolfe:2000rf}, while
electromagnetically induced effects were considered in
\cite{Neufeld:1995mu,Cirigliano:2001mk}. 
Here we follow the formalism outlined in Ref.~\cite{Ecker:1999kr}, 
treating strong and electromagnetic effects simultaneously.   
In the LO mass eigenfield basis, the NLO inverse propagator 
(an eight-by-eight matrix) can be written as follows: 
\bea
\widehat{\Delta} (q^2) ^{-1} &=& q^2\, {\bf 1}  - \widehat{M}^2 - 
\widehat{\Pi} 
(q^2)\\
\widehat{\Pi} (q^2) &=& \widehat{C}\, q^2  + \widehat{D} \ , \nn
\eea
where $\widehat{M}^2$ is the diagonal LO mass matrix and $\widehat{C}$, 
$\widehat{D}$
are symmetric matrices, which are diagonal except for their 
restriction to the $(\pi^0, \eta)$  subspace. 
The relation between the LO and NLO mass eigenfields 
(collected in a vector $\phi_a$) can be  summarized as follows: 
\beq
\phi_{\rm LO} = \left( {\bf 1}  + \frac{\widehat{C}}{2} +  \widehat{W} 
\right) \, 
\phi_{\rm NLO} \ , 
\label{eq:NLOmixing1}
\eeq
where $\widehat{W}$ is an antisymmetric matrix, non-vanishing only in the
$(\pi^0, \eta)$ subspace.
Except for the reduction to this subspace, Eq.~(\ref{eq:NLOmixing1}) 
is just the familiar field renormalization, with wave function 
renormalization given by $Z_{i} = 1 + \widehat{C}_{ii}$.
Focusing on the $(\pi^0, \eta)$ sector, we note that $\widehat{W}$ is
characterized by a single entry called
$\varepsilon^{(4)}$ \cite{Ecker:1999kr}. This quantity is UV finite 
and represents a natural 
generalization to $\cO(p^4)$ of the tree-level mixing angle 
$\varepsilon^{(2)}$. Explicitly, the relation between the 
$(\pi^0, \eta)$ mass eigenfields at LO and NLO is  
given by
{\renewcommand{\arraystretch}{1.5}
\beq
\left( \begin{array}{c} \pi^0 \\ \eta  \end{array} \right)_{\rm LO} = 
\left( 
{\renewcommand{\arraystretch}{2.1}
\begin{array}{cc} 1 + \displaystyle\frac{\widehat{C}_{\pi^0 \pi^0}}{2} 
  &  - \varepsilon^{(4)} +  \displaystyle\frac{\widehat{C}_{\eta 
\pi^0}}{2}  \\ 
\varepsilon^{(4)} + \displaystyle\frac{ \widehat{C}_{\eta \pi^0}}{2} 
 & 1  + \displaystyle\frac{\widehat{C}_{\eta \eta}}{2}   \end{array} 
 }\right) \, 
\left( \begin{array}{c} \pi^0 \\ \eta  \end{array} \right)_{\rm NLO} ~.  
\label{eq:NLOmixing2}
\eeq
}
Eqs.~(\ref{eq:LOmixing}) and (\ref{eq:NLOmixing2}) give the full relation 
between the original fields $(\pi_3, \pi_8)$ and the NLO mass eigenfields.
We do not report here the factors $\widehat{C}_{ab}$, as they are UV 
divergent and make sense only in combination with other terms 
in the full amplitudes. We do report, however, the expression 
for $\varepsilon^{(4)}$ because the replacement 
$\varepsilon^{(2)} \rightarrow \varepsilon^{(2)} +  \varepsilon^{(4)}$
in Eqs.~(\ref{eq:LOamp1},\ref{eq:LOamp2}) gives rise to 
a well defined (UV finite) subset of the NLO corrections. 
Breaking up $\varepsilon^{(4)}$ into contributions from 
strong (S) and electromagnetic (EM) isospin breaking, one gets 
\bea
\varepsilon^{(4)} &=& \varepsilon^{(4)}_{\rm S} + 
\varepsilon^{(4)}_{\rm EM} \nl
\varepsilon^{(4)}_{\rm S} &=& 
- \frac{2 \, \varepsilon^{(2)}}{3 (4 \pi F)^2 (M_{\eta}^2 - 
M_{\pi}^2)} 
  \bigg\{ (4 \pi)^2 \, 64 \left[3 L_7 + 
L_8^r (\nu_\chi) \right] 
(M_K^2 - M_\pi^2)^2 
\nl
&& {} -  M_\eta^2 (M_K^2 - M_\pi^2) \log \frac{M_\eta^2}{\nu_\chi^2}
 +  M_\pi^2 (M_K^2 - 3 M_\pi^2) \log \frac{M_\pi^2}{\nu_\chi^2}  \nn \\
&& {} - 2 M_K^2 (M_K^2 - 2 M_\pi^2) \log \frac{M_K^2}{\nu_\chi^2} 
- 2 M_K^2 (M_K^2 - M_\pi^2) \bigg\} \\
\varepsilon^{(4)}_{\rm EM} &=&   
\frac{2 \, \sqrt{3} \, \alpha }{108 \, \pi \, (M_\eta^2 
-M_\pi^2)}  
  \bigg\{ 
-  9 M_K^2   Z \left(\log \frac{M_K^2}{\nu_\chi^2} + 1 \right) \nl
&& {} + 2  M_K^2  (4 \pi)^2 \Big[ 2 U_{2}^{r} (\nu_\chi) + 
3 U_{3}^{r} (\nu_\chi) \Big] \nl
&& {} 
+  M_\pi^2  (4 \pi)^2 \Big[ 2 U_{2}^{r} (\nu_\chi) + 
3 U_{3}^{r} (\nu_\chi) - 6 U_{4}^{r} (\nu_\chi) \Big]  \bigg\}
\nn ~. 
\eea
The electromagnetic LECs $U_i$ are linear combinations of the $K_i$
(defined in Sec.~\ref{subsec:Ki}).

\subsection{Remaining NLO contributions: a guided tour}

\begin{figure}[tb]
\centering
\epsfig{figure=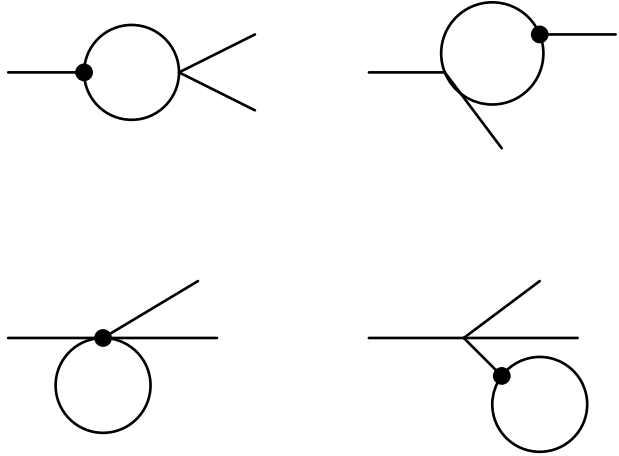,height=6cm}
\caption{Topologies for purely mesonic loop diagrams contributing to 
 $K \to \pi \pi$: the filled circles indicate $\Delta S=1$ vertices of
 lowest order. Wave function renormalization diagrams are not shown.}
\hfill 
\label{fig:meson}
\end{figure}

\begin{figure}[bht]
\centering
\epsfig{figure=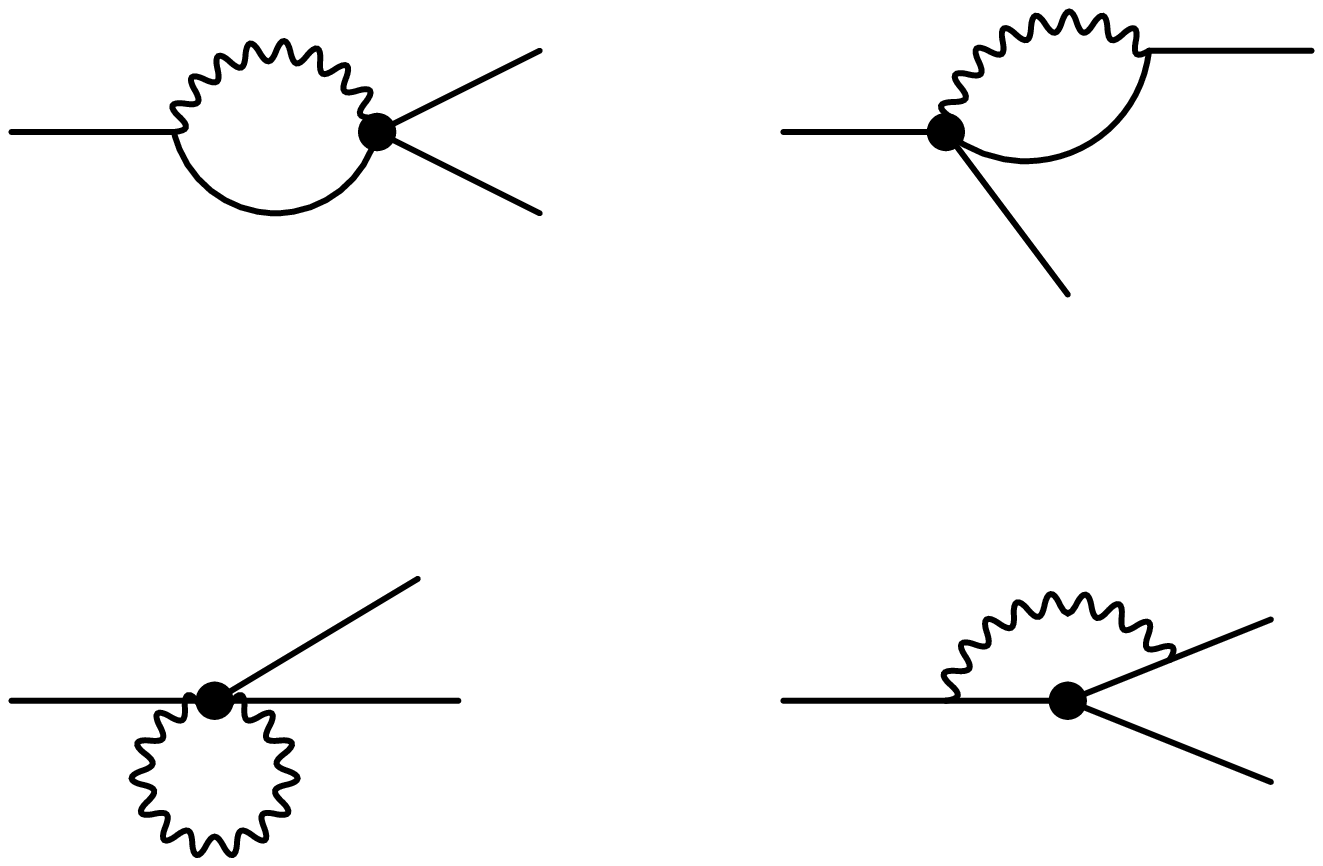,height=6cm}
\caption{Topologies for meson-photon loop diagrams contributing to 
 $K \to \pi \pi$: the filled circles indicate $\Delta S=1$ octet 
vertices of lowest order.}\hfill
\label{fig:photon}
\end{figure}

Having dealt with the propagator corrections in the previous 
section, we now describe the remaining  contributions 
to the $K \rightarrow \pi \pi$ amplitudes at NLO, starting 
with the one-loop terms.
There are two main classes of contributions: loops involving 
only pseudoscalar mesons (Fig.~\ref{fig:meson}) 
and loops involving virtual photons (Fig.~\ref{fig:photon}). 
In the isospin limit, contributions to the amplitudes arise  from
the topologies of Fig.~\ref{fig:meson}, by inserting the LO weak vertices 
proportional to $G_8$ or $G_{27}$ in the Lagrangians 
(\ref{eq:Lweak}) and (\ref{eq:Lelweak}).
Given the large suppression of $G_{27}/G_{8}$, 
we consider in this work only isospin breaking effects generated through
the octet component of the effective Lagrangian. We are therefore interested
in the terms proportional to $\varepsilon^{(2)} G_{8}$ (strong
isospin breaking) and $e^2 G_{8}$ (electromagnetic isospin breaking).

Strong isospin breaking terms ($\varepsilon^{(2)} G_{8}$) at NLO 
come from several sources:    
\begin{itemize} 
\item Explicit terms $\sim (m_u - m_d)$ in the strong vertices 
of Fig.~\ref{fig:meson}, obtained by expressing 
${\cal L}_{\rm strong}$ in terms of the LO mass eigenfields. 
\item Mass corrections in the internal propagators, 
for which we use the LO diagonal form (and the corresponding 
mass relations of Eq.~(\ref{eq:treemass})). 
\item Mass corrections arising when external momenta are taken 
on-shell (using again Eq.~(\ref{eq:treemass})).  
\end{itemize} 
The combination of these effects leads in principle to 
non-linear contributions in the isospin breaking parameter. 
We have chosen to expand the final expressions for the amplitudes 
to first order in $\varepsilon^{(2)}$ .  

Electromagnetic isospin breaking terms ($e^2 G_{8}$) at NLO  
can be naturally divided into three categories: 
\begin{itemize}
\item $e^2 Z G_{8}$: these arise exactly in the same way as 
 the strong isospin breaking terms (see discussion above).  
\item $e^2 G_8 g_{\rm ewk}$: these arise from insertions of the 
$g_{\rm ewk}$ vertices of ${\cal L}_{\rm EW}$ in the 
topologies of Fig.~\ref{fig:meson}, keeping all other 
contributions (masses and strong vertices) in the isospin limit. 
\item $e^2 G_{8}$: these arise from the photonic diagrams of 
Fig.~\ref{fig:photon}, using the LO weak vertices of 
${\cal L}_{\rm weak}$ proportional to $G_{8}$. 
This class of contributions to $\cA_{+-}$ is infrared divergent. 
We regulate the infrared divergence by means of a fictitious photon 
mass $M_\gamma$. The cancellation of infrared divergences 
only happens when one considers an inclusive sum of 
$K \rightarrow \pi \pi$ and $K \rightarrow \pi \pi \gamma $
decay rates and we postpone details on this point to 
Sec.~\ref{sec:pheno1}. At this stage, we split 
the photonic correction to $\cA_{+-}$  into an ``infrared component'' 
$\cA_{+-}^{\rm IR}(M_\gamma)$ (to be treated in combination with 
real photons) and a structure dependent part $\cA_{+-}^{(\gamma)}$, 
which is infrared finite and has to be used 
together with the non-photonic amplitudes in Eq.~(\ref{eq:intro1}).
Clearly, an arbitrary choice appears here as one can shift infrared
finite terms from $\cA_{+-}^{(\gamma)}$ to $\cA_{+-}^{\rm
IR}(M_\gamma)$.  This also implies that the isospin amplitudes 
all depend on this choice. The observables, however, are only  
affected by this ambiguity at order $\alpha^2$. 
$\cA_{+-}^{\rm IR}(M_\gamma)$ has the following structure:
\beq
\cA_{+-}^{\rm IR}(M_\gamma) = \sqrt{2}\, G_{8} F \left(M_K^2 - 
M_\pi^2 \right) \, \alpha \, B_{+-} (M_\gamma) \ ,  
\label{Eq:tour1}
\eeq
in terms of the function $B_{+-} (M_\gamma)$  reported in 
App.~\ref{app:loops}. 
\end{itemize}

\begin{figure}[htb]
\centering
\epsfig{figure=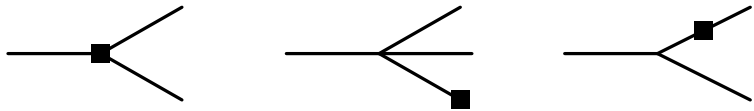,height=1.4cm}
\caption{Diagrams for NLO local contributions: the filled square
  denotes a NLO vertex.}
\hfill 
\label{fig:NLOct}
\end{figure}

This concludes our description of one-loop contributions to $K
\rightarrow \pi \pi$ amplitudes. The NLO local contributions arise
from tree-level graphs with insertions of one NLO vertex  
and any number of LO vertices, 
according to the topologies depicted in Fig.~\ref{fig:NLOct}.
 
\subsection{Structure of the amplitudes}

Having identified the various diagrammatic contributions to the
physical amplitudes, we now introduce a general parametrization that
explicitly separates isospin conserving and
isospin breaking parts and allows to keep track of the various
sources of isospin breaking. 
Let $n$ be the label for any amplitude. Then, including the leading 
isospin breaking corrections (proportional to $G_8$), one has:
\bea
\cA_n &=& G_{27} \, F_\pi \, \bigg( M_K^2 - M_\pi^2 \bigg) \,  \cA_{n}^{(27)} 
 \label{eq:structure1}  \\
&+& 
G_{8} \, F_\pi \,  \Bigg\{ 
\left( M_K^2 - M_\pi^2 \right)  
\bigg[ \cA_{n}^{(8)} +  \varepsilon^{(2)} \, \cA_{n}^{(\varepsilon)} \bigg] 
-   e^2 \, F_\pi^2 \,  \bigg[ \cA_{n}^{(\gamma)} + Z \,  \cA_{n}^{(Z)} + 
 g_{\rm ewk} \,  \cA_{n}^{(g)} \bigg]  \Bigg\}~. \nn
\eea 
The meaning of the amplitudes $\cA_n^{(X)}$ can be inferred from the
superscript $X$.  $\cA_{n}^{(8)}, \cA_{n}^{(27)}$
represent the octet and 27-plet amplitudes in the isospin limit.
$\cA_{n}^{(\varepsilon)}$ represents the effect of strong
isospin breaking, while the electromagnetic contribution is split
into a part induced by photon loops $\cA_{n}^{(\gamma)}$ and the parts
induced by insertions of $Z$ and $g_{\rm ewk}$ vertices 
($\cA_{n}^{(Z)}$ and $\cA_{n}^{(g)}$, respectively).

At the order we are working, each of the amplitudes $\cA_n^{(X)}$ 
can be decomposed as follows: 
\beq
\cA_{n}^{(X)} = \left\{ \begin{array}{lll}
a_n^{(X)} \, \left[ 1  + \Delta_{L} \cA_{n}^{(X)} + \Delta_{C} \cA_{n}^{(X)}
\right]  & \ \ \  \mbox{if} \ \ \  &  a_n^{(X)} \neq 0 \\
  &   &     \label{eq:structure2}  \\
\qquad \quad \quad \ \Delta_{L} \cA_{n}^{(X)} + \Delta_{C} \cA_{n}^{(X)} &
\ \ \ \mbox{if} \ \ \ & 
 a_n^{(X)} = 0
\end{array}
\right.
\eeq
with
\bea
a_n^{(X)}  & \mbox{ : } &  \mbox{ LO contribution 
[Eqs.~(\ref{eq:LOamp1},\ref{eq:LOamp2})] } 
%F \rightarrow  F_\pi   
\nn \\
\Delta_L \cA_n^{(X)} & \mbox{ : } &  
\mbox{ NLO loop correction} \nn \\
\Delta_C \cA_n^{(X)} &  \mbox{ : } & 
\mbox{ NLO local correction} ~. \nn 
\eea

The amplitudes $a_{n}^{(X)}$, $\Delta_L \cA_n^{(X)}$ and $\Delta_C
\cA_{n}^{(X)}$ are dimensionless and we have chosen to normalize the
NLO contributions to the LO contributions whenever possible.
Moreover, in Eq.~(\ref{eq:structure1}) we have traded the constant
$F$ for $F_\pi$, the physical pion decay constant at NLO. The
relation between the two is given explicitly by \cite{gl85a,
Knecht:1999ag}
\bea
F &=& F_\pi \;\Bigg\{ 1 - \frac{4}{F^2} \Bigg[ L_{4}^{r} (\nu_\chi)  
\left( M_\pi^2 + 2 M_K^2 \right) + L_{5}^{r}(\nu_\chi)  M_\pi^2 \Bigg] 
%\right.
\nn \\ 
&+ & 
%\left.  
\frac{1}{2 (4 \pi)^2 F^2} \left[ 2 M_\pi^2 \log \frac{M_\pi^2}
{\nu_\chi^2} + 
M_K^2 \log \frac{M_K^2}{\nu_\chi^2} \right] 
%\right.
\nn \\
&+ & 
%\left. 
   \frac{2 \varepsilon^{(2)}}{\sqrt{3}} \left( M_K^2 - M_\pi^2 
\right)  \, \left[ \frac{8 L_4^r (\nu_\chi)  }{F^2} - 
\frac{1}{2 (4 \pi)^2 F^2} 
\left( 1 +  \log \frac{M_K^2}{\nu_\chi^2} \right)  \right] 
\Bigg\} \ .
\eea 

Both  $\Delta_L \cA_n^{(X)}$ and $\Delta_C \cA_{n}^{(X)}$ 
individually are UV divergent and scale dependent. Only in their  
sum the UV divergence and the scale dependence cancel, 
providing a valuable check on the calculation.  
The explicit form of the various loop contributions is given in 
App.~\ref{app:loops} while the local amplitudes 
are reported in the next subsection. 

\subsection{Local amplitudes}

The NLO $ K \rightarrow \pi \pi $ local amplitudes receive
contributions from the NLO couplings $L_i$, $N_i$, $D_i$, 
$K_i$, $Z_i$ in the effective Lagrangians of Sec.~\ref{sec:general}. 
Following Ref.~\cite{PPS01}, it is convenient 
to define the combinations
\bea
\widetilde{\Delta}_C &=&
 - \frac{M_K^2}{F^2} \left(  4 \, L_5^r  +  32 \, L_4^r \right) 
-  \frac{M_\pi^2}{F^2} \left( 12 \, L_5^r  + 16 \, L_4^r \right)  \nn \\
\widetilde{\Delta}_C^{({\rm ew})} &=&
- \frac{M_K^2}{F^2} \left(  4 \, L_5^r  +  48 \, L_4^r \right)
-  \frac{M_\pi^2}{F^2} \left( 20 \, L_5^r  + 24 \, L_4^r \right) \ .
\eea
In terms of the quantities defined above, the finite parts of the NLO
local amplitudes have the form reported below.  In this section we use
the notation $D_i, N_i, Z_i$ as a shorthand for the ratios of NLO to
LO chiral couplings $(g_8 D_i)/g_8, (g_8 N_i)/g_8, (g_8 Z_i)/g_8$.
\vspace{1.0cm}

\begin{center}
%\subsubsection{
{\bf $\Delta I = 1/2$ amplitudes}
%}
\end{center}
\bea
\Delta_C \cA_{1/2}^{(27)} & =& \widetilde{\Delta}_C
+ {M_K^2\over F^2}\,\left ( D_4^r  - D_5^r - 9 \, D_6^r + 4 \, D_7^r  \right)
\nn \\  
& + &  {2 \, M_\pi^2 \over F^2} \,\left( 
-6 \, D^r_1 - 2\, D^r_2 + 2\, D^r_4 + 6 \, D^r_6 + D^r_7 \right) 
\nn \\
\Delta_C \cA_{1/2}^{(8)} & = &  \widetilde{\Delta}_C
- \frac{2\, M_K^2}{F^2} \,\left(  -  N_5^{r} +
       2\, N_7^{r} - 2\, N_8^{r} -  N_9^{r}    \right)
\nn  \\
& - &  \frac{2\,M_\pi^2}{F^2} \,\left(  
       - 2\, N_5^{r} - 4\, N_7^{r} -  N_8^{r}
      + 2\, N_{10}^{r} + 4\, N_{11}^{r} + 2\, N_{12}^{r}
       \right)  
\nn \\ 
%\eea
%\bea
\Delta_C \cA_{1/2}^{(\varepsilon)} & = &  \widetilde{\Delta}_C  - 
\frac{ \left( M_K^2 -  M_\pi^2 \right)}{F^2} \, 
\left( 96 L_4^r +  32\, \left( 3 L_7^r + L_8^r \right) \right)
\nn \\
&-&  \frac{2\, M_K^2}{F^2} \,\left(
	  N_5^{r}
       + 6\, N_6^{r} + 12\, N_7^{r} - 8\, N_8^{r}
       -  N_9^{r}
	- 4\, N_{10}^{r}
       - 8\, N_{12}^{r} - 12\, N_{13}^{r} 
 \right)
\nn  \\
&+&  \frac{2\, M_\pi^2}{F^2} \,\left( 
 14\, N_5^{r} + 6\, N_6^{r}
       + 24\, N_7^{r} - 5\, N_8^{r} 
       - 26\, N_{10}^{r} -
       24\, N_{11}^{r} - 10\, N_{12}^{r} -
       12\, N_{13}^{r} 
\right)
\nn \\ 
%\eea
%\bea
\Delta_C \cA_{1/2}^{(Z)} & = &  \widetilde{\Delta}_C^{({\rm ew})} 
- \frac{4\, M_K^2}{F^2} \,\left( 2\, N_7^{r} - N_8^{r} -  N_9^{r} \right)
+ \frac{2\, M_\pi^2}{F^2} \,
\left(  2\, N_5^{r} + 4\, N_7^{r} +  N_8^{r} \right)
\nn \\ 
%\eea
%\bea
\Delta_C \cA_{1/2}^{(g)} & = &  \widetilde{\Delta}_C^{({\rm ew})}  
\nn \\
%\eea       
%\bea
\Delta_C \cA_{1/2}^{(\gamma)} & = &    
\frac{ 2 \sqrt{2}}{3} \, \left. \Bigg[
 \frac{M_K^2}{F^2} \left(   6\, U_1^r + 4\, U_2^r +  U_3^r \right) 
-  \frac{M_\pi^2}{6 \, F^2} \left( 36\, U_1^r + 22\, U_2^r + 3\, U_3^r +
 2\, U_4^r \right)
\right. \nn \\
&+&  \left. \frac{ \left( M_K^2 - M_\pi^2 \right)}{ 6 \, F^2}  \, \left(
     - 8\, Z_3^r + 24\, Z_4^r - 9\, Z_5^r
       - 6\, Z_7^r + 3\, Z_8^r + 3\, Z_9^r
     + 2\, Z_{10}^r - 2\, Z_{11}^r - 2\, Z_{12}^r
         \right) \right.   
\nn \\
 &+& \left. \frac{ M_K^2}{ F^2} \, \left(2\, Z_1^r +  4\, Z_2^r   \right) 
+ \frac{ M_\pi^2}{F^2} \, \left( 4\, Z_1^r + 2\, Z_2^r  -  Z_6^r \right) 
\right.   \Bigg]
\label{eq:I1/2} 
\eea
\\[0.2cm]

\begin{center}
%\subsubsection{
{\bf $\Delta I = 3/2$ amplitudes}
%}
\end{center}
\bea
\Delta_C \cA_{3/2}^{(27)} &  =&  \widetilde{\Delta}_C
+{M_K^2 \over F^2}\,\left( D^r_4 - D^r_5 + 4 \, D^r_7 \right) 
+ { 2 \, M_\pi^2 \over F^2} \,\left ( -2\, D^r_2 + 2\, D^r_4 + D^r_7 \right)
\nn \\
\Delta_C \cA_{3/2}^{(\varepsilon)} & = &  \widetilde{\Delta}_C  - 
\frac{ 32 \,  \left( M_K^2 -  M_\pi^2 \right)}{F^2} \, 
  \left( 3 L_7^r + L_8^r \right)
\nn \\
&-&  \frac{2\, M_K^2}{F^2}\, \left( 
	 N_5^{r} + 6\, N_6^{r} - 2\, N_8^{r} -  N_9^{r}
	- 4\, N_{10}^{r}
       - 8\, N_{12}^{r} - 12\, N_{13}^{r} 
       \right) 
\nn \\ 
&+&  \frac{2\, M_\pi^2}{F^2} \,\left(
	2\, N_5^{r} + 6\, N_6^{r} +  N_8^{r} 
        - 2\, N_{10}^{r} - 10\, N_{12}^{r}
       - 12\, N_{13}^{r} 
\right)
\nn \\
%\eea
%\bea
\Delta_C \cA_{3/2}^{(Z)} & = &  \widetilde{\Delta}_C^{({\rm ew})}
+  \frac{ M_K^2}{ 5  \, F^2}\, \left( 
	12\, N_5^{r} - 16\, N_7^{r} +
       20\, N_8^{r} + 8 \, N_9^{r} \right)
\nn \\ 
&+& \frac{ M_\pi^2}{5 \,  F^2} \,\left(  
	 8\, N_5^{r} + 16\, N_7^{r} + 10\, N_8^{r} +
        12\, N_9^{r} \right)
\nn \\
%\eea
%\bea
\Delta_C \cA_{3/2}^{(g)} & = &  \widetilde{\Delta}_C^{({\rm ew})}  
\nn \\
%\eea
%\bea
\Delta_C \cA_{3/2}^{(\gamma)} & = &    
\frac{2}{3} \,\left. \Bigg[
- \frac{M_K^2}{F^2} \, {4 \over 5}\,  U_3^r 
- \frac{ M_\pi^2}{F^2} \left( {2 \over 3} \, U_2^r + {1 \over 5} \, U_3^r 
- {2 \over 3} \, U_4^r \right)
\right. \nn \\
&+& \left. \frac{\left(M_K^2 - M_\pi^2 \right)}{3 \, F^2} \, 
\left( - 4\, Z_3^r + {24 \over 5}\, Z_4^r - 3 \, Z_8^r - 3 \, Z_9^r -  
2 \,  Z_{10}^r - {8 \over 5} \, Z_{11}^r - {8 \over 5} \, Z_{12}^r 
\right)  \right. \nn \\
&+& \left. 
\frac{M_K^2}{F^2} \,\left(  2\, Z_1^r + 4\, Z_2^r - Z_6^r  \right)
+ \frac{M_\pi^2}{F^2} \,\left( 4\, Z_1^r +  2\, Z_2^r  \right) 
\right. \Bigg]
\label{eq:I3/2}
\eea
\\[0.2cm]

\begin{center}
{\bf $\Delta I = 5/2$ amplitudes}
%}
\end{center}
\bea
\Delta_C \cA_{5/2}^{(Z)} & = &  \frac{4}{3} \, \left[  
 \frac{\left( M_K^2 - M_\pi^2 \right)}{ 5  \, F^2} \, 
\left(  - 12 \, N_5^{r} - 24 \, N_7^{r} +
    12 \,  N_9^{r} \right) \right]
\nn \\
\Delta_C \cA_{5/2}^{(\gamma)} & = &    \frac{2}{3} \, \left[  
  \frac{\left( M_K^2 - M_\pi^2 \right)}{ 15 \, F^2} \, 
\left( - 18 \,  U_3^{r} + 36 \, Z_4^{r} + 18 \,  Z_{11}^{r} + 18 \, 
Z_{12}^{r} \right) \right]
\label{eq:I5/2}
\eea

\section{LECs at leading order in $1/N_c$}
\label{sec:LECs}
\renewcommand{\theequation}{\arabic{section}.\arabic{equation}}
\setcounter{equation}{0}
Owing to the presence of very different mass scales 
($M_\pi < M_K < \Lambda_\chi \ll M_W$), the gluonic
corrections to the underlying flavour-changing transition
are amplified by large logarithms. 
The short-distance logarithmic corrections can be summed up
with the use of the operator product expansion \cite{WI:69}
and the renormalization group \cite{RGroup}, all the way down to
scales $\mu_{\rm SD} < m_c$.
One gets in this way an effective $\Delta S=1$ Lagrangian, defined in the
three-flavour theory \cite{GLAM:74,VZS:75,GW:79,NLO:93},
\beq\label{eq:Leff}
 {\cal L}_{\mathrm eff}^{\Delta S=1}= - \frac{G_F}{\sqrt{2}}
 V_{ud}^{\phantom{*}}\,V^*_{us}\,  \sum_i  C_i(\mu_{\rm SD}) 
\; Q_i (\mu_{\rm SD}) \; ,
\eeq
which is a sum of local four-fermion operators $Q_i$,
constructed with the light degrees of freedom ($m<\mu_{\rm SD}$), 
modulated by Wilson coefficients $C_i(\mu_{\rm SD})$ which are functions of the
heavy masses ($M>\mu_{\rm SD}$) and CKM parameters:
\bea
C_i (\mu_{\rm SD}) &=&  z_i (\mu_{\rm SD}) \ + \  \tau 
\, y_i (\mu_{\rm SD}) \\
\tau &=&  - \displaystyle\frac{V_{td}^{\phantom{*}}\,
V^*_{ts}}{V_{ud}^{\phantom{*}}\,V^*_{us}} \nn \ . 
\eea
The low-energy electroweak chiral Lagrangian arises from the bosonization
of the short-distance Lagrangian (\ref{eq:Leff}) 
below the chiral symmetry breaking scale $\Lambda_\chi$.
Chiral symmetry fixes the allowed operators, at a given order in momenta,
but the calculation of the corresponding CHPT couplings is a difficult
non-perturbative dynamical question, which
requires to perform the matching between the two effective field theories.

The $1/N_c$ expansion provides a systematic approximation scheme to
this problem. At leading order in $1/N_c$ the matching between the
three-flavour quark theory and CHPT can be done exactly because the
T-product of two colour-singlet quark currents factorizes.  Since
quark currents have well-known realizations in CHPT the
hadronization of the weak operators $Q_i$ can then be done in a 
straightforward way.  As a result, the electroweak chiral couplings
depend upon strong and electromagnetic low-energy constants of order
$p^2, p^4, p^6$ and $e^2 p^2$, respectively.

\subsection{Weak couplings of $\cO (G_F p^2), \cO (e^2 G_8 p^0)$}

At lowest-order [$\cO(G_F p^2)$, $\cO(e^2 G_8 p^0)$],
the chiral couplings of the nonleptonic electroweak 
Lagrangians~(\ref{eq:Lweak}) and (\ref{eq:Lelweak})
have the following large-$N_c$ values:
\bea
g_8^\infty\, &=&\, -{2\over 5}\,C_1(\mu_{\rm SD})+{3\over 5}\,
C_2(\mu_{\rm SD})+C_4(\mu_{\rm SD})
- 16\, L_5 \, B(\mu_{\rm SD})\, C_6(\mu_{\rm SD})\;  \nl
g_{27}^\infty\, &=&\, {3\over 5}\, [C_1(\mu_{\rm SD})+
C_2(\mu_{\rm SD})]\; \label{eq:c2} \\
(e^2 g_8\, g_{\rm ewk})^\infty\, &=& \, -3\, B(\mu_{\rm SD})\, 
C_8(\mu_{\rm SD}) 
- \frac{16}{3} \,  B(\mu_{\rm SD})\, C_6 (\mu_{\rm SD}) \, 
e^2 \, (K_9 - 2 K_{10}) \nn ~.
\eea

The operators $Q_i$ ($i\not=6,8$) factorize into products of 
left- and right-handed vector currents, 
which are renormalization-invariant quantities.
Thus, the large-$N_c$ factorization of these operators
does not generate any scale dependence. 
The only anomalous dimensions that survive for $N_c\to\infty$
are the ones corresponding to $Q_6$ and $Q_8$ \cite{BBG87}.
These operators  factorize into colour-singlet
scalar and pseudoscalar currents, which are $\mu_{\rm SD}$ dependent.
The CHPT evaluation of the scalar and pseudoscalar currents provides,
of course, the right $\mu_{\rm SD}$ dependence, since only physical observables
can be realized in the low-energy theory. What one actually
finds is the chiral realization of the renormalization-invariant
products $m_q \;\bar{q}(1,\g_5) q$. This generates the factors
\begin{eqnarray} 
B(\mu_{\rm SD})\,\equiv\, \left({B_0^2\over F^2}\right)^\infty\, &=&
 \left[{M_K^2\over (m_s + m_d)(\mu_{\rm SD})\, F_\pi}\right]^2 
\Biggl[ 1  -\,\fr{16 M_K^2}{F_\pi^2}\, (2 L_8-L_5)  \nl
&&  +\, {8 M_\pi^2\over F_\pi^2}\,  L_5\, +\, 
\fr{8 (2 M_K^2+M_\pi^2)}{F_\pi^2}\, (3 L_4-4 L_6)\Biggr]
\label{eq:B0_comp}  
\end{eqnarray} 
in Eq.~\eqn{eq:c2}, which exactly cancel 
\cite{BBG87,dR:89,PI:89,PR:91,JP:94}
the $\mu_{\rm SD}$ dependence of $C_{6,8}(\mu_{\rm SD})$ at large $N_c$.
There remains a dependence at next-to-leading order. 

Explicitly, the large-$N_c$ expressions imply\footnote{
According to the discussion presented in the following subsections, we  
use here $L_5^r (M_\rho) = (1.0 \pm 0.3) \cdot 10^{-3}$ 
and $(K_9^r  - 2 K_{10}^r) (M_\rho) = -(9.3 \pm 4.6) \cdot 10^{-3}$. 
}
\bea
g_8^\infty\, &=&\, \left(1.10 \pm 0.05_{ (\mu_{\rm SD}) }
 \pm 0.08_{ (L_5)}  
 \pm 0.05_{(m_s)}  %my add
 \right) \nl &&\mbox{} + \ 
\tau \, \left(0.55 \pm 0.15_{(\mu_{\rm SD})}  \pm 0.20_{ (L_5)} 
\,{{}^{+0.25}_{-0.16}}_{(m_s)}  %my add
\right)  \nonumber\\[10pt]
g_{27}^\infty\, &=&\, 0.46 \pm 0.01_{  (\mu_{\rm SD})}    
\label{eq:gewk} \\[10pt]
(g_8\ g_{\rm ewk})^\infty\, &=&\, \left(-1.37  \pm 0.86_{ (\mu_{\rm SD})}
\pm 0.25_{ (K_i)} 
\,{{}^{+0.57}_{-0.35}}_{(m_s)}  %my add
\right) \nl &&\mbox{} - \ 
\tau \, \left(21.7 \pm 4.5_{ (\mu_{\rm SD})} \pm  1.0_{(K_i)} 
\,{{}^{+9.1}_{-5.6}}_{(m_s)}  %my add
\right) \ ,\nn
\eea
where the first uncertainty has been estimated by varying the
renormalization scale $\mu_{\rm SD}$ between 0.77 and 1.3 GeV, 
the second one reflects the error on 
the strong LECs of order $p^4$ and $e^2 p^2$,
and the third indicates   
the uncertainty induced by $m_s$ \cite{ms} which has been taken in the 
range \cite{PPS01} $(m_s+m_d)(\mu_{\rm SD}=1 {\rm GeV})= (156\pm 25)$~MeV.
While the CP-odd component of $g_{\rm ewk}$ is 
dominated by the electroweak penguin contribution 
(proportional to $\tau \, y_8 (\mu_{\rm SD})$), 
the CP-even part receives contributions of similar size from 
both strong ($Q_6$) and electroweak ($Q_8$) penguin operators. 
Its large uncertainty within this approach reflects the GIM mechanism
($z_8 (\mu_{\rm SD} >  m_c)=0 $).  
For the CP-even component, there exists an independent estimate,  
consistent with the one given here within the large uncertainties:
\begin{equation} 
\frac{  {\rm Re} \, (g_8 g_{\rm ewk}) }{ {\rm Re} \, g_8} 
= \left\{ \begin{array}{ll}
-0.99 \pm 0.30 \qquad \qquad \qquad \qquad  \qquad \cite{Cirigliano:1999hj} \\
-1.24 \pm  0.77_{(\mu_{\rm SD}) } \pm 0.40_{(L_5,K_i)}
\qquad \mbox{Eq.~(\ref{eq:gewk})}
\end{array} \right. ~.
\end{equation} 
In this work we shall always use the latter value, in order 
to perform a consistent analysis at leading order in $1/N_c$. 

Finally, the large-$N_c$ matching also produces the so-called weak
mass term (see Appendix~\ref{app:NLOLag} for notations):
\beq
\cL_{\rm wmt} =  G_8 '  F^4  \, \langle \lambda \,  \chi_+^{U} 
\rangle + \ {\rm h.c.} \ ,  
\eeq
with 
\bea
G_{8} ' & = & -{G_F \over \sqrt{2}}\,  V_{ud}^{\phantom{*}} V_{us}^*  
\  g_{8} '  \   \nn  \\
(g_8 ')^\infty\, &=&\, 
- 16\, \left( L_8 + \frac{1}{2} \, H_2 \right) \, 
B(\mu_{\rm SD})\, C_6(\mu_{\rm SD}) \ . 
\eea
We eliminate this term with an appropriate field redefinition
\cite{Bernard,CreLeu,kmw90}, of the form 
\beq 
U \ \rightarrow  \ e^{i \alpha} \, U \, e^{i \beta} \ , 
\eeq
where the chiral rotation parameters ($\alpha$ and $\beta$) are 
proportional to $G_8 '$.   
When applied to the strong effective Lagrangians of order $p^4$ and
$e^2 p^2$, the above redefinition generates monomials of the NLO
Lagrangians of order $G_8 p^4$ and $e^2 G_8 p^2$.  
The corresponding contributions to the couplings $g_8 N_i$ (of the
form $L_n \times (L_8 \,  + \, 1/2 \,  H_2)$) and $g_8 Z_i$ (of the form $K_n
\times (L_8 \,  +  \, 1/2  \, H_2)$) need to be added to the results obtained 
by direct matching at large-$N_c$.   
The complete results (reported in the next section)
are independent of the unphysical LEC $H_2$ of $\cO(p^4)$ \cite{gl85a}.

\subsection{Weak couplings of $\cO (G_F p^4), \cO (e^2 G_8 p^2) $}

The large-$N_c$ matching at the next-to-leading chiral order fixes
the couplings $G_8 N_i$, $G_{27} D_i$ and $G_8 Z_i$ of the 
nonleptonic weak and electroweak
Lagrangians (\ref{eq:Lweak}) and (\ref{eq:Lelweak}).  The operators
$Q_3$ and $Q_5$ start to contribute at $\cO(G_F p^4)$, while the
electroweak penguin operators $Q_7,\, Q_9$ and $Q_{10}$ make their
first contributions at $\cO(e^2 G_8 p^2)$.  The  contributions
from the operator $Q_6$ at $\cO(G_8 p^4)$ involve the  strong 
CHPT Lagrangian of $\cO(p^6)$ \cite{bce99} (to avoid confusion with the
Wilson coefficients $C_i$, the corresponding
dimensionless couplings \cite{bce00} are denoted
here as $X_i$).  With the definitions
\bea
\widetilde{C}_1 (\mu_{\rm SD}) &=&  
%\left( 
-\frac{2}{5} C_1 (\mu_{\rm SD}) + 
\frac{3}{5} C_2 (\mu_{\rm SD}) + C_4 (\mu_{\rm SD}) 
%\right) 
\nn \\
\widetilde{C}_2 (\mu_{\rm SD}) &=&  
%\left( 
+\frac{3}{5} C_1 (\mu_{\rm SD}) 
-\frac{2}{5} C_2 (\mu_{\rm SD}) + C_3 (\mu_{\rm SD}) - C_5 (\mu_{\rm SD}) ~, 
%\right) 
\eea 
the non-vanishing couplings contributing to $K\to \pi\pi$
amplitudes are:
\bea
(g_{27}\, D_4) &=&  4\  L_5 \, g_{27}^\infty 
\nl
(g_8 \, N_5) &=& - 2 \, L_5 \, \widetilde{C}_1 (\mu_{\rm SD}) 
%\nn \\
%&+& 
+ C_6 (\mu_{\rm SD}) \, B(\mu_{\rm SD}) \left[ 
- 16 \,  X_{14}
+ 32 \,  X_{17}
- 24 \,  X_{38}
- 4 \,  X_{91}
\right] \nl
(g_8 \, N_6) &=& 4 \, L_5 \,   \widetilde{C}_1 (\mu_{\rm SD}) + 
C_6 (\mu_{\rm SD}) \, B(\mu_{\rm SD}) \left[ 
- 32 \,  X_{17}
- 32 \,  X_{18}
+ 32  \,  X_{37}
+ 16  \,  X_{38}
\right] \nl
(g_8 \, N_7) &=& 2 \, L_5 \,  \widetilde{C}_1 (\mu_{\rm SD}) + 
C_6 (\mu_{\rm SD}) \, B(\mu_{\rm SD}) \left[
- 32 \,  X_{16}
- 16 \,  X_{17}
+ 8  \,  X_{38} 
\right] \nl
(g_8 \, N_8) &=& 4 \, L_5 \,   \widetilde{C}_1 (\mu_{\rm SD}) + 
C_6 (\mu_{\rm SD}) \, B(\mu_{\rm SD}) \left[ 
- 16 \,  X_{15}
- 32 \,  X_{17}
+ 16 \,  X_{38}
\right]  \nl
(g_8 \, N_9) &=& 
C_6 (\mu_{\rm SD}) \, B(\mu_{\rm SD}) \left[ 
- 64 \, L_5 \,  L_8 
- 8 \,  X_{34}
+ 8  \,  X_{38}
+ 4 \,  X_{91}
\right]  \label{eq:NiNc} \\
(g_8 \, N_{10}) &=& 
C_6 (\mu_{\rm SD}) \, B(\mu_{\rm SD}) \left[ 
- 48 \,  X_{19}
- 8 \,  X_{38}
- 2 \,  X_{91}
- 4 \,  X_{94}
\right] \nl
(g_8 \, N_{11}) &=& 
C_6 (\mu_{\rm SD}) \, B(\mu_{\rm SD}) \left[ 
- 32 \,  X_{20}
+ 4 \,  X_{94}
\right] \nl
(g_8 \, N_{12}) &=& 
C_6 (\mu_{\rm SD}) \, B(\mu_{\rm SD}) \left[ 
 128 \, L_8 \, L_8 
%\right. 
%\nl
%& & \left. 
+ 16 \,  X_{12}
- 16 \,  X_{31}
+ 8 \,  X_{38}
- 2 \,  X_{91}
- 4 \,  X_{94}
\right] \nl
(g_8 \, N_{13}) &=& 
C_6 (\mu_{\rm SD}) \, B(\mu_{\rm SD}) \left[
256  \, L_7 \,  L_8 
%\right.
%\nl
%& & \left. 
- \frac{32}{3} \,  X_{12}
- 16 \,  X_{33}
+ 16 \,  X_{37}
+ \frac{4}{3} \,  X_{91}
+ 4 \,  X_{94}
\right] ~.
% \label{eq:NiNc} 
\nn 
\eea
Bosonization of the four-quark operators $Q_i$ in (\ref{eq:Leff}) 
leads to the following expressions\footnote{$Z_{13},Z_{14},Z_{15}$ 
do not contribute to $K  \to \pi\pi$ amplitudes.} for the LECs $Z_i$: 
\bea
(g_8 \, Z_{1}) &=& 
\widetilde{C}_1 (\mu_{\rm SD}) \, 
\left( \frac{1}{3} \, K_{12} -  K_{13} \right) 
+ 64\, C_6 (\mu_{\rm SD}) \, B(\mu_{\rm SD}) \, L_{8} \, \bigg[  
-\frac{1}{3} \, K_{9}  + \frac{5}{3} \, K_{10}  + K_{11}   
\bigg] 
\nl
& & - 24 \, \frac{C_8 (\mu_{\rm SD}) \, B (\mu_{\rm SD})}{e^2} \, L_8 
\nl
(g_8 \, Z_{2}) &=& 
\frac{4}{3} \, \widetilde{C}_{1} (\mu_{\rm SD}) \, K_{13} 
- \frac{4}{3}\, 64\,  C_6 (\mu_{\rm SD}) \, B(\mu_{\rm SD}) \,  
%\bigg[  
\left( K_{10} + K_{11} \right) \, L_8
%\bigg]
\nl
(g_8 \, Z_{3}) &=&   \widetilde{C}_{1} (\mu_{\rm SD}) \, K_{13}  
- 64 \,  C_6 (\mu_{\rm SD}) \, B(\mu_{\rm SD}) \, 
\left( K_{10} + K_{11} \right)  \, L_8 
\nl
(g_8 \, Z_{4}) &=& 
-  \widetilde{C}_{1} (\mu_{\rm SD}) \, K_{13}   + 
 64\, C_6 (\mu_{\rm SD}) \, B(\mu_{\rm SD}) \, L_{8}\, 
\left(K_{10} + K_{11} \right)  
\nl
(g_8 \, Z_{5}) &=& 
\frac{4}{3} \, \widetilde{C}_1 (\mu_{\rm SD}) \, 
\left( 4 \, K_1 + 3 \, K_{5} + 3 \, K_{12} 
\right) 
- \frac{64}{3} \,  C_6 (\mu_{\rm SD}) \, B(\mu_{\rm SD}) \, 
\left( 2 K_{7} + K_{9} \right)  \, L_5 
+ \frac{C_{10} (\mu_{\rm SD})}{e^2}
\nl
(g_8 \, Z_{6}) &=& 
\widetilde{C}_1 (\mu_{\rm SD}) \,  
\left( -\frac{2}{3} \,  ( K_{5} + K_{6} )  + 2  \, (K_{12} + K_{13}) \right)
 \nl
& & - \frac{32}{3} \,  C_6 (\mu_{\rm SD}) \, B(\mu_{\rm SD}) \, 
\left( K_{9} + K_{10} + 3 K_{11} \right)  \, L_5
%\nn \\
%& &  
- 12 \, \frac{C_8 (\mu_{\rm SD}) \, B (\mu_{\rm SD})}{e^2} \, L_5 
\label{ZiNc} \\
(g_8 \, Z_{7}) &=& 
\widetilde{C}_1 (\mu_{\rm SD}) \,  
\left( 8 \, K_2 + 6 \, K_{6}  - 4 \, K_{13} \right) 
- 32 \,  C_6 (\mu_{\rm SD}) \, B(\mu_{\rm SD}) \, 
\left(2 K_{8} + K_{10} + K_{11} \right)  \, L_5 
\nl
(g_8 \, Z_{8}) &=& 
 \widetilde{C}_1 (\mu_{\rm SD}) \, 
\left( \frac{8}{3} \, K_{3} + 4 \, K_{12}  \right) + 
\frac{4}{3} \left( \widetilde{C}_1 (\mu_{\rm SD}) +  
\widetilde{C}_2 (\mu_{\rm SD}) \right) \, K_5 
+ \frac{3}{2 e^2} 
%\frac{
 \left( C_{9} (\mu_{\rm SD}) + C_{10} (\mu_{\rm SD}) \right) 
%}{e^2}
\nl
(g_8 \, Z_{9}) &=& 
- \frac{4}{3} \,  \widetilde{C}_1 (\mu_{\rm SD}) \, \left( K_4 + K_{12} + 
K_{13} \right) + 
\frac{4}{3} \left( \widetilde{C}_1 (\mu_{\rm SD}) +  
\widetilde{C}_2 (\mu_{\rm SD}) \right) \, K_5 
- \frac{3}{2} \frac{C_{7} (\mu_{\rm SD})}{e^2}
\nn \\
(g_8 \, Z_{10}) &=& 
- 2 \ \widetilde{C}_1 (\mu_{\rm SD}) \, K_{13} + 
4 \, \left( \widetilde{C}_1 (\mu_{\rm SD}) +  \widetilde{C}_2 (\mu_{\rm 
SD}) 
\right) \, K_6 
\nl
(g_8 \, Z_{11}) &=& 
2 \,  \widetilde{C}_1 (\mu_{\rm SD}) \, \left( K_4 + K_{13} \right) 
\nl
(g_8 \, Z_{12}) &=& 
- 4 \,  \widetilde{C}_1 (\mu_{\rm SD}) \, K_3 
\nl
(g_8 \, Z_{13}) &=& 
\frac{2}{3} \, \widetilde{C}_1 (\mu_{\rm SD}) \, \left( K_5 - K_{12} -
K_{13} \right)
+ \frac{64}{3} \,  C_6 (\mu_{\rm SD}) \, B(\mu_{\rm SD}) \, 
\left( K_{10} + K_{11} \right)  \, L_5 
\nl
(g_8 \, Z_{14}) &=& 
\widetilde{C}_1 (\mu_{\rm SD}) \, \left( - 2 \, K_{6} + 4 \, K_{13} 
\right)
- 32 \,  C_6 (\mu_{\rm SD}) \, B(\mu_{\rm SD}) \, 
\left( K_{10} + K_{11} \right)  \, L_5 
\nl
(g_8 \, Z_{15}) &=&
\widetilde{C}_1 (\mu_{\rm SD}) \, \left( - 2 \, K_{6} + 4 \, K_{13} 
\right)
- 32 \,  C_6 (\mu_{\rm SD}) \, B(\mu_{\rm SD}) \, 
\left( K_{10} + K_{11} \right)  \, L_5 ~. \nn 
\eea
We recall here that a matching ambiguity arises when working to
next-to-leading order in the chiral expansion and at leading order 
in $1/N_c$: we cannot identify at which value of the chiral
renormalization scale $\nu_\chi$ the large-$N_c$ estimates for 
the LECs apply. This turns out to be a major uncertainty in this 
approach. In order to account for this uncertainty, we
vary the chiral renormalization scale between 0.6 and 1 GeV.
The corresponding changes in the amplitudes are sub-leading effects in
$1/N_c$ and we take them as indication of the uncertainty associated
with working at leading order in $1/N_c$.

Finally, from the above expressions we see that in order to estimate
the weak NLO LECs at leading order in $1/N_c$, one requires as input
several combinations of strong LECs of order $p^4, p^6$ and $e^2
p^2$.  Below we summarize our knowledge of the needed parameters.

\subsection{Strong couplings of $\cO (p^4)$}
\label{sec:LECp4}

It is well known that the limit $N_c\to\infty$ provides an excellent 
description of the ${\cal O}(p^4)$ CHPT couplings at 
$\nu_\chi\sim M_\rho$
\cite{ap02}.
The leading-order contribution of $Q_6$ involves the LEC $L_5$.
The large-$N_c$ value
of this coupling can be estimated from 
resonance exchange \cite{egpr89}. Within the single-resonance
approximation (SRA) \cite{ap02,PPR:98}, taking $F=F_\pi$ and $M_S =
1.48$~GeV~\cite{cenp03a}, one finds $L_5^\infty = F^2/(4 M_S^2) = 1.0
\cdot 10^{-3}$. In our analysis we assign a $30 \% $ error to this
parameter so that the adopted range for $L_5$ reaches at
the upper end the value implied by the $p^4$ fit and at the lower end
the value obtained in the $p^6$ fit of Ref.~\cite{Amoros:2001cp}.
The combination $(2 L_8 - L_5)^\infty$ can also be
determined through resonance exchange. The only non-zero
contribution comes from the exchange of pseudoscalar
resonances. Within the SRA one gets \cite{ap02}:
\beq
(2 L_8 - L_5)^\infty \: = \: - {F^2\over 8 M_P^2}\:\approx\: 
- {F^2\over 16 M_S^2}\: = \: -\frac{1}{4}\, L_5^\infty
\: = \: -0.25\cdot 10^{-3} \, .
\eeq

The factor $B(\mu_{\rm SD})$ in (\ref{eq:B0_comp}) and the
$\cO(p^4)$ corrections $\widetilde\Delta_C$, $\widetilde\Delta_C^{({\rm 
ew})}$
and $\Delta_C\cA^{(\epsilon)}_n$
introduce additional dependences on the 
strong chiral couplings $L_4,L_6$ and $(3 L_7 + L_8)$. At large $N_c$, 
$L_{4}^\infty =L_{6}^\infty =0$ and
\beq\label{eq:L78}
(3 L_7 + L_8)^\infty \, = \,
- {(4 M_K^2 - 3 M_\eta^2 - M_\pi^2) F_\pi^2\over 24\, (M_\eta^2 - M_\pi^2)^2}
- \frac{1}{4}\, (2 L_8-L_5)^\infty
\, = \, - 0.15\cdot 10^{-3} \, .
\eeq
The same numerical estimate is obtained within the SRA, taking for
$L_7^\infty$ the known contribution from $\eta_1$ exchange \cite{egpr89}.

\subsection{Strong couplings of  $\cO (p^6)$} 

A systematic analysis of the LECs of $\cO (p^6)$ 
is still missing. Resonance contributions to some of the 
$X_i$ have been studied in Refs.~\cite{Amoros:2001cp,Knecht:2001xc}.    

Resonance dominance (that can be justified within large-$N_c$
QCD) implies that the  LECs of $\cO (p^6)$ occurring in the
bosonization of the penguin operator $Q_6$ are determined by scalar
exchange.  The mass splitting in the lightest scalar nonet strongly
influences those LECs.  

% We have adopted the following strategy
We have estimated the relevant $X_i$ with the scalar resonance
Lagrangian discussed in Ref.~\cite{cenp03a} (setting $g_4^S=0$).  The
relevant resonance parameters in the nonet limit are $c_d, c_m, M_S$,
and $e_m^S$, the latter governing the mass splitting within the scalar
nonet.  We use $c_m=c_d=F_\pi/2$, as determined from short-distance
constraints \cite{ap02}, and
\beq
M_S = 1.48 \ {\rm GeV}, \qquad e_m^S = 0.2 
\eeq
from a phenomenological analysis of mass
spectra \cite{cenp03a} (these numbers correspond to scenario A of
Ref.~\cite{cenp03a}).  Even within resonance saturation, this is not 
a complete calculation of the relevant LECs of $\cO(p^6)$ but we
expect it to capture the most significant physics. We refrain from 
reporting explicit numerics for the individual LECs here. Numerical
values for the relevant combinations are reported in the next section.

Finally, one can include nonet breaking effects within the framework 
of Ref.~\cite{cenp03a}. In the chiral resonance
Lagrangian, these effects are needed in order to understand the
scalar mass spectrum (the coupling $k_m^S$ and $\gamma_S$ of
Ref.~\cite{cenp03a}). Once the resonances are integrated out, nonet
breaking effects, sub-leading in $1/N_c$, appear in the $X_i$ and 
therefore in the weak LECs $g_8 N_i$.  Although this is
far from being a complete analysis of sub-leading corrections it
gives already an indication of their size.  For all the quantities of
physical interest, inclusion of $k_m^S$ and $\gamma_S$  produces
shifts within our estimate of $1/N_c$ corrections based on varying the
chiral renormalization scale (see discussion above).  

\subsection{Strong couplings of $\cO (e^2 p^2)$}

\label{subsec:Ki}
Four combinations of the $K_i$ appear directly in the 
local amplitudes $\Delta_C A_{n}^{(\gamma)}$ of $\cO(e^2 G_8 p^2)$: 
\bea
U_1 &=& K_1 + K_2      \nl
U_2 &=& K_5 + K_6      \nl
U_3 &=& K_4 - 2 K_3    \nl
U_4 &=& K_{9} + K_{10}   ~. 
\label{eq:uicomb}
\eea 
Within our large-$N_c$ estimates, also other combinations of $K_i$ 
appear through the couplings $g_8 g_{\rm ewk}$ and $g_8 Z_i$.  
The ones relevant for $K \rightarrow 
\pi \pi$ decays are $ K_{7},K_{8},K_{9},K_{10},K_{11},K_{12},K_{13}$.
It turns out that all the relevant combinations can be obtained 
from existing estimates \cite{BP:97,MOU:97}, which we now briefly review. 

The LECs $K_i$ can be expressed as convolutions of a QCD
correlation function with the electromagnetic propagator. Therefore,
their calculation involves an integration over the internal momenta of
the virtual photon, which makes reliable numerical estimates difficult
even at large $N_c$.  In contrast to the strong LECs
$L_i^r$, the dependence of the $U_i^r$ on the CHPT 
renormalization scale $\nu_\chi$ is already present at leading
order in $1/N_c$.  In addition, the $K_i$ depend also on the
short-distance QCD renormalization scale $\mu_{\rm SD}$ and on the 
gauge parameter $\xi$. Whenever numerical estimates are reported in
the following, they refer to the Feynman gauge ($\xi =1$) and 
$\mu_{\rm SD} = 1$GeV.

A first attempt to estimate the couplings $K_i$, using the
extended Nambu-Jona-Lasinio model at long distances, has found the
results \cite{BP:97}:
\bea\label{eq:BP:97}
\left[ 3\, U_1^r + U_2^r\right] (\nu_\chi =M_\rho)& = & 
(2.85\pm 2.50)\cdot 10^{-3}
\nl
\left[ U_1^r + 2\, K_{11}^r\right] (\nu_\chi =M_\rho)& = & 
-(2.5\pm 1.0)\cdot 10^{-3}
\nl
U_4^r(\nu_\chi =M_\rho)& = & (2.7\pm 1.0)\cdot 10^{-3}\nl
K_{10}^r(\nu_\chi =M_\rho)& = & (4.0\pm 1.5)\cdot 10^{-3}
\, .
\eea
The last two equations imply (adding the errors linearly):
\beq
K_{9}^r(\nu_\chi =M_\rho) =  -(1.3\pm 2.5)\cdot 10^{-3}\, .\nonumber
\eeq
Moreover, in the limit $N_c\to\infty$, one has the relation \cite{BP:97}
\beq 
U_3^r\: =\: 2\, U_1^r 
\eeq
and the couplings $K_7, K_8$ are subleading. We therefore take 
$K_{7,8}^r(M_\rho) = 0$.

The remaining couplings needed were obtained at large $N_c$ in
Ref.~\cite{MOU:97} through the evaluation of the relevant correlation
functions in terms of narrow hadronic resonances.  Within the SRA, one 
gets \cite{MOU:97}:
\bea\label{eq:MOU:97}
K_{11}^r &=& {1\over 8\, (4\pi)^2}\;\left\{
-(\xi+3)\,\ln{\left({\mu_{\rm SD}^2\over M_V^2}\right)}
+ \left(\xi-\frac{3}{2}\right)\,\ln{\left({\nu_\chi^2\over M_V^2}\right)}
-\xi -\frac{27}{4} + \frac{33}{2}\,\ln{2}\right\}\, \nl
K_{12}^r &=& {1\over 4\, (4\pi)^2}\;\left\{
(\xi- {3 \over 2})\,\ln{\left({\nu_\chi^2\over M_V^2}\right)}
- \xi \,\ln{\left({\mu_{\rm SD}^2\over M_V^2}\right)}
-\xi -\frac{17}{4} + \frac{9}{2}\,\ln{2}\right\} \nl
K_{13}^r &=& {3 \over 4\, (4\pi)^2}\;\left\{1 + ( 1 - \xi) \left[
{1 \over 12} + {1 \over 2} \, \ln\left({M_V^2 \over 2 \nu_\chi^2}
\right) \right] \right\} \ . 
\eea
Taking $\mu_{\rm SD}=1$ GeV, $\nu_\chi=M_V$ and $\xi=1$,  this gives
$K_{11}^r = 1.26\cdot 10^{-3}$,  $K_{12}^r = -4.2\cdot 10^{-3}$ and 
$K_{13}^r = 4.7 \cdot 10^{-3}$. 
Inserting the SRA  prediction from (\ref{eq:MOU:97}) 
into (\ref{eq:BP:97}), we get:
\beq
U_1^r(\nu_\chi =M_\rho)\, =\, -5.0\cdot 10^{-3}\; , \qquad\qquad
U_2^r(\nu_\chi =M_\rho)\, =\, 17.9\cdot 10^{-3}\; .
\eeq
A direct evaluation of $U_1^r$ and $U_2^r$ is in principle
possible within the SRA \cite{MOU:97}. However, it requires 
an analysis of resonance couplings beyond the known
results of Ref.~\cite{egpr89}.

\section{Numerical results}
% : (``Isospin'' basis)
\label{sec:num}
\renewcommand{\theequation}{\arabic{section}.\arabic{equation}}
\setcounter{equation}{0}

We are now in the position to quantify the size of NLO contributions
to the relevant isospin amplitudes, due to both chiral loops and local
couplings in the effective theory.  The master formulas for the
amplitudes at NLO are given in Eq.~(\ref{eq:structure1}) and
(\ref{eq:structure2}).  They are organized in such a way as to easily
identify the distinct sources of IB and to separate the LO from the
NLO contributions in the chiral expansion.  In
Tables~\ref{tab:A12ratios}, \ref{tab:A32ratios} and
\ref{tab:A52ratios} we report explicit results for the isospin
amplitudes $\cA_n$, $n=1/2,3/2,5/2$, quoting for each component the
following quantities: 
\begin{itemize}
\item The LO contributions  $a_{n}^{(X)}$.
\item The NLO loop corrections  $\Delta_{L} \cA_{n}^{(X)}$, 
consisting of absorptive and dispersive components.
The dispersive component depends on the chiral renormalization scale 
$\nu_\chi$ (fixed at $0.77$ GeV). 
\item The NLO local contributions to the CP-even and CP-odd
amplitudes, denoted respectively by $[\Delta_{C} \cA_{n}^{(X)}]^+$ and
$[\Delta_{C} \cA_{n}^{(X)}]^-$. 
Our estimates of $[\Delta_{C} \cA_{n}^{(X)}]^{\pm}$ at the scale 
$\nu_\chi = 0.77$ GeV are based on the leading $1/N_c$ approximation. 
We discuss below the uncertainty associated with this method. 
\end{itemize}
The definition of  $[\Delta_{C} \cA_{n}^{(X)}]^{\pm}$ is:
\begin{equation}
[\Delta_{C} \cA_{n}^{(X)}]^+ \ = \ \left\{ 
\begin{array}{ccc} 
\displaystyle\frac{ \real \left( G_{27} \ \Delta_{C} \cA_{n}^{(27)} 
\right)}{  \real ( G_{27} )}  &   &  X = 27    \\ 
\displaystyle\frac{ \real \left( G_8 g_{\rm ewk} \ \Delta_{C} \cA_{n}^{(g)}
 \right)}{ \real ( G_8 g_{\rm ewk} )}  &   &  X = g    \\ 
\displaystyle\frac{ \real  \left( G_8 \ \Delta_{C} \cA_{n}^{(X)} \right)}{ 
\real ( G_8 )}  &   &  X = 8, Z, \epsilon, \gamma     
\end{array}
\right. 
\end{equation}
\begin{equation}
[\Delta_{C} \cA_{n}^{(X)}]^- \ = \ \left\{ 
\begin{array}{ccc} 
\displaystyle\frac{ \imag \left( G_{27} \ \Delta_{C} \cA_{n}^{(27)} 
\right)}{ \imag ( G_{27} )}  &   &  X = 27    \\ 
\displaystyle\frac{ \imag \left( G_8 g_{\rm ewk} \ \Delta_{C} \cA_{n}^{(g)}
 \right)}{ \imag ( G_8 g_{\rm ewk} )}  &   &  X = g    \\ 
\displaystyle\frac{ \imag \left( G_8 \ \Delta_{C} \cA_{n}^{(X)} \right)}{ 
\imag ( G_8 )}  &   &  X = 8, Z, \epsilon, \gamma ~.    
\end{array}
\right. 
\end{equation}

The uncertainty in $[\Delta_C \cA_{n}^{(X)}]^{\pm}$ has two sources, related
to the procedure used to estimate the NLO local couplings (see
Sec.~\ref{sec:LECs}), and we quote them separately in the tables. 
The first one corresponds to the short-distance input, essentially  
the renormalization scale used to evaluate the Wilson coefficients.  
We estimate this uncertainty by varying the scale $\mu_{\rm SD}$ 
between 0.77 GeV and 1.3 GeV. 
The second one derives from working at leading order in the 
large-$N_c$ expansion. At this order, there is 
a matching ambiguity because we do not know at 
which value of the chiral scale the estimates apply. 
Therefore,  we vary the chiral renormalization scale $\nu_\chi$ 
between 0.6 and 1 GeV. 
The results show that the second uncertainty (long-distance) dominates 
over the first one (short-distance) in most cases.  Moreover, one 
should keep in mind that the errors quoted for the $[\Delta_C 
\cA_{n}^{(X)}]^{\pm}$ are strongly correlated. In phenomenological 
applications we shall take such correlations into account. 
\begin{table}[!ht] 
\begin{center}
\caption{Numerics for $\cA_{1/2}$:  $a_{1/2}^{(X)}$, 
 $\Delta_{L} \cA_{1/2}^{(X)} $, 
$\Delta_{C} \cA_{1/2}^{(X)} $ }  
\label{tab:A12ratios}
\vspace{0.3cm}
%\begin{tiny}
\begin{tabular}{|c|c|c|c|c|}\hl \hl 
\hl  
(X) & $a_{1/2}^{(X)}$ &
$\Delta_{L} \cA_{1/2}^{(X)} $ &
$[\Delta_{C} \cA_{1/2}^{(X)}]^+ $  & 
$[\Delta_{C} \cA_{1/2}^{(X)}]^- $ \\
\hl
\hl  
\hl
(27)  & 
${\sqrt{2} \over 9}$   & 
1.02 + 0.47 i    & 
0.01 $\pm$ 0 $\pm$ 0.60   & 
0.01 $\pm$ 0 $\pm$ 0.60   \\
\hl
(8)  &  
$\sqrt{2}$   & 
0.27 +  0.47  i   &
0.03 $\pm$ 0.01 $\pm$ 0.05    & 
0.17 $\pm$ 0.01 $\pm$ 0.05    
\\
\hl
($\varepsilon$) &  
$- {2 \sqrt{2} \over 3 \sqrt{3}}$   & 
0.26 + 0.47  i & 
$-$ 0.17 $\pm$ 0.03 $\pm$ 0.05   & 
1.56 $\pm$ 0.06 $\pm$ 0.05  \\ 
\hl
($\gamma$) &  
--      & 
$-$ 1.38   & 
$-$ 0.30 $\pm$ 0.05 $\pm$ 0.30 &   
$-$ 12.6 $\pm$ 2.5 $\pm$ 0.30  \\
\hl
(Z)  & 
${4 \sqrt{2} \over 3}$  & 
$-$ 1.06 + 0.79 i   &
$-$ 0.08 $\pm$ 0.01 $\pm$ 0.18    &
0.17 $\pm$ 0.01 $\pm$ 0.18     \\
\hl
(g)  & 
$ {2 \sqrt{2} \over 3}$  & 
0.27 +  0.47 i & 
$-$ 0.15 $\pm$ 0 $\pm$ 0.05     & 
$-$ 0.15 $\pm$ 0 $\pm$ 0.05    \\
\hl 
\end{tabular}
\end{center} 
\end{table}
\begin{table}[!ht] 
\begin{center}
\caption{Numerics for $\cA_{3/2}$:  $a_{3/2}^{(X)}$, 
 $\Delta_{L} \cA_{3/2}^{(X)} $, 
$\Delta_{C} \cA_{3/2}^{(X)} $ }  
\label{tab:A32ratios}
\vspace{0.3cm}
%\begin{tiny}
\begin{tabular}{|c|c|c|c|c|}\hl \hl 
\hl  
(X) & $a_{3/2}^{(X)}$ &
$\Delta_{L} \cA_{3/2}^{(X)} $ &
$[\Delta_{C} \cA_{3/2}^{(X)}]^+ $  & 
$[\Delta_{C} \cA_{3/2}^{(X)}]^- $ \\
\hl
\hl
\hl  
\hl  
(27)  & 
${10 \over 9}$   & 
$-$ 0.04 $-$ 0.21 i    & 
0.01 $\pm$ 0 $\pm$ 0.05   & 
0.01 $\pm$ 0 $\pm$ 0.05   \\
\hl
($\varepsilon$) &  
${4 \over 3 \sqrt{3}}$   & 
$-$ 0.69 $-$ 0.21  i & 
$-$ 0.15 $\pm$ 0.02 $\pm$ 0.50   & 
1.74 $\pm$ 0.06 $\pm$ 0.50  \\ 
\hl
($\gamma$) &  
--      & 
$-$ 0.47   & 
0.59 $\pm$ 0.02 $\pm$ 0.10 &   
1.70 $\pm$ 0.35 $\pm$ 0.10   \\
\hl
(Z)  & 
${4 \over 3}$  & 
$-$ 0.86 $-$ 0.78 i   &
0.02 $\pm$ 0.01 $\pm$ 0.30    &
0.16 $\pm$ 0.01 $\pm$ 0.30     \\
\hl
(g)  & 
$ {2 \over 3}$  & 
$-$ 0.50 $-$ 0.21 i & 
$-$ 0.15 $\pm$ 0 $\pm$ 0.20     & 
$-$ 0.15 $\pm$ 0 $\pm$ 0.20    \\
\hl 
\end{tabular}
\end{center} 
\end{table}

\begin{table}[!ht] 
\begin{center}
\caption{Numerics for $\cA_{5/2}$:  $a_{5/2}^{(X)}$, 
 $\Delta_{L} \cA_{5/2}^{(X)} $, 
$\Delta_{C} \cA_{5/2}^{(X)} $ }  
\label{tab:A52ratios}
\vspace{0.3cm}
%\begin{tiny}
\begin{tabular}{|c|c|c|c|c|}\hl \hl 
\hl  
(X) & $a_{5/2}^{(X)}$ &
$\Delta_{L} \cA_{5/2}^{(X)} $ &
$[\Delta_{C} \cA_{5/2}^{(X)}]^+ $  & 
$[\Delta_{C} \cA_{5/2}^{(X)}]^- $ \\
\hl
\hl
\hl  
($\gamma$) &  
--      & 
$-$ 0.51   & 
$-$ 0.20 $\pm$ 0 $\pm$ 0.10 &   
$-$ 0.11 $\pm$ 0.01 $\pm$ 0.10 \\
\hl
(Z)  & 
--  & 
$-$ 0.93 $-$ 1.15 i   &
$-$ 0.14 $\pm$ 0.01 $\pm$ 0.40    &
0.01 $\pm$ 0.01 $\pm$ 0.40     \\
\hl 
\end{tabular}
\end{center} 
\end{table}

Some comments on the numerical results are now in order.  From chiral
power counting, the expected size of NLO corrections is at the level
of $\sim 0.2-0.3$, reflecting $M_K^2/(4 \pi F_\pi)^2 \simeq 0.2$.
This estimate sets the reference scale in the following discussion.

The following pattern seems to emerge from our results.  On one hand,
whenever the absorptive loop correction is small, the dispersive
correction is dominated by the local contribution. Therefore, it is
rather sensitive to the chiral renormalization scale and to the values 
of LECs.  In these cases, the size of NLO corrections is
rather uncertain, at least within the approach we follow here in
evaluating the LECs. Extreme examples of this
behaviour are provided by $\Delta_C \cA_{1/2}^{(27)}$ (of little
phenomenological impact) and $\Delta_C \cA_{3/2}^{(\epsilon)}$ (which
is instead quite relevant phenomenologically).

On the other hand, whenever the absorptive loop correction is large,
the dispersive component is dominated by the non-polynomial part of
the loops and it is relatively insensitive to the chiral
renormalization scale and to the values of LECs. In all
relevant cases we have checked that the absorptive component of
$\Delta_L \cA_n^{(X)}$ is consistent with perturbative unitarity.
Therefore we conclude that in these cases the size of NLO corrections
is rather well understood, being determined by the physics of final
state interactions.  Typical examples of this behaviour are given by
$\Delta_L \cA_{1/2,3/2}^{(Z)}$, which have an important
phenomenological impact. 

We conclude this section with some remarks on apparently anomalous 
results.
\begin{itemize}
\item $\Delta_L \cA_{1/2,3/2}^{(Z)}$ is $O (1)$. 
As discussed above, the physics underlying this result is well
understood, being related to the absorptive cut in the amplitude.  
The key point is that this feature is absent at LO in the chiral
expansion. It first shows up at NLO, setting the natural size of
the loop corrections.  NNLO terms in the chiral expansion are then
expected to scale as NLO $\times (0.2-0.3)$, since corrections to the
absorptive cut behave this way.  Therefore $\Delta_L
\cA_{1/2,3/2}^{(Z)} \sim O (1)$ does not imply a breakdown of the
chiral expansion. 

\item $[\Delta_C \cA_{1/2,3/2}^{(\epsilon)}]^{-}$ is $O (1)$.  This
result is determined by the large numerical coefficients multiplying
the couplings $N_{6,7,8,13}^{r}$, which turn out to have natural size
within the leading $1/N_c$ approximation. We observe, however, that in
the case of the $\Delta I = 3/2$ amplitude (phenomenologically
relevant) the leading $1/N_c$ approximation is afflicted by a large
uncertainty due to high sensitivity to $\nu_\chi$.  This
uncertainty mitigates the apparent breakdown of chiral power counting.
\item Yet another surprising result is the one for $[\Delta_C
\cA_{1/2}^{(\gamma)}]^-$. The underlying reason is in the large size
of the CP violating component of the Wilson coefficients $C_{9,10}$.
Again, the operators $Q_{9,10}$ only make their first contribution at
NLO in the chiral expansion.
\end{itemize}

\section{Phenomenology I:  CP conserving amplitudes}
\label{sec:pheno1}
\renewcommand{\theequation}{\arabic{section}.\arabic{equation}}
\setcounter{equation}{0}
This section is devoted to a phenomenological analysis of $K
\rightarrow \pi \pi$ decays including all sources of isospin breaking.
The theoretical parametrization of the amplitudes is based on the NLO
CHPT  analysis discussed in the previous sections.  Our goal is to
extract information on the pure weak amplitudes (or equivalently on
the couplings $g_8$ and $g_{27}$) and to clarify the role of isospin
breaking in the observed $K \rightarrow \pi \pi$ rescattering phases.
All along we keep track of both experimental errors and
the theoretical uncertainties related to our estimates of the NLO
couplings at leading order in the $1/N_c$ expansion.

\subsection{Including the radiative modes}

When considering electromagnetic effects at first order in $\alpha$,
only an inclusive sum of $K \rightarrow \pi \pi$ and $K \rightarrow
\pi \pi \gamma$ widths is theoretically meaningful (free of IR
divergences) and experimentally observable.  

We denote the appropriate observable widths by 
$\Gamma_{n [\gamma]} (\omega)$ for $n=+-, 00, +0 $. These widths
depend in general on
the amount of radiative events included in the data sample, according
to specific experimental cuts on the radiative mode. This dependence 
is compactly represented by the parameter $\omega$.
Denoting by  $\cA_{n}$ the IR finite amplitudes as defined in 
Eq.~(\ref{eq:intro3}), the relevant decay rates can be written as 
\begin{equation}
\Gamma_{n [\gamma]} (\omega) = \frac{1}{2 \sqrt{s_n}} \, 
\left| \cA_{n} \right|^2  \, \Phi_{n} \, G_{n} (\omega) \ . 
\label{eq:pheno1}
\end{equation} 
Here $\sqrt{s_n}$ is the total cms energy (the appropriate kaon
mass) and $\Phi_n$ is the appropriate two-body phase space.
The infrared factors $G_n (\omega)$ are defined as
\begin{equation}
G_n (\omega) = 1 + \frac{\alpha}{\pi} \, \bigg[ 
2 \pi \,   \real B_{n} (M_\gamma) + I_{n} ( M_\gamma ; \omega) \bigg]
\ .
\label{eq:pheno2}
\end{equation}
Note that $G_{n} (\omega)$ is different from 1 only in the $K^0
\rightarrow \pi^+ \pi^-$ and $K^+ \rightarrow \pi^+ \pi^0$ modes.  The
factor $B_{n} (M_\gamma)$ arises from the IR divergent loop amplitude
(its definition for $n=+-$ is given in Eq.~(\ref{Eq:tour1})), while
$\alpha/\pi\, I_{n} (M_\gamma ; \omega)$ is the $K \rightarrow \pi \pi
\gamma$ decay rate normalized to the non-radiative rate. The latter
term depends on the treatment of real photons (hence on $\omega$) and
is infrared divergent.  The combination of IR divergences induced by
virtual and real photons cancels in the sum, leaving the $\omega$
dependent factor $G_{n} (\omega)$.

We discuss here in some detail the expression for $G_{+-} (\omega)$,
which plays an important phenomenological role.  On the other hand, the
inclusion of $G_{+0} (\omega)$ only produces an effect of order
$\alpha \cA_{3/2}$ (or $\alpha G_{27}$) and therefore represents a
sub-leading correction. Its numerical effect will be taken into
account, following the analysis of Ref.~\cite{Cirigliano:2000zw}. 
The explicit form of $B_{+-} (M_\gamma)$, the virtual photon
contribution to $G_{+-} (\omega)$, can be found in
Eq.~(\ref{eq:appD1}). The real photon contribution $\alpha/\pi\, I_{n}
(M_\gamma ; \omega)$ arises from the decay $K^0 (P) \rightarrow \pi^+
(p_+) \pi^- (p_-) \gamma (k)$ and it has the form
\beq
I_{+-} (M_\gamma ; \omega) = \frac{2}{M_K^2   
\sqrt{1 -  \frac{4 M_\pi^2}{M_K^2}} } \, 
\int_{s_{-}(\omega)}^{s_{\rm max}} \ d s \, f_{+-} (s ; M_\gamma) 
\label{eq:pheno3}
\eeq
where 
\bea
s &=& (p_+  + p_-)^2~,   \qquad s_{\rm min} = 4 M_\pi^2 ~, 
\qquad  s_{\rm max} = (M_K - M_\gamma)^2  
\\
f_{+-}(s ; M_\gamma) &=& M_\pi^2 \, \left( \frac{1}{X_+} - \frac{1}{X_-} 
\right) + \frac{s - 2 M_\pi^2}{M_K^2 - s - M_\gamma^2} \, \log \left(
\frac{X_+}{X_-} \right) 
\\
X_{\pm} &=& \frac{1}{2} \left( M_K^2 - s - M_\gamma^2 \right)  \pm 
\frac{1}{2} \sqrt{1 -  \frac{4 M_\pi^2}{s}} 
\lambda^{1/2} (M_K^2,s,M_\gamma^2) \\
\lambda(x,y,z) &=& x^2 + y^2 + z^2 - 2 (x y + x z + y z) ~.
\eea
The infrared divergence comes from the upper end of the integration in
the dipion invariant mass ($s \sim s_{\rm max}$).  We have verified by
analytic integration in the range $M_K (M_K - 2 \omega) < s < s_{\rm
max}$ that $I_{+-} (M_\gamma ; \omega)$ has the correct $M_\gamma$
dependence to cancel the infrared singularity generated by virtual
photons.  For $\omega / M_K \ll 1$,
the analytic expression of $I_{+-} (M_\gamma ; \omega)$ can be found
in Eq.~(21) of Ref.~\cite{Cirigliano:2000zw}.  
The corresponding function  $G_{+-} (\omega)$ 
is plotted in Fig.~2 of Ref.~\cite{Cirigliano:2000zw}. 

Recently, the KLOE collaboration has reported a high-precision
measurement of the ratio $\Gamma_{+-}/\Gamma_{00}$ \cite{KLOE},
where
the result refers to the fully inclusive treatment of radiative
events.  In order to use the KLOE measurement in our analysis, we need
to calculate the fully inclusive rate (no cuts on the $\pi \pi \gamma$
final state). We have done this by numerical integration of
Eq.~(\ref{eq:pheno3}) and we find
\beq G_{+-} \Big|_{\rm inclusive}
= 1 \ + \ 0.67 \cdot 10^{-2} \ .
\label{eq:pheno4}
\eeq

\subsection{Constraints from measured branching ratios}

CP conserving $K \rightarrow \pi \pi$ phenomenology is based on the
following input from Eq.~(\ref{eq:pheno1}) for $n=+-, 00, +0 $:
\begin{equation}
\left| \cA_{n} \right| =  \left( \frac{2 \, \sqrt{s_n}  \, 
\Gamma_{n}}{G_{n} \, \Phi_{n} } \right)^{1/2}  \equiv  C_n \ .  
\end{equation}
It is convenient to express these equations in terms of the isospin 
amplitudes $A_0, A_2 ,A_2^+$ and the phase shift $\chi_0 - \chi_2$ 
(as defined in Eq.~(\ref{eq:intro2})). 
With $r=(C_{+-}/C_{00})^2$ one obtains\footnote{
Note that in the last equation one has to use $A_2$ and not $A_2^+$ (as
done in the isospin conserving analyses). For this reason the
extraction of the phase shift is related to the $\Delta I = 5/2$
amplitude.} 
\begin{eqnarray}
A_{2}^+ &=&  \frac{2}{3} \, C_{+0} \nonumber \\
(A_0)^2  + (A_2)^2 &=& \frac{2}{3} \, C_{+-}^2  + \frac{1}{3} \, C_{00}^2 
\label{eq:masterph} \\
\displaystyle\frac{A_2}{A_0} \cos(\chi_0 - \chi_2) &=&
\displaystyle\frac{r-1 + (\frac{A_2}{A_0})^2(2 r - \frac{1}{2})}
{\sqrt{2}(1 + 2 r)} ~. \nonumber
\end{eqnarray} 
In general, in the presence of isospin breaking, these three independent
experimental constraints are not sufficient to fix the three isospin
amplitudes ($A_0, A_2, A_2^+$ or $\cA_{1/2}, \cA_{3/2}, \cA_{5/2}$)
plus the phase difference ($\chi_0 - \chi_2$). In the previous
sections, we have shown how CHPT relates the amplitude $\cA_{5/2}$
to $\cA_{1/2}$, thus effectively reducing the number of independent
amplitudes. Including also all other isospin breaking
effects, we can extract the couplings $g_8$ and $g_{27}$ from 
Eqs.~(\ref{eq:masterph}). 

\subsection{CHPT  fit to  $K \rightarrow \pi \pi$ data}
\label{subsec:fit}

Using Eqs.~(\ref{eq:masterph}) as starting point, we perform a fit to
$g_8$, $g_{27}$ and the phase difference $\chi_0 - \chi_2$.  In order
to do so, we employ the CHPT parametrization for the $A_I$.  The
detailed relations between $\cA_{1/2,3/2,5/2}$ (presented in
Secs.~\ref{sec:chptLO}, \ref{sec:chptNLO}) and $A_{0}, A_{2},
A_{2}^+$, to first order in isospin breaking, are reported in
Section~\ref{sec:pheno2a}.
We leave the phase difference $\chi_0 - \chi_2$ as a free parameter 
because one-loop CHPT  fails in reproducing the strong s-wave phase 
shifts. 

Apart from $g_8$ and $g_{27}$, the amplitudes $A_n$ depend on the LO
couplings $g_8 g_{\rm ewk}, Z$ and on a large set of NLO couplings.
Given our large-$N_c$ estimates  for $g_8 g_{\rm ewk}$ and for the
ratios of NLO over LO couplings $(g_8 N_i)/g_8 \, ,\, \cdots $ (see
Sec.~\ref{sec:LECs}), we study the constraints imposed on $g_8,
g_{27}$ by the experimental branching ratios.  In this process we keep
track of the theoretical uncertainty induced by the use of a specific
approximation in estimating the relevant LECs (leading order in the
large-$N_c$ expansion).  In practice, this reduces to studying the
dependence of the amplitude (and of the output values of $g_8$ and
$g_{27}$) on two parameters: the short distance scale ($\mu_{\rm SD}$) 
and the chiral renormalization scale ($\nu_\chi$).

In summary, the experimental input to the fit is given by the three
partial widths $\Gamma_{+-,00,+0}$ (kaon lifetimes and branching
ratios) \cite{PDG02} and by the new KLOE measurement for
$\Gamma_{+-}/\Gamma_{00}$ \cite{KLOE}.  The theoretical input is given
by the NLO CHPT amplitudes as well as the estimates for $g_{\rm ewk}$
and the NLO couplings.  As primary output we report ${\rm Re} \, g_8,
{\rm Re} \, g_{27}$ and $\chi_{0} - \chi_{2}$.
Derived quantities of interest for
phenomenological applications will be reported subsequently.

(1) Using the {\bf NLO isospin conserving} amplitudes (IC-fit), 
we find
\beq
\begin{array}{ccclcc}
{\rm Re}\, g_8 &=&   
3.665  & \pm\,  0.007 \,  ({\rm exp}) & \pm\,  0.001  \, (\mu_{\rm SD})  
 &  \pm\,   0.137 \, (\nu_\chi) \\
{\rm Re}\,  g_{27} &=&  
0.297 & \pm\,  0.001  \, ({\rm exp}) &     &  \pm\,   0.014 \, 
(\nu_\chi)  \\  
\chi_0 - \chi_2  &=&  48.6  & \pm\,  2.6 \, ({\rm exp})~.  &     &      
\end{array}
\label{fit1}
\eeq 
Using instead the tree-level (LO) amplitudes in the isospin limit would lead 
to  ${\rm Re}\, g_8 = 5.09 \pm 0.01$ and 
${\rm Re}\, g_{27}= 0.294 \pm 0.001$.  
This result is in qualitative agreement with Ref.~\cite{kmw91}: NLO  
chiral corrections enhance the $I=1/2$ amplitude by roughly $30 \%$. 

(2) Using the full {\bf NLO isospin breaking} amplitudes (IB-fit), 
we find\footnote{ The uncertainty in $\varepsilon^{(2)} = (1.061 \pm
0.083) \cdot 10^{-2}$ ~\cite{Leutwyler96} produces errors one order of
magnitude smaller than the smallest uncertainty quoted above.}
\beq
\begin{array}{ccclll}
{\rm Re}\,  g_8 &=&   3.650  & \pm\,  0.007 \,  ({\rm exp}) & \pm\,  0.001  
\, (\mu_{\rm SD})   &  \pm\,   0.143 \, (\nu_\chi) \\
{\rm Re}\,  g_{27} &=&  0.303 & \pm\,  0.001  \, ({\rm exp}) & \pm\,  0.001  
\, (\mu_{\rm SD})  &  \pm\,   0.014 \, (\nu_\chi)  \\  
\chi_0 - \chi_2  &=&  54.6  & \pm\,  2.2 \, ({\rm exp})  &
    & \pm\,  0.9  \, (\nu_\chi) ~.      
\end{array}
\label{fit2}
\eeq 
In the IB case, a tree-level (LO) fit leads to 
${\rm Re}\, g_8 = 5.11 \pm 0.01$ and ${\rm Re}\, g_{27}= 0.270  \pm  0.001$.   

A few remarks are in order:  
\begin{itemize}

\item Using NLO amplitudes, both $g_8$ and $g_{27}$ receive small
shifts after inclusion of IB corrections. While this could be expected
for $g_8$, it results from a cancellation of different effects in the 
case of $g_{27}$ (at tree level the inclusion of isospin 
breaking reduces $g_{27}$ by roughly $10 \%$ ). 
Note also that competing loop effects reduce the
$\nu_{\chi}$-dependence of ${\rm Re}\, g_{27}$ (IB-fit) to only $\pm
0.002$.  As a more realistic estimate of the long-distance error we
have chosen to quote the $\nu_\chi$-dependence induced by each one of
the competing effects (for example the isospin-conserving loops).

\item The output for $g_8$ and $g_{27}$ is sensitive to the input used
for the strong LECs $L_i$ of $\cO(p^4)$.  The results of
Eqs.~(\ref{fit1}) and
(\ref{fit2}) correspond to the central values quoted in
Sec.~\ref{sec:LECs}.  We have repeated the fit with non-central
input and have found the variations in $g_8$ and $g_{27}$ to be 
below $5 \% $.

\item In obtaining the results in Eqs.~(\ref{fit1}) and (\ref{fit2})
we have used the large-$N_c$ predictions for the ratios $(g_8 N_i)/g_8$.
We have also employed the alternative procedure of using large $N_c$
directly for the couplings $g_8 N_i$. In this case we find $g_8=3.99$
and $g_{27}=0.289 $, reflecting the change in size of the $p^4$ local
amplitudes. All other quantities of phenomenological interest are
stable under this change in the fitting procedure.

\item Some derived quantities of phenomenological interest are
the ratios of isospin amplitudes: ${\rm Re}A_0/{\rm Re}A_2$, 
${\rm Re}A_0/{\rm Re} A_2^+$, 
$f_{5/2} \equiv {\rm Re}A_2/{\rm Re}A_2^+ - 1$. From our fit we find
\beq
\begin{array}{cccccl}
\left[ \displaystyle\frac{{\rm Re} A_0}{{\rm Re} A_2} 
\right]_{\rm IB-fit} &=&  
20.33  & \pm\,  0.07  \, ({\rm exp}) & \pm\, 0.01 \,  (\mu_{\rm SD}) &  
\pm\,   0.47 \, (\nu_\chi)     \\
\left[ \displaystyle\frac{ {\rm Re} A_0}{ {\rm Re} 
A_2^+} \right]_{\rm IB-fit} &=&  
22.09  & \pm\,  0.09 \, ({\rm exp}) & 
 \pm\, 0.01 \,  (\mu_{\rm SD}) &  \pm\,   0.05 \, (\nu_\chi)    \\
\left[ f_{5/2} \right]_{\rm IB-fit} &=&  \Big(
8.6  & \pm\,  0.03 \, ({\rm exp})  & \pm\, 0.01 \, (\mu_{\rm SD}) &  
\pm\,   2.5 \, (\nu_\chi)  \Big) \cdot 10^{-2}    \ . 
\end{array}
\label{fit5}
\eeq 
In the absence of isospin breaking, one finds instead $f_{5/2}=0$ 
and ${\rm Re}A_0/{\rm Re}A_2 = 22.16 \pm 0.09$. 

\end{itemize}

\subsection{Isospin breaking in the phases}

This section is devoted to understanding isospin breaking in the
rescattering phases of $K \rightarrow \pi \pi$.  If isospin is
conserved Watson's theorem predicts $\chi_0 - \chi_2 = \delta_0 -
\delta_2 \sim 45^{\circ}$.  For a long time this prediction has not 
been fulfilled by the data, as one typically encountered 
$\chi_0 - \chi_2  \sim 60^{\circ}$.  

The situation has recently improved.  Using the KLOE data 
\cite{KLOE} and 
working in the isospin limit, our fit (\ref{fit1}) gives
$\chi_0 - \chi_2 \sim 49^{\circ}$ and so there seems to be no more
phase problem.  However, the inclusion of isospin breaking appears to 
reintroduce the issue. In order to understand what is going on, we 
analyse in detail the various factors determining $\chi_0 - \chi_2$.

The last of Eqs.~(\ref{eq:masterph}) can be rewritten as
\begin{equation} 
\displaystyle\frac{A_2^+}{A_0} \cos(\chi_0 - \chi_2) =
\displaystyle\frac{r-1 + (\frac{A_2}{A_0})^2(2 r - \frac{1}{2})}
{\sqrt{2}\, (1 + 2 r)(1+f_{5/2})} ~.
\label{coschi}
\end{equation} 
Let us first calculate the right-hand side without an $I=5/2$ amplitude
($f_{5/2}=0$). 
With the old value $r = 1.1085$ as input 
(PDG2000 \cite{PDG00}), one obtains
\begin{equation} 
\displaystyle\frac{A_2^+}{A_0} \cos(\chi_0 - \chi_2) = 0.02461~.
\end{equation} 
With $A_2^+/A_0 = 0.045$ (based on the IC-fit),    
this leads to the standard puzzle that 
$\chi_0 - \chi_2$ is much bigger than 45$^\circ$:
\begin{equation} 
\chi_0 - \chi_2 = 57^\circ~.
\end{equation}

What could be the reasons for this discrepancy of about 30 \%
($\cos 45^\circ/\cos 57^\circ=1.30 $)? 
Let us consider several effects:
\begin{itemize} 
\item First of all, the right-hand side of (\ref{coschi}) has changed with
the recent KLOE result \cite{KLOE} $r = 1.1345$ to give
\begin{equation} 
\displaystyle\frac{A_2^+}{A_0} \cos(\chi_0 - \chi_2) = 0.02987~.
\end{equation} 
This is a sizable correction of about 18 \% and it goes more than 
half-way in the right direction to decrease the phase difference.

\item Taking into account isospin breaking  introduces an $I=5/2$
amplitude via the ratio $f_{5/2}$ in Eq.~(\ref{coschi}). 
According to our results (\ref{fit5}),
\begin{equation} 
f_{5/2} =  \Big( 6.5 \,({\rm loops}) \ + \ 2.1 \, ({\rm local}) \ 
\pm \ 2.5 (\nu_{\chi})
\Big) 
 \cdot 10^{-2}~, 
\label{f52}
\end{equation}
increasing again the discrepancy.
This value is dominated by loop 
contributions.  
Even if one changes the sign of the relevant
combination of LECs, $f_{5/2}$ would still be positive. Note that 
(\ref{f52}) amounts to a correction of $\sim$ 8 \% (in the ``wrong''
direction). As already noted by Wolfe and Maltman \cite{Wolfe:1999br}, 
it seems impossible to solve the phase problem with a reasonable 
choice of counterterms. 
\item Finally, the ``infrared factor'' for the $+-$ mode must be taken
into account. This is straightforward with the inclusive measurement
of KLOE. We find $r = 1.127$, which increases again the discrepancy. 
Including also the effect of $f_{5/2}$, we obtain 
\begin{equation} 
\displaystyle\frac{A_2^+}{A_0} \cos(\chi_0 - \chi_2) = 0.02611~, 
\end{equation} 
leading to $\chi_0 - \chi_2 = (54.6 \pm 2.4)^\circ$.  
\end{itemize} 

Before addressing the question whether this result is
in disagreement with the $\pi \pi$ phase shift 
prediction~\cite{cgl01} 
$\delta_0 - \delta_2 = (47.7 \pm 1.5)^\circ$,
it is mandatory to study the effect of isospin breaking on the phases
themselves. We use the general decomposition 
\beq  
\chi_{I} \ = \ \delta_{I} \, + \, \gamma_{I}  \qquad (I=0,2)  \ ,   
\eeq  
where 
%$\delta_{I}$ is the relevant s-wave $\pi \pi$ phase 
%shift at $\sqrt{s} = M_K$ and 
$\gamma_{I}$ represents an isospin 
breaking correction.  The $\gamma_{I}$ are related to isospin 
breaking dynamics 
in $\pi \pi$ rescattering as well as to the presence  
of radiative channels~\cite{Cirigliano:2000zw,Gardner:2001rv}. 
Since the analysis of Ref.~\cite{Cirigliano:2000zw},   
new information on radiative corrections in $\pi\pi$ scattering 
has become available, allowing for a reevaluation of 
$\gamma_0 - \gamma_2$.

\subsection{Optical theorem and $\gamma_0 - \gamma_2$}

The $K \rightarrow \pi \pi$ amplitudes at NLO in CHPT 
allow for a perturbative evaluation of $\gamma_{0,2}$.  We
find\footnote{The results depend on the ratio $g_8 / g_{27}$, for
which we use the IB-fit output.} 
\begin{eqnarray}
\gamma_0 &=& (-0.18 \pm 0.02)^{\circ}  \nn \\
\gamma_2 &=&  (3.0 \pm 0.4)^{\circ}~, \label{eq:pertphase} 
\end{eqnarray} 
where the error is obtained by varying the chiral renormalization
scale $\nu_{\chi}$ as in the main fit. Setting the NLO local
terms to zero would lead to results within the range quoted in
Eq.~(\ref{eq:pertphase}). 
This evaluation incorporates the constraints of the optical
theorem at leading order in perturbation theory.  In practice, this   
only reflects the $\cO (e^2 p^0)$ mixing
between the $I=0$ and $I=2$ $\pi \pi$ channels, completely missing both 
higher-order corrections and the new physical effect due to 
the radiative channel $\pi \pi \gamma$. 
In order to improve upon these perturbative results, a more general 
analysis of the optical theorem for $K^0 \rightarrow \pi \pi$
amplitudes is required. We shall follow here the approach of 
Ref.~\cite{Cirigliano:2000zw}, except for a few details. 
The main novelty lies in the final stage, in which one needs an explicit 
calculation of isospin breaking effects in $\pi \pi$ scattering: 
we use the results obtained at $\cO(e^2 p^2)$ in CHPT 
in Refs.~\cite{ku98,kn02}.

We now summarize the steps involved in the optical theorem analysis of
Ref.~\cite{Cirigliano:2000zw}, relegating some technical details to 
App.~\ref{app:optical}. For this section, CP is assumed to be 
conserved.  

\begin{enumerate}

\item The first step is to work out the consequences of the optical
theorem for $K^0 \rightarrow \pi \pi$ amplitudes, considering the
following intermediate states: $\pi^+ \pi^-, \pi^0 \pi^0$ and $\pi^+
\pi^- \gamma$.  For the radiative amplitudes describing $K^0
\rightarrow \pi^+ \pi^- \gamma$ and $\pi^+ \pi^- \gamma \rightarrow
\pi \pi$ we use the leading Low parametrization, thus neglecting
possible structure dependent terms\footnote{This is known to be an
excellent approximation for $K^0 \rightarrow \pi^+ \pi^- \gamma$.}. 
In this approximation the radiative amplitudes are known in terms of 
the non-radiative ones. 
Under the assumptions listed above, and collecting the $K^0
\rightarrow \pi \pi$ amplitudes in a two-component vector $\cA$, 
the optical theorem has the  following form:
\beq
{\cal A}bs \, \cA = \beta \, \left( {\cal T}^{\dagger} + {\cal R} 
\right) \, \cA 
\label{eq:opt1}
\eeq 
where $\beta = \sqrt{1 - 4 M_{\pi^0}^2 /M_K^2}$, while ${\cal T}$ and
${\cal R}$ are two-by-two matrices: ${\cal T}$ is related to the s-wave
projection of the $\pi \pi$ T-matrix, while ${\cal R}$ encodes the effect
of both radiative modes and phase space corrections induced by mass
splitting. \\
The explicit form of Eq.~(\ref{eq:opt1}) is best derived by working with
$\pi \pi$ states in the charge basis ($\pi^+ \pi^-, \pi^0 \pi^0$)
where it is more transparent to deal properly with IR 
singularities and phase space corrections.  Special care is needed in
removing the IR and Coulomb singularities from the amplitudes
$\cA_{+-}$, ${\cal T}_{+-,00}$ and ${\cal T}_{+-,+-}$. 
This step involves an arbitrary choice, which only affects the 
intermediate states of the analysis but not the final results.  
We adopt the following prescription\footnote{The $\pi \pi$ amplitudes 
are functions of two of the 
three  Mandelstam variables $s,t,u$.  In the following 
we set $s=M_K^2$ and trade the other independent variable for the 
cms scattering angle $\theta$. Moreover, the explicit dependence on 
$\cos \theta$ is suppressed in order to keep the expressions compact.}:
\bea
A (K^0 \rightarrow \pi^+ \pi^-) &=& \cA_{+ -}  \ \ \ \  \exp{ \{ 
\alpha \, B_{\pi \pi} \} } \nl
A (\pi^0 \pi^0 \rightarrow \pi^+ \pi^-) &=& 
 T_{+-,00} \, \, \ \exp{ \{ \alpha \, B_{\pi \pi} \} } \\
A (\pi^+ \pi^- \rightarrow \pi^+ \pi^-) &=& 
 T_{+-,+-}  \  \exp{ \{ \alpha \left( 2 \, B_{\pi \pi} +  C_{\pi \pi}
\right) \} }~, \nn 
\eea
where the infrared singularity is separated in the factors $B_{\pi \pi}$
and $C_{\pi \pi}$, whose form is reported in App.~\ref{app:optical}.
These factors depend only on the charges and kinematical configuration
of the external particles. 

\item In order to make contact with standard treatments, it is 
convenient to represent Eq.~(\ref{eq:opt1}) in the ``isospin'' basis for
$\pi \pi$ amplitudes. Explicit relations between charge and isospin
amplitudes are reported in Ref.~\cite{Cirigliano:2000zw}. 
In the isospin basis the matrices have the form
\beq
{\cal T} = \left(  \begin{array}{ccc} 
{\cal T}_{00} & & {\cal T}_{0 2} \\
 & & \\
{\cal T}_{2 0} & & {\cal T}_{22} 
\end{array}
\right)  \, ~,
\quad
{\cal R} = \left(  \begin{array}{ccc} 
\displaystyle{2 \over 3}\, (\Delta_{+-} {\cal T}_{00}^{*} + \delta_{+-}) & &
\displaystyle{\sqrt{2} \over 3}\, 
(\Delta_{+-} {\cal T}_{00}^{*} + \delta_{+-})  \\
 & & \\
\displaystyle{\sqrt{2} \over 3}\, 
(\Delta_{+-} {\cal T}_{22}^{*} + \delta_{+-}) & &
\displaystyle{1 \over 3}\, (\Delta_{+-} {\cal T}_{22}^{*} + \delta_{+-})   \\
\end{array}
\right) ~, 
\eeq
where, using the notation 
\bea
\langle f \rangle \, \equiv \, \int_{-1}^{+1} \, d(\cos \theta) \ 
f (\cos \theta) \ , 
\eea
the various quantities have the following structure:
\bea
 {\cal T}_{\rm a b} &=& \displaystyle{1 \over 64 \pi} \, 
\langle T_{\rm a b} \rangle \  \nl
\Delta_{+-} &=& - {2 (M_{\pi^\pm}^2 - M_{\pi^0}^2) \over \beta^2 M_K^2}  + 
2 \alpha\, {\rm Re} (B_{\pi \pi}) + { e^2 \over \Phi_{+-} } 
\int \, d \Phi_{+- \gamma} \  f_{1}^{\rm rad} \\
\delta_{+-} &=&  {\alpha \over 32 \pi} \, 
\langle T_{+-,+-} \cdot C_{\pi \pi} \rangle 
+ { \alpha \over 4  \Phi_{+-} } 
\int \, d \Phi_{+- \gamma} \  T_{+-,+-} \cdot f_{2}^{\rm rad}  \ .\nn 
\eea
${\cal T}_{\rm ab}$ is the s-wave projection of the ${\rm b} \rightarrow
{\rm a}$ $\pi \pi$ scattering amplitude.  The factor $\Delta_{+-}$
receives a contribution from phase space corrections (pion mass
splitting), one from virtual photons ($B_{\pi \pi}$) and one from
real photons ($f_{1}^{\rm rad}$). Likewise, $\delta_{+-}$ reflects
both virtual corrections ($C_{\pi \pi}$) and  real photon effects
($f_{2}^{\rm rad}$).  The definition of the phase space factors $d
\Phi_{+-}$ and $ d \Phi_{+- \gamma}$, as well as of $f_{1,2}^{\rm
rad}$ is reported in App.~\ref{app:optical}.  We remark here
that $\Delta_{+-}$ and $\delta_{+-}$ are free of infrared
singularities, as these cancel in the sum of real and virtual photon
contributions.

\item At this point one needs a general parametrization of the matrix
${\cal T}_{I J}$, the T-matrix restricted to the dimension-two
subspace of $\pi \pi$ channels.  Assuming T-invariance (but not
unitarity of the S-matrix restricted to this subspace), an explicit
form is given by
\beq
{\cal T} = \frac{1}{\beta} \, \left(  \begin{array}{ccc} 
\displaystyle{ (\eta_0\, e^{2 i \delta_0} - 1) \over 2 i}
 & & a \, e^{i (\delta_0 + \delta_2)}  \\
 & & \\
a \, e^{i (\delta_0 + \delta_2)}  & & 
\displaystyle{ (\eta_2\, e^{2 i \delta_2} - 1) \over 2 i}
\end{array}
\right)  \,
\eeq
in terms of five parameters (two phase shifts, two inelasticities 
and one off-diagonal amplitude).
If one assumes that only one extra state couples to the ones considered 
here (namely the $\pi^+ \pi^- \gamma$ state), then the inelasticities 
are correlated, as noted in Ref.~\cite{Gardner:2001rv}. 
Since our subsequent discussion does not depend on the inelasticities, 
we do not elaborate further on this point.  

\item The next step is to assume an ansatz for the $K \rightarrow \pi
\pi$ amplitudes of the type
\beq
\cA_{I} = A_{I} \, e^{ i (\delta_I + \gamma_I) } \qquad (I=0,2) 
\eeq
and work out the constraints imposed upon $\gamma_I$ by 
Eq.~(\ref{eq:opt1}). 
Solving for $\sin \gamma_{0}$ and $\sin \gamma_{2}$ 
to first order in $\alpha$ and taking into account 
$A_2/A_0 \ll 1$, one finds \cite{Cirigliano:2000zw} 
\bea
\sin\gamma_0 &=& \beta \left(  {\rm Re} ({\cal R}_{00})  - \tan \delta_0 \, 
{\rm Im}  ({\cal R}_{00}) \right)  \simeq {\cal O} (\alpha \sin \delta_0) 
\nonumber \\ 
\sin\gamma_2 &=& \beta \, {A_0 \over A_2 }\;
\left[ |{\cal T}_{20}|  +  \frac{1}{\cos \delta_2} \left( 
 {\rm Re} ({\cal R}_{20}) \cos\delta_0  - {\rm Im} ({\cal R}_{20})
\sin\delta_0 \right) 
\right]~.
\label{eq:optsol}
\eea
The key feature of Eq.~(\ref{eq:optsol}) is that in the expression 
for $\gamma_2$ the isospin breaking effects get once again enhanced 
by the factor $A_0/A_2 \sim 22$. 

\item The final step consists in evaluating ${\cal T}_{02}$, $\Delta_{+-}$ 
and $\delta_{+-}$, for which we need an explicit expression for the
$\pi \pi$ amplitudes with isospin breaking \cite{ku98,kn02}, as well as 
explicit expressions of $B_{\pi \pi}$ and $C_{\pi \pi}$.  
The details of the calculations cannot be given in a concise way 
and we report here only the results. \\
For ${\cal T}_{02}$  we find
\beq
{\cal T}_{02} = \frac{\sqrt{2}\, (M_{\pi^\pm}^2 - M_{\pi^0}^2)}{24 \pi 
F_{\pi}^2} \, \bigg( 1 + \Delta^{e^2 p^2}_{02} \bigg) ~.
\eeq
Using the results of Refs.~\cite{ku98,kn02} and their estimate 
of the relevant LECs $K_i$, we obtain
\beq
\Delta^{e^2 p^2}_{02} = (0.78 \pm 0.83) + i \, 0.54 
\eeq
where we have added the various errors in quadrature. \\
For the radiative factors the calculation cannot be done 
in a fully analytic form and we employ Monte Carlo integration 
to deal with the real-photon contribution to $\delta_{+-}$. 
We find
\bea
\Delta_{+-} &=&  - 0.81 \cdot 10^{-2} \nl
\delta_{+-} &=&   0.09 \cdot 10^{-2}~. 
\eea
Using these input values in Eq.~(\ref{eq:optsol}), we arrive at
\bea
\gamma_0 &=&  -0.2^{\circ} \nl
\gamma_2 &=&  (6 \pm 3)^{\circ} ~.
\eea
\end{enumerate}

The conclusion is that the optical theorem estimate of 
$\gamma_0 - \gamma_2$ agrees roughly with the perturbative estimate
(\ref{eq:pertphase}) and that it tends to worsen the discrepancy 
between the theoretical prediction of $\delta_0 - \delta_2$~\cite{cgl01}  
and its phenomenological extraction from $K \rightarrow \pi \pi$ 
decays. Explicitly one has
\bea
(\delta_0 - \delta_2)_{\pi\pi \rightarrow \pi\pi} 
%\quad \ \, 
&=& (47.7 \ \pm \ 1.5)^\circ \nl
(\delta_0 - \delta_2)_{K \rightarrow \pi \pi} &=& (60.8 \ \pm \ 
2.2 \,({\rm exp}) \ \pm  \ 0.9 \, (\nu_\chi) \ \pm \ 
3.0 \, (\gamma_{2}) )^\circ ~. 
\label{eq:phasediff}
\eea
%%%%%%%%%%%%%%%%%%%%%%%%%%%%%%%%%%%%%%%%%%%%%%%%%
%%% Adding errors in quadrature 6.1 ---> 3.8  %%%
%%%%%%%%%%%%%%%%%%%%%%%%%%%%%%%%%%%%%%%%%%%%%%%%%
Although the precise KLOE measurement \cite{KLOE} of the ratio of 
$K_S \to \pi\pi$ rates has considerably improved the situation we still
obtain a difference of about $13^\circ$ for the phase shift difference in
the isospin limit between the two determinations in
Eq.~(\ref{eq:phasediff}). The theoretical error is much bigger in the 
present case due to uncertainties in the NLO LECs. However, we observe 
that more than half of this difference is due to the $\Delta I=5/2$
loop amplitude that depends only on the well-established lowest-order
electromagnetic LEC $Z$ in the Lagrangian (\ref{eq:Lelm}). In order to
obtain a phase shift difference in the isospin limit below $50^\circ$,
the local amplitude with $\Delta I=5/2$ would have to be more than
twice as big and of opposite sign. While such an explanation cannot be
totally excluded at this time, the discrepancy in the two entries of 
Eq.~(\ref{eq:phasediff}) certainly warrants further study.

The $\Delta I=5/2$ amplitude induced by isospin violation in the octet
amplitude is small because it only arises at NLO and it is of purely
electromagnetic origin. One may wonder whether isospin violation in
the 27-plet amplitude, which occurs already at leading order, could
compete. Whereas isospin violating contributions to the  $\Delta
I=1/2, 3/2$ amplitudes proportional to $G_{27}$ are certainly
negligible, the effect on the $\Delta I=5/2$ amplitude is worth
investigating.

It is straightforward to calculate isospin violation in the LO 
27-plet amplitudes in (\ref{eq:LOamp2}) due to mass differences
and \pe~mixing. We are only interested in the resulting $\Delta I=5/2$ 
amplitude, entirely due to the quark mass difference:
\begin{equation} 
\cA_{5/2}^{(27)} = \displaystyle\frac{2 \sqrt{3}}{9}\, G_{27} F_\pi
(M_K^2 - M_\pi^2)\, \varepsilon^{(2)}~.
\end{equation} 
This amplitude may now be compared to the corresponding 5/2
amplitude in (\ref{eq:structure1}):  
\begin{equation} 
\cA_{5/2} = - e^2 G_8 F_\pi^3\, (\cA_{5/2}^{(\gamma)} 
            + Z \cA_{5/2}^{(Z)})~.
\end{equation} 
With the numerical information of Table \ref{tab:A52ratios} and 
Eq.~(\ref{fit2}), we obtain for the ratio
\begin{equation}
\displaystyle\frac{\cA_{5/2}^{(27)}}{ {\cal D}isp ~\cA_{5/2}}
 \simeq 6 \cdot 10^{-2}~.
\end{equation}
The conclusion is that the 27-plet contribution to the $\Delta I=5/2$ 
amplitude is of the same sign and only about 6 $\%$ of the octet
contribution. The impact on the $\Delta I=1/2, 3/2$ amplitudes is of
course much smaller still. Isospin violation in the 27-plet amplitude
can safely be neglected.

\section{Phenomenology II:  CP violation}
\label{sec:pheno2}
\renewcommand{\theequation}{\arabic{section}.\arabic{equation}}
\setcounter{equation}{0}

The main contents of this section have already been published in 
Ref.~\cite{cenp03b}. They are included here for completeness.

\subsection{Isospin violation and $\epsp$}
\label{sec:pheno2a}

The direct CP  violation parameter $\epsp$ is given by
\begin{equation} 
\epsp  = - \frac{i}{\sqrt{2}} \, e^{i ( \chi_2 - \chi_0 )} \, 
\frac{\real A_{2}}{ \real A_{0}} \,  
\left[ 
\frac{\imag A_{0}}{ \real A_{0}} \, - \,  
\frac{\imag A_{2}}{ \real A_{2}} \right]  
\ . 
\label{eq:cp1}
\eeq
The expression (\ref{eq:cp1}) is valid to first order in CP violation.
Since $\imag A_{I}$ is CP odd the quantities $\real A_{I}$ and 
$\chi_{I}$ are only needed in the CP limit ($I=0,2$).

To isolate the isospin breaking corrections in $\epsp$, we write the
amplitudes $A_0,A_2$ more explicitly as
\begin{eqnarray}  
A_0 \,e^{i \chi_0 } &=&   \cA_{1/2}^{(0)} + \delta \cA_{1/2} 
\nn \\
 A_2 \,e^{i \chi_2 }&=& \cA_{3/2}^{(0)} + \delta \cA_{3/2}+ 
\cA_{5/2} ~,
\label{eq:A02}   
\end{eqnarray}
where the superscript $(0)$ denotes the isospin limit and 
$\delta \cA_{1/2,3/2}$, $\cA_{5/2}$ are first order in isospin
violation. In the limit of isospin conservation, the amplitudes
$\cA_{\Delta I}$ would be generated by the $\Delta I$ component of 
the electroweak effective Hamiltonian.

To the order we are considering, the amplitudes $\cA_{\Delta I}$ have both
absorptive and dispersive parts. To disentangle the (CP conserving)
phases generated by the loop amplitudes from the CP violating phases
of the various LECs, we express our results explicitly in terms of 
${\cal D}isp ~\cA_{\Delta I}$ and ${\cal A}bs ~\cA_{\Delta I}$. 
Writing Eq.~(\ref{eq:A02}) in the generic form
\begin{equation} 
A_I \,e^{i \chi_I} =  \cA_{n} \equiv 
\cA_{n}^{(0)} + \delta \cA_{n} ~,
\label{eq:form1}  
\end{equation}  
we obtain to first order in CP violation:
\begin{eqnarray} 
\real A_I &=& \displaystyle\sqrt{ (\real [{\cal D}isp \,  \cA_n ])^2 +
 (\real [{\cal A}bs \,  \cA_n ])^2 } \no \\*
\imag A_I &=&  ( \real A_I )^{-1} \ 
\left(  \imag [{\cal D}isp \,  \cA_n ] \,  \real [{\cal D}isp \,  \cA_n ]  + 
\imag [{\cal A}bs \,  \cA_n ] \, \real [{\cal A}bs \,  \cA_n ] \right) \\
e^{i \chi_I } &=& ( \real A_I )^{-1} \ \left( 
\real [{\cal D}isp \,  \cA_n ] +  i \, \real [{\cal A}bs \,  \cA_n ]
 \right)~. \no 
\end{eqnarray} 
Using the second equality in Eq.~(\ref{eq:form1}), one can now
expand $\real A_I$ and $\imag A_I$ to first order in
isospin breaking. With the notation $| \cA^{(0)}_n | =  
\sqrt{ (\real [{\cal D}isp \,  \cA_n^{(0)}  ])^2 +
(\real [{\cal A}bs \,  \cA_n^{(0)} ])^2 } $, we find 
\begin{eqnarray}  
\real A_I  &=&     \left| \cA_{n}^{(0)} \right| +
\left| \cA_{n}^{(0)} \right|^{-1}  
\left( \real  [ {\cal D}isp \, \cA_{n}^{(0)} ]~ 
\real [{\cal D}isp \, \delta \cA_{n}] + 
\real [ {\cal A}bs \,\cA_{n}^{(0)}]~ 
\real [{\cal A}bs \, \delta \cA_{n}]
\right) \qquad \label{eq:ReA0} \\
\imag A_{I}   &=& \left| \cA_{n}^{(0)} \right|^{-1}
\left\{\imag [ {\cal D}isp \,\cA_{n}^{(0)} ] \,
\real [ {\cal D}isp \, \cA_{n}^{(0)} ] + 
\imag [ {\cal A}bs \,\cA_{n}^{(0)}]  \,
\real[ {\cal A}bs \,  \cA_{n}^{(0)} ] \right\} \nn \\
&+& \left|  \cA_{n}^{(0)} \right|^{-1}
\left\{\imag [ {\cal D}isp \,\delta\cA_{n} ] \,
\real [ {\cal D}isp \,\cA_{n}^{(0)} ] + 
\imag [ {\cal D}isp \,\cA_{n}^{(0)} ] \,
\real[ {\cal D}isp \,  \delta\cA_{n} ] \right. \nn\\
&& \qquad \qquad +  \,  \left. \imag [  {\cal A}bs \,\delta\cA_{n} ] \,
\real[ {\cal A}bs \,  \cA_{n}^{(0)} ] + 
\imag [ {\cal A}bs \,\cA_{n}^{(0)} ] \,
\real [ {\cal A}bs \, \delta\cA_{n} ] \right\} \nn\\
&-& \left| \cA_{n}^{(0)} \right|^{-3}
\left\{ \real [ {\cal D}isp \, \cA_{n}^{(0)}] \, 
\real [ {\cal D}isp \, \delta \cA_{n}] + 
\real [ {\cal A}bs \, \cA_{n}^{(0)}] \, 
\real[ {\cal A}bs \, \delta \cA_{n} ] \right\} \, \times  \nn\\
&& \left\{\imag [ {\cal D}isp \,\cA_{n}^{(0)} ] \,
\real [ {\cal D}isp \, \cA_{n}^{(0)} ] + 
\imag [ {\cal A}bs \,\cA_{n}^{(0)}]  \,
\real[ {\cal A}bs \,  \cA_{n}^{(0)} ] \right\}~, 
\label{eq:ImA0}
\end{eqnarray} 
where the first term in each equation above represents  
$\real A_I^{(0)}$ and $\imag A_{I}^{(0)}$, respectively. 

We now turn to the different sources of isospin violation in the
expression (\ref{eq:cp1}) for $\epsp$. We disregard the phase which
can be obtained from the $K \to \pi\pi$ branching ratios. The same 
branching ratios are usually employed to extract the ratio 
$\omega_S=\real A_{2}/ \real A_{0}$ assuming isospin conservation. 
Accounting for isospin violation via the general parametrization
(\ref{eq:intro3}), one is then really evaluating 
$\omega_+ = \real A_{2}^{+}/\real A_{0}$ rather than $\omega_S$.
The two differ by a pure $\Delta I=5/2$ effect:
\bea
\omega_S &=& \omega_+ \, \left( 1 +  f_{5/2} \right) \\
f_{5/2} &=& \frac{\real A_{2}}{\real A^{+}_{2}} - 1  
 \ .   
\label{cp4}
\eea
Because $\omega_+$ is directly related to branching ratios it proves
useful to keep $\omega_+$ in the normalization of $\epsp$, introducing 
the $\Delta I=5/2$ correction $f_{5/2}$ \cite{Cirigliano:2000zw}.

Since $\imag A_2$ is already first order in isospin breaking
the formula for $\epsp$ takes the following form, with all first-order
isospin violating corrections made explicit:
\begin{equation} 
\epsp = - \frac{i}{\sqrt{2}} \, e^{i ( \chi_2 - \chi_0 )} \, 
\omega_+  \,   \left[ 
\frac{\imag A_{0}^{(0)}}{ \real A_{0}^{(0)}} \, 
(1 + \Delta_0 + f_{5/2}) - \frac{\imag A_{2}}{ \real
  A_{2}^{(0)}} \right] \ , 
\label{eq:cpiso}
\eeq
where 
\begin{equation} 
\Delta_0 = \frac{\imag A_0}{\imag A_0^{(0)}} 
\frac{\real A_0^{(0)}}{\real A_0}  - 1  ~.
\end{equation}
With the help of Eqs.~(\ref{eq:ReA0})  and
(\ref{eq:ImA0}), one obtains
\begin{eqnarray}  
\imag A_0^{(0)} &=&  \left| \cA_{1/2}^{(0)} \right|^{-1} \, 
\left\{\imag [ {\cal D}isp \,\cA_{1/2}^{(0)} ] \,
\real [ {\cal D}isp \, \cA_{1/2}^{(0)} ] + 
\imag [ {\cal A}bs \,\cA_{1/2}^{(0)}]  \,
\real[ {\cal A}bs \,  \cA_{1/2}^{(0)} ] \right\}  
\label{eq:ima00} \\
\imag A_2 \ \,  &=&  \left| \cA_{3/2}^{(0)} \right|^{-1} \, 
\left\{\imag [ {\cal D}isp \, 
\left( \delta \cA_{3/2} + \cA_{5/2} \right) ]  \,
\real [ {\cal D}isp \, \cA_{3/2}^{(0)} ] \right.  \nn \\
&& \left. \qquad \quad + \   
\imag [ {\cal A}bs \left( \delta  \cA_{3/2} + \cA_{5/2} \right) ]  \,
\real[ {\cal A}bs \,  \cA_{3/2}^{(0)} ] \right\}   \\
\Delta_0 &=& -2 \left| \cA_{1/2}^{(0)} \right|^{-2} \, 
\left( \real  [ {\cal D}isp \, \cA_{1/2}^{(0)} ]~ 
\real [{\cal D}isp \, \delta \cA_{1/2}] + 
\real [ {\cal A}bs \,\cA_{1/2}^{(0)}]~ 
\real [{\cal A}bs \, \delta \cA_{1/2}]
\right) \quad\nn  \\
&+& \left[ \imag [ {\cal D}isp \,\cA_{1/2}^{(0)} ] \,
\real [ {\cal D}isp \, \cA_{1/2}^{(0)} ] + 
\imag [ {\cal A}bs \,\cA_{1/2}^{(0)}]  \,
\real[ {\cal A}bs \,  \cA_{1/2}^{(0)} ] \right]^{-1} \, \times \nn  \\ 
&& \left\{\imag [ {\cal D}isp \,\delta\cA_{1/2} ] \,
\real [ {\cal D}isp \,\cA_{1/2}^{(0)} ] + 
\imag [ {\cal D}isp \,\cA_{1/2}^{(0)} ] \,
\real[ {\cal D}isp \,  \delta\cA_{1/2} ] \right. \nn \\
&&  +  \,  \left. \imag [  {\cal A}bs \,\delta\cA_{1/2} ] \,
\real[ {\cal A}bs \,  \cA_{1/2}^{(0)} ] + 
\imag [ {\cal A}bs \,\cA_{1/2}^{(0)} ] \,
\real [ {\cal A}bs \, \delta\cA_{1/2} ] \right\} \\ 
f_{5/2} &=& \frac{5}{3} \left| \cA_{3/2}^{(0)} \right|^{-2} \, 
\left\{\real [ {\cal D}isp \,\cA_{3/2}^{(0)} ] \,
\real [ {\cal D}isp \, \cA_{5/2} ] + 
\real [ {\cal A}bs \,\cA_{3/2}^{(0)}]  \,
\real[ {\cal A}bs \,  \cA_{5/2} ] \right\}
\label{eq:f52}.   
\end{eqnarray} 
These expressions are general results to first order in CP and isospin
violation but they are independent of the chiral expansion. Working
strictly to a specific chiral order, these formulas 
simplify considerably. We prefer to keep them in their general form
but we will discuss later the numerical differences between the
complete and the systematic chiral expressions. The differences are
one indication for the importance of higher-order chiral corrections.

Although $\imag A_2$ is itself first order in isospin breaking we now
make the usual (but scheme dependent) separation of the
electroweak penguin contribution to $\imag A_2$ from the
isospin breaking effects generated by other four-quark operators: 
\begin{equation} 
\imag A_2 = \imag A_2^{\rm emp} \ + \ \imag  A_2^{\rm non-emp} \ .  
\end{equation}    
In order to perform the above separation within the CHPT approach,
we need to identify the electroweak penguin contribution to a given
low-energy coupling.  In other words, we need a matching procedure
between CHPT and the underlying theory of electroweak and strong
interactions.  Such a matching procedure is given here by working at
leading order in $1/N_c$. 
Then, the electroweak LECs of $\cO(G_8 e^2 p^n)$ ($n=0,2$) in $\imag
A_2^{\rm non-emp}$ must be calculated by setting to zero the Wilson
coefficients $C_7,C_8,C_9,C_{10}$ of electroweak penguin operators.

Splitting off the electromagnetic penguin contribution to $\imag A_2$
in this way, we can now write $\epsp$ in a more familiar way as

\begin{equation}  
\epsp = - \displaystyle\frac{i}{\sqrt{2}} \, e^{i ( \chi_2 - \chi_0 )} \, 
\omega_+  \,   \left[ 
\displaystyle\frac{\imag A_{0}^{(0)} }{ \real A_{0}^{(0)} } \, 
(1 - \Omega_{\rm eff}) - \displaystyle\frac{\imag A_{2}^{\rm emp}}{ \real
  A_{2}^{(0)} } \right]  
\label{eq:cpeff}
\end{equation} 
where 
\begin{eqnarray}  
\Omega_{\rm eff} &=& \Omega_{\rm IB} - \Delta_0 - f_{5/2}  
\label{eq:omegaeff} \\
\Omega_{\rm IB} &=& \displaystyle\frac{\real A_0^{(0)} }
{ \real A_2^{(0)} } \cdot \displaystyle\frac{\imag A_2^{\rm non-emp} }
{ \imag A_0^{(0)} } ~.
\end{eqnarray}  
The quantity $\Omega_{\rm eff}$ includes all effects to leading order
in isospin breaking and it generalizes the more traditional parameter
$\Omega_{\rm IB}$. Although $\Omega_{\rm IB}$ is in principle
enhanced by the large ratio $\real A_0^{(0)}/ \real A_2^{(0)}$ the
actual numerical analysis shows all three terms in
(\ref{eq:omegaeff}) to be relevant when both
strong and electromagnetic isospin violation are included.

\subsection{Numerical results}

We present numerical results for the following two cases:
\begin{enumerate} 
\item[i.] We calculate $\Omega_{\rm eff}$ and its components for 
  $\alpha=0$, i.e., we keep only terms proportional to the quark mass 
  difference (strong isospin violation). In this case, there is a clean
  separation of isospin violating effects in $\imag A_{2}$.
   We compare the
  lowest-order result of $\cO(m_u - m_d)$ with the full result of 
  $\cO[(m_u - m_d)p^2]$.
\item[ii.] Here we include electromagnetic corrections, comparing
  again $\cO(m_u - m_d,e^2 p^0)$ with $\cO[(m_u - m_d)p^2,e^2 p^2]$. In
  this case, the splitting  between $\imag A_2^{\rm emp}$ and 
  $\imag  A_2^{\rm non-emp}$ is performed at leading order in
  $1/N_c$.
\end{enumerate}

The LO entries depend on $\real g_8/\real g_{27}$ as well as on 
$\imag (g_8 g_{\rm ewk})/\imag g_8$.  Subleading effects in $1/N_c$ 
are known to
be sizable for the LECs of leading chiral order.  We will therefore
not use the large-$N_c$ values for $\real g_8$, $\real g_{27}$ in the
numerical analysis but instead determine these couplings from our fit
to the $K \to \pi\pi$ branching ratios.  

The other combination of interest is the ratio $\imag (g_8 g_{\rm
ewk})/\imag g_8$. In this case, existing calculations beyond
factorization~\cite{nonfact} suggest that the size of $1/N_c$ effects
is moderate, roughly $- (30 \pm 15)\%$.  As a consequence, it turns
out that the dominant uncertainty comes from the input parameters in
the factorized expressions.  Finally, one also needs the ratio $\imag
(g_8 g_{\rm ewk})^{\rm non-emp}/\imag g_8$: in this case, leading
large-$N_c$ implies $-3.1 \pm 1.8$ (error due to input parameters),
while the calculation of Ref.~\cite{Cirigliano:fv} gives $ -1.0 \pm
0.5 $.  Given the overlap between the two ranges and the large error
in the large-$N_c$ result, we use in the numerics the range 
implied by leading large-$N_c$.

At NLO the quantities we need to evaluate depend on the ratio of
next-to-leading to leading-order LECs. In Table \ref{tab:tab1}, we use
the leading $1/N_c$ estimates for the ratios $G_8 N_i/G_8$, \dots
The final error for each of the quantities $\Omega_{\rm IB}$,
$\Delta_0$, $f_{5/2}$ and $\Omega_{\rm eff}$ is obtained by adding in
quadrature the LO error and the one associated to weak LECs at NLO.
Moreover, only $f_{5/2}$ and $\real A_0^{(0)} / \real A_2^{(0)}$
depend on the ratio $g_8/g_{27}$. In these cases we rely on the
phenomenological value implied by our fit.  Some of the errors in
Table \ref{tab:tab1} are manifestly correlated, e.g., in the LO column
for $\alpha \neq 0$.

\renewcommand{\arraystretch}{1.1}
\begin{table}[ht]
\begin{center}
\begin{tabular}{|c|cc|cc|}
\hline
 & & & & \\
 & \multicolumn{2}{c|}{ $\alpha=0$}& \multicolumn{2}{c|}{ $\alpha \neq
 0$}  \\[5pt]
 &  \mbox{     } LO \mbox{     }  & \mbox{     } LO+NLO \mbox{     } & 
\mbox{     }  LO \mbox{     } & \mbox{     } LO+NLO \mbox{     } \\[5pt]
\hline
 & & & & \\
$\Omega_{\rm IB}$ & $11.7$ & $15.9 \pm 4.5$ & $ 18.0 \pm 6.5 $  & 
$ 22.7  \pm  7.6 $ \\[5pt]
$\Delta_0$ & $-$ $0.004$ & $-$ $0.41 \pm 0.05$ & $8.7 \pm 3.0$ & 
$ 8.4  \pm 3.6 $ \\[5pt]
$f_{5/2}$ & $0$ & $0$ & $ 0  $ & 
$ 8.3  \pm 2.4 $ \\[10pt]
\hline 
 & & & & \\
\mbox{   } $\Omega_{\rm eff}$ \mbox{   } & $11.7$ & $16.3 \pm 4.5$ & 
$9.3 \pm  5.8 $   &  $ 6.0  \pm  7.7$  \\[10pt]
\hline
\end{tabular}
\end{center}
\caption{Isospin violating corrections for $\epsp$ in units of
  $10^{-2}$. The first two columns refer to strong isospin violation
  only ($m_u \neq m_d$), the last two contain the complete results
  including electromagnetic corrections. LO and NLO denote leading and
  next-to-leading orders in CHPT. 
The small difference between the value of $f_{5/2}$ reported here and the 
one  in Eq.~(\ref{fit5}) is due to higher-order effects in isospin breaking 
(absent in this table according to Eq.~(\ref{eq:f52})).   
}
\label{tab:tab1}
\end{table}

The NLO results are obtained with the full expressions (\ref{eq:ima00}), 
\dots, (\ref{eq:f52}). Using instead the simplified expressions
corresponding to a fixed chiral order, the modified results are found
to be well within the quoted error bars. We therefore expect our errors 
to account also for higher-order effects in the chiral expansion.

We have also employed an alternative procedure for estimating the
non-leading weak LECs. In contrast to the previous analysis, we now
apply large $N_c$ directly to the LECs $G_8 N_i$, \dots This amounts
to assuming that the failure of large $N_c$ for $G_8$ is specific to
the leading chiral order and that the non-leading LECs are not
significantly enhanced compared to the large-$N_c$ predictions. Of
course, this implies that the local amplitudes of $\cO(G_8 p^4)$ are
less important than in the previous case. Consequently, the fitted
value for $g_8$ comes out quite a bit bigger than in 
Eq.~(\ref{fit2}), 
whereas $g_{27}$ gets smaller (see Sec.~\ref{subsec:fit}). 
However, the isospin
violating ratios in Table \ref{tab:tab1} are very insensitive to those
changes. Not only are the numerical values in this case well within
the errors displayed in Table \ref{tab:tab1} but they are in fact very
close to the central values given there.

Finally, we have investigated the impact of some subleading effects
in $1/N_c$ \cite{cenp03a}. Although this is not meant to be a
systematic expansion in $1/N_c$, the nonet breaking terms considered
in \cite{cenp03a} may
furnish another indication for the intrinsic uncertainties of some
of the LECs. The size of those terms depends on the assignment of
isosinglet scalar resonances.  Since nonet breaking effects are large
in the scalar sector they affect most of the entries in Table
\ref{tab:tab1} in a non-negligible way, although always within the
quoted uncertainties.  Employing scenario A for the lightest scalar
nonet \cite{cenp03a}, $\Omega_{\rm eff}$ in (\ref{eq:cpeff}) decreases
from $6.0 \cdot 10^{-2}$ to $-1.4 \cdot 10^{-2}$.

The lessons to be drawn from our analysis of isospin violating
corrections for $\epsp$ are straightforward. Separate parts of those
corrections turn out to be sizable. A well-known example are the
contributions of strong isospin violation to \pe ~mixing where the sum
of $\eta$ and $\eta^\prime$ exchange generates an $\Omega_{\rm IB}$ of
the order of 25 $\%$ \cite{etaetap}. However, already at the level of
\pe ~mixing alone, a complete calculation at next-to-leading order
\cite{Ecker:1999kr} produces a destructive interference in
$\Omega_{\rm IB}$.  
Additional contributions to the $K \to \pi\pi$
amplitudes from strong isospin violation at next-to-leading order
essentially cancel out. The final result  $\Omega_{\rm IB}= (15.9 \pm 4.5) 
\cdot 10^{-2}$  
is consistent  within errors  with the findings of  Ref.~\cite{Wolfe:2000rf}.
Inclusion of electromagnetic effects slightly increases
$\Omega_{\rm IB}$ and generates sizable $\Delta_0$ and $f_{5/2}$, which
interfere destructively with $\Omega_{\rm IB}$ to produce the final 
result $\Omega_{\rm eff}= (6.0 \pm 7.7)\cdot 10^{-2}$. 

It turns out that $\Delta_0$ is largely dominated by electromagnetic 
penguin contributions. In those theoretical calculations of $\epsp$
where electromagnetic penguin contributions are explicitly included
one may therefore drop $\Delta_0$ to a very good approximation.
Finally, if all electromagnetic corrections 
are included in theoretical calculations of $\imag A_0 / \real A_0$, 
$\imag A_2 / \real A_2$ and $\real A_2 / \real A_0$, 
$\Omega_{\rm eff}$ is essentially given by $\Omega_{\rm IB}$. In this 
case, $\Omega_{\rm eff}= (16.3 \pm 4.5)\cdot 10^{-2}$ is practically 
identical to the result based on \pe ~mixing only
\cite{Ecker:1999kr}. 

\section{Conclusions}
\label{sec:concl}
\renewcommand{\theequation}{\arabic{section}.\arabic{equation}}
\setcounter{equation}{0}

In most processes isospin violation induces a small effect on physical
amplitudes.  In $K \rightarrow \pi \pi$ decays, however, it is 
amplified by the $\Delta I=1/2$ rule: isospin breaking admixtures of
the dominant $\Delta I=1/2$ amplitudes can generate sizable
corrections to $\Delta I > 1/2$ amplitudes.  Understanding isospin
violation is crucial for a quantitative analysis of the $\Delta I=1/2$
rule itself and for a theoretical estimate of $\epsp$.
 
The theoretical description of $K$ decays involves a delicate
interplay between electro-weak and strong interactions in the
confinement regime. Chiral perturbation theory provides a
convenient framework for a systematic low-energy expansion of the
relevant amplitudes.
In this paper we have performed the first complete analysis of isospin
violation in $ K \rightarrow \pi \pi $ decays induced by the dominant
octet operators to NLO in CHPT.  We have reported explicit expressions
for loop and counterterm amplitudes, verifying cancellation of
ultraviolet divergences at NLO.

On the phenomenological side, the main features/results of this work are:
\begin{enumerate} 
%(1) 
\item
We have included for the first time both strong and
electromagnetic isospin violation in a joint analysis. 
%(2) 
\item
Nonleptonic weak amplitudes in CHPT depend on a number of
low-energy constants: we have used leading large-$N_c$
estimates for those constants which cannot be obtained by a fit to $K
\rightarrow \pi \pi$ branching ratios (i.e., all NLO couplings and the 
electroweak coupling of order $e^2 G_8 p^0$). Uncertainties within
this approach arise
from (i) input parameters in the leading $1/N_c$ expressions as well
as from (ii) potentially large subleading effects in $1/N_c$.  We have
discussed the impact of both (i) and (ii) on the relevant quantities. 

%(3)
\item
Using this large-$N_c$ input, we have
performed a fit to the CP-even component of the couplings $g_8$ and
$g_{27}$, both without and with inclusion of isospin breaking.  We
find that in general the inclusion of NLO effects (loops and
counterterms) has a significant impact on the output.  The main
outcome of the NLO fit is that both $g_8$ and $g_{27}$ are only mildly
affected by isospin breaking (e.g., $g_{27}$ gets shifted upwards by
only $2 \% $).  While this result is fully expected for $g_8$, in the
case of $g_{27}$ it arises from non-trivial cancellations between LO
and NLO corrections. 
For the ratio measuring the $\Delta I= 1/2$ enhancement in $K^0$
decays we find $\real A_0 / \real A_2 = 20.3 \pm 0.5$, compared to 
$22.2 \pm 0.1$ in the isospin limit.

\item  
Using as input a NLO calculation of electromagnetic corrections to
$\pi \pi$ scattering~\cite{ku98,kn02}, we have used the optical
theorem to study the effect of isospin breaking on the 
final-state-interaction
phases~\cite{Cirigliano:2000zw}.  According to our analysis,
isospin breaking leads to a discrepancy between the
theoretical prediction~\cite{cgl01} of $\delta_0 - \delta_2$ from
pion-pion scattering and its
phenomenological extraction from $K \rightarrow \pi \pi$
(see also Refs. \cite{Cirigliano:2000zw,gv00}).  Before
drawing a definite conclusion about the possible presence of an 
additional $\Delta I =5/2$ amplitude, more work is necessary to
understand this discrepancy.

\item
We have studied the effect of isospin violation on the direct
CP violation observable $\epsp$.  In this case isospin breaking
affects the destructive interference between the two main
contributions to $\epsp$ from normal and electromagnetic penguin
operators.
Apart from the traditional term $\Omega_{\rm IB}$, 
%describing strong penguin contributions to ${\rm Im} A_2$, 
we have identified and
studied the effect of isospin violation in the ratio $\imag A_0 /
\real A_0$, parametrized by the quantity $\Delta_0$ and the purely
electromagnetic $\Delta I=5/2$ amplitude.  Both $\Delta_0$ and the
$\Delta I=5/2$ contribution $f_{5/2}$ interfere
destructively with $\Omega_{\rm IB}$ to yield a final value
$\Omega_{\rm eff}= (6.0 \pm 7.7) \cdot 10^{-2}$ for the overall
measure of isospin violation in $\epsp$. 
If electromagnetic penguin contributions are included in theoretical
calculations of $\imag A_0 / \real A_0$, $\Delta_0$ can be dropped in
$\Omega_{\rm eff}$ to a very good approximation. Finally, if all
electromagnetic corrections are included in $\imag A_0 / \real A_0$,
$\imag A_2 / \real A_2$ and $\real A_2 / \real A_0$, $\Omega_{\rm
eff}$ is essentially determined by $\Omega_{\rm IB}$ and is
practically identical to the result based on $\pi^0-\eta$ mixing only.

\end{enumerate}

\vfill
\section*{Acknowledgements}
\noindent  

We thank Hans Bijnens, J\"urg Gasser and Ignazio Scimemi for useful 
exchanges, Gino Isidori and Guido M\"uller for collaboration at
earlier stages of this project, and J\"urg Gasser and Eduardo de
Rafael for collaboration at an even earlier stage. We are also
grateful to Ximo Prades for reminding us to consider the weak mass 
term in the effective octet Lagrangian. The work of V.C. and 
A.P. has been supported in part by MCYT, Spain (Grant
No. FPA-2001-3031) and by ERDF funds from the European Commission. 
In the final stage of this work  V. C. has been supported by a 
Sherman Fairchild fellowship from Caltech.

\newpage

\appendix 

\newcounter{zaehler}
\renewcommand{\thesection}{\Alph{zaehler}}
\renewcommand{\theequation}{\Alph{zaehler}.\arabic{equation}}

\setcounter{equation}{0}
\addtocounter{zaehler}{1}
\section{NLO  effective Lagrangians}
\label{app:NLOLag}

In this appendix we collect the relevant parts of the NLO Lagrangians. 

First we recall our notation. The gauge-covariant derivative of the
matrix field $U$ is denoted $D_\mu U$, the external scalar field
$\chi$ accounts for explicit symmetry breaking through the quark
masses, the matrix $\lambda = (\lambda_6 - i \lambda_7)/2$ 
projects onto the $\bar s\to \bar d$ transition and 
$Q= {\rm diag}(2/3,-1/3,-1/3)$ is the quark charge matrix. For
compactness of notation, we introduce the definitions
\begin{eqnarray} 
\chi_+^U = U^\dg \chi + \chi^\dg U ~, & \quad  \quad &
\chi_-^U = U^\dg \chi - \chi^\dg U \nl
& Q_U = U^\dg Q\, U ~. &
\end{eqnarray}

Starting with the strong Lagrangian (\ref{eq:Lstrong}), we have the
familiar terms \cite{gl85a}
\begin{eqnarray} 
 \sum_i\; L _i\ O^{p^4}_i &=&  L_4 \,\langle D_\mu U^\dg D^\mu U\rangle 
\langle \chi_+^U \rangle  + L_5 \,\langle D_\mu U^\dg 
D^\mu U \chi_+^U \rangle \nn \\[-3mm]
 &+&   L_7 \,\langle \chi_-^U \rangle^2  
 + L_8 \,\langle \chi^\dg U \chi^\dg U + \chi U^\dg \chi U^\dg\rangle
+ \dots
\label{eq:Lstrong4}
\end{eqnarray}
For the explicit form of the strong Lagrangian of $\cO(p^6)$ we refer
to Ref.~\cite{bce99}.

The electromagnetic Lagrangian (\ref{eq:Lelm}) 
is explicitly given by \cite{urech95}
\begin{eqnarray} 
 \sum_i\; K _i O^{e^2 p^2}_i  &=&  
K_1 \, \langle D_\mu U^\dagger D^\mu U \rangle \langle Q^2 \rangle + 
K_2 \, \langle D_\mu U^\dagger D^\mu U \rangle \langle Q Q_U \rangle 
\nn \\[-3mm]
&+& 
K_3 \, ( \langle D_\mu U^\dagger Q U \rangle^2 +  
\langle D_\mu U Q U^\dagger  \rangle^2 ) + 
K_4 \, \langle D_\mu U^\dagger Q U \rangle 
\langle D_\mu U Q U^\dagger \rangle 
\nn \\
&+& 
K_5 \, \langle \{ D_\mu U^\dagger, D^\mu U \} Q^2 \rangle +
K_6 \, \langle D_\mu U^\dagger D^\mu U  Q U^\dagger Q U + 
D_\mu U  D^\mu U^\dagger  Q U Q U^\dagger  \rangle 
\nn \\
&+&
K_7 \, \langle  \chi_+^U \rangle \langle  Q^2 \rangle +
K_8 \, \langle  \chi_+^U \rangle \langle  Q Q_U \rangle 
\nn \\
&+&
K_9 \, \langle  (\chi^\dagger U + U^\dagger \chi)  Q^2
+ (\chi U^\dagger + U \chi^\dagger)  Q^2 \rangle  
\nn \\
&+&
K_{10} \, \langle  (\chi^\dagger U + U^\dagger \chi)  Q U^\dagger Q U 
+ (\chi U^\dagger + U \chi^\dagger)  Q U Q U^\dagger \rangle  
\nn \\
&+&
K_{11} \, \langle  (\chi^\dagger U - U^\dagger \chi)  Q U^\dagger Q U 
+ (\chi U^\dagger - U \chi^\dagger)  Q U Q U^\dagger \rangle  
\nn \\
&+&
K_{12} \, \langle  D_\mu U^\dagger [D^\mu Q_R,Q] U + 
D_\mu U  [D^\mu Q_L,Q] U^\dagger \rangle 
\nn \\
&+&
K_{13} \, \langle  D^\mu Q_R U  D_{\mu} Q_L  U^\dagger \rangle \ ,  
\label{eq:Lurech}
\end{eqnarray}
with ($l_\mu$ and $r_\mu$ denote external spin-1 fields)
\beq
D_{\mu} Q_L = \partial_\mu Q  - i\, [l_\mu, Q] ~, \qquad 
D_{\mu} Q_R = \partial_\mu Q  - i\, [r_\mu, Q]  \ . 
\eeq

Turning to the nonleptonic weak Lagrangian, we first display the octet
couplings in the notation of Ref.~\cite{ekw93}:
\begin{eqnarray} 
\sum_i\; N_i O^8_i &=&  N_5\; \langle \lambda \{\chi_+^U, 
D_\mu U^\dg D^\mu U\} \rangle +
N_6\; \langle \lambda D_\mu U^\dg U\rangle \langle U^\dg D^\mu U
\chi_+^U \rangle \nn \\[-3mm]
 &+& N_7\; \langle \lambda \chi_+^U \rangle \langle 
D_\mu U^\dg D^\mu U \rangle +
N_8\; \langle \lambda D_\mu U^\dg D^\mu U\rangle \langle \chi_+^U
\rangle \nl
& + & N_9\; \langle \lambda [\chi_-^U, D_\mu U^\dg D^\mu U]
\rangle + N_{10}\; \langle \lambda (\chi_+^U)^2\rangle 
 +  N_{11}\; \langle \lambda \chi_+^U \rangle
\langle \chi_+^U \rangle \nl 
& + & N_{12}\; \langle \lambda (\chi_-^U)^2\rangle 
 +  N_{13}\; \langle \lambda \chi_-^U \rangle \langle \chi_-^U
\rangle + \dots  \label{eq:Loctet4}
\end{eqnarray} 

The corresponding 27-plet couplings \cite{kmw90} are 

\bea
\sum_i
D_i O^{27}_i &=& D_1  \ 
t_{ij,kl} \,  \langle \lambda_{ij} \chi_+^U \rangle 
\langle \lambda_{kl} \chi_+^U \rangle  + 
D_2 \
t_{ij,kl} \,  \langle \lambda_{ij} \chi_-^U \rangle 
\langle \lambda_{kl} \chi_-^U \rangle  \nn \\[-3mm]
&-& D_4 \ 
t_{ij,kl} \,  \langle \lambda_{ij} U^\dagger D_{\mu} U  \rangle 
\langle \lambda_{kl}  \{ \chi_+^U, U^\dagger D^{\mu} U \} \rangle
\nn  \\
&+& D_5 \
t_{ij,kl} \,  \langle \lambda_{ij} U^\dagger D_{\mu} U  \rangle 
\langle \lambda_{kl} [ \chi_-^U, U^\dagger D^{\mu} U ] \rangle
\nn  \\
&+& D_6 \ 
t_{ij,kl} \, \langle \lambda_{ij} \chi_+^U \rangle 
\langle \lambda_{kl} D_\mu U^\dagger D^\mu U \rangle  
\nn  \\
&-& D_7 \
t_{ij,kl} \, \langle \lambda_{ij} U^\dagger D_{\mu} U  \rangle
\langle \lambda_{kl} U^\dagger D^{\mu} U  \rangle 
\langle \chi_+^U \rangle 
\label{eq:L274}
\eea
with $(\lambda_{ij})_{ab}= \delta_{i a} \delta_{j b}$.  The 
non-zero components of $t_{ij,kl}$ are given by ($i,j,k,l=1,2,3$)
\bea
t_{21,13} &=& t_{13,21} = t_{23,11} = t_{11,23}=\frac{1}{3}   
\nn \\  
t_{22,23} &=& t_{23,22} = t_{23,33} = t_{33,23}=-\frac{1}{6} \ .   
\eea

Finally, the relevant part of the electroweak Lagrangian of
$O(e^2 G_8 p^2)$ \cite{eimnp00} is 
\begin{eqnarray} 
\sum_i\; Z_i O^{EW}_i &=& Z_1 \langle \lambda \{ Q_U, \chi_+^U \}
\rangle + Z_2 \langle \lambda Q_U \rangle \langle \chi_+^U \rangle 
+ Z_3 \langle \lambda Q_U \rangle \langle \chi_+^U Q_U \rangle 
\nn \\[-3mm] 
&+& Z_4 \langle \lambda \chi_+^U \rangle \langle Q Q_U \rangle 
+ Z_5 \langle \lambda D_\mu U^\dg D^\mu U \rangle 
+ Z_6 \langle \lambda \{Q_U,D_\mu U^\dg D^\mu U\} \rangle \nl
&+& Z_7 \langle \lambda D_\mu U^\dg D^\mu U \rangle \langle Q Q_U \rangle
+ Z_8 \langle \lambda D_\mu U^\dg U \rangle \langle Q U^\dg D^\mu U 
\rangle \label{eq:Lew4} \\
&+& Z_9 \langle \lambda D_\mu U^\dg U \rangle \langle Q_U U^\dg 
D^\mu U \rangle
+ Z_{10} \langle \lambda D_\mu U^\dg U \rangle \langle \{Q,Q_U\}
U^\dg D^\mu U \rangle \nl
&+& Z_{11} \langle \lambda \{Q_U,U^\dg D_\mu U\} \rangle 
\langle Q D^\mu U^\dg U \rangle 
+ Z_{12} \langle \lambda \{Q_U,U^\dg D_\mu U\} \rangle \langle 
Q_U D^\mu U^\dg U \rangle + \dots \nn
\end{eqnarray}

\setcounter{equation}{0}
\addtocounter{zaehler}{1}
\section{Explicit form of NLO  loop amplitudes }
\label{app:loops}

In this appendix we report explicit expressions for the NLO loop
corrections $\Delta_L \cA_{n}^{(X)}$ appearing in the master formulas
of Eqs.~(\ref{eq:structure1}) and (\ref{eq:structure2}).

\subsection{Photonic amplitudes}
Let us start with the terms arising from exchange of virtual photons.
The amplitude $\cA_{+-}$ is infrared divergent and is regulated by
introducing a fictitious photon mass $M_\gamma$.  Moreover, it is
convenient to work with a subtracted amplitude, after removing the
infrared component $ \cA_{+-}^{\rm IR}$ (see 
discussion in Sec.~\ref{sec:chptNLO}).   
The function $B_{+-} (M_\gamma)$ appearing in our
definition of the infrared-divergent amplitude $ \cA_{+-}^{\rm IR}$ 
(see Eq.~(\ref{Eq:tour1})) is given by 
\bea
{B}_{+-} ( M_\gamma) &=& {1 \over 4 \pi} \bigg[
2 a (\beta) \log {M_\pi^2 \over M_\gamma^2} +
{1 + \beta^2 \over 2 \beta} h ( \beta)
+ 2 + \beta \log {1 + \beta \over 1 - \beta}
\nonumber \\
& & \phantom{xxxxx} 
+ i \pi \left( {1 + \beta^2 \over \beta}
\log { M_K^2 \beta^2 \over M_\gamma^2} - \beta \right) \bigg] \ \ ,
\label{eq:appD1} 
\eea
where
\bea
\beta &=& (1 - 4 M_\pi^2 /M_K^2 )^{1/2} \nl
a (\beta) &=& 1 + {1 + \beta^2 \over 2 \beta}
\log {1 - \beta \over 1 + \beta} \nl
h ( \beta) &=& \pi^2 + \log {1 + \beta \over 1 - \beta}
\log {1 - \beta^2  \over 4 \beta^2 } + 2 f \left(
{1 + \beta \over 2 \beta} \right) - 2 f
\left( { \beta - 1 \over 2 \beta} \right) 
\label{eq:appD2} \nl
f(x) &=& - \int_0^x dt ~ {1\over t} \log|1 - t| ~.
\eea
The amplitudes $\Delta_L \cA_{n}^{(\gamma)}$ are given by 
\bea
\Delta_L \cA_{1/2}^{(\gamma)} &=&  
\frac{\sqrt{2}}{(4 \pi F_\pi)^2} \left[- \frac{14}{3} M_\pi^2   
+ 2 M_K^2 \left(1 + \log 
\frac{M_\pi^2}{\nu_\chi^2} \right) \right] + 
\frac{ 4 \sqrt{2} M_K^2}{F_\pi^2} \Lambda (\nu_\chi)   \\
\Delta_L \cA_{3/2}^{(\gamma)} &=&  
\frac{1}{(4 \pi F_\pi)^2} \left[ -\frac{14}{3} M_\pi^2 
+ \frac{4}{5} (M_K^2 + \frac{3}{2} M_\pi^2)  \left(1 + \log 
\frac{M_\pi^2}{\nu_\chi^2} \right) \right] \nn \\
 &+& 
\frac{8}{5 F_\pi^2} (M_K^2 + \frac{3}{2} M_\pi^2) 
\Lambda (\nu_\chi)   \\
\Delta_L \cA_{5/2}^{(\gamma)} &=&  \frac{6}{5} \, 
\frac{M_K^2 - M_\pi^2}{(4 \pi F_\pi)^2} 
\left(1 + \log  \frac{M_\pi^2}{\nu_\chi^2} \right)  + 
\frac{12 (M_K^2 - M_\pi^2)}{5 F_\pi^2} \Lambda (\nu_\chi)  ~. 
\eea
The divergent factor $\Lambda (\nu_\chi)$ is defined in 
Eq.~(\ref{eq:div}).

\subsection{Non-photonic amplitudes}

The mesonic loop corrections can be expressed in terms of the following 
basic function (and its derivatives):
\bea
J (p^2, M_1^2, M_2^2 ) &=& \frac{1}{i} \int 
\displaystyle\frac{d^d k}{(2 \pi)^d} \, 
\displaystyle\frac{1}{ 
\left[ k^2 - M_1^2 \right] \, \left[ (k - p)^2 - M_2^2 \right] } \nl
&=& \bar{J} (p^2, M_1^2, M_2^2) + J (0, M_1^2, M_2^2)  \ .  
\eea
The subtraction term is given by 
\bea
J (0, M_1^2, M_2^2) &=& \displaystyle\frac{ 
M_1^2 \, T (M_1^2)  - M_2^2 \, T (M_2^2) }{
M_1^2 - M_2^2}  \ - \ 2 \, \Lambda (\nu_\chi)  \\
T (M^2) &=& - \displaystyle\frac{1}{(4 \pi)^2} \, 
\log \frac{M^2}{\nu_\chi^2} \ . \label{eq:TM}
\eea
Expansion around the neutral meson masses generates terms 
involving derivatives of the function  $\bar{J} (p^2, M_1^2, M_2^2)$. 
In order to deal with such terms we use the notation 
\bea 
\bar{J}^{(1,0,0)} (p^2, M_1^2, M_2^2 ) &=& \displaystyle\frac{
\partial }{\partial p^2} \,  \bar{J} (p^2, M_1^2, M_2^2) \nl
\bar{J}^{(0,1,0)} (p^2, M_1^2, M_2^2 ) &=& \displaystyle\frac{
\partial }{\partial M_1^2} \,  \bar{J} (p^2, M_1^2, M_2^2) \nl
\bar{J}^{(0,0,1)} (p^2, M_1^2, M_2^2 ) &=& \displaystyle\frac{
\partial }{\partial M_2^2} \,  \bar{J} (p^2, M_1^2, M_2^2)  \ . 
\eea
We report below the explicit form of the relevant functions. 
For this purpose we define
\beq
\lambda (t,x,y)  = 
%%   t^2 + x^2 + y^2   - 2 ( x y + x t + y t )  =  
\left[ t - \left(\sqrt{x} + \sqrt{y} \right)^2 \right] \, 
\left[ t - \left(\sqrt{x} - \sqrt{y} \right)^2 \right]    
\eeq
and 
\beq 
\Sigma_{12}  =  M_1^2 + M_2^2~, \qquad \Delta_{12}  =  M_1^2 -
M_2^2    \ .  
\eeq
Then 
\bea
\bar{J} (p^2,M_1^2,M_2^2) & = &
\frac{1}{32 \pi^2} \Bigg[ 2 + 
\frac{\Delta_{12}}{p^2} 
\log \frac{M_2^2}{M_1^2} - \frac{\Sigma_{12}}{\Delta_{12}}
\log \frac{M_2^2}{M_1^2}   
\nn \\
&-&   \frac{\lambda^{1/2} (p^2,M_1^2,M_2^2)}{p^2} \times
\log  \left( \frac{[p^2 + \lambda^{1/2} (p^2,M_1^2,M_2^2)]^2 - 
\Delta_{12}^2}{[p^2 - 
\lambda^{1/2} (p^2,M_1^2,M_2^2)]^2 - \Delta_{12}^2} \right) \Bigg] 
\nn \\
\bar{J} (p^2,M^2,M^2) & = & \displaystyle\frac{1}{16 \pi^2} \, 
\left[ 2 - \sigma \log  \left(\frac{ \sigma + 1}{\sigma - 1} \right) 
\right] , 
\quad \sigma \equiv  \sqrt{ \lambda \left( 1,M^2 / p^2,
M^2 / p^2 \right) } \ .
\eea
The relevant derivative functions are reported below (recalling the
symmetry property\footnote{In the equal mass case we adopt the definition
$$ 
\bar{J}^{(0,0,1)} (p^2,M^2,M^2) \equiv \lim_{M_2 \rightarrow M} \, 
\bar{J}^{(0,0,1)} (p^2,M_2^2,M^2) = 
\frac{1}{2} \frac{\partial}{\partial M^2} \bar{J} (p^2,M^2,M^2)
\ .  $$} 
\\ $\bar{J}^{(0,1,0)} (p^2,M_1^2,M_2^2) =
\bar{J}^{(0,0,1)} (p^2,M_2^2,M_1^2)$):
\bea
\bar{J}^{(1,0,0)} (p^2,M_1^2,M_2^2) & = & \frac{1}{32 \pi^2} \, \Bigg\{
-\displaystyle\frac{2}{p^2} 
-\displaystyle\frac{\Delta_{12}}{(p^2)^2} 
\log \frac{M_2^2}{M_1^2} 
\nn \\
 & - &
\displaystyle\frac{ (p^2 \, \Sigma_{12}  - \Delta_{12}^2)}{
(p^2)^2  \, \lambda^{1/2} (p^2,M_1^2,M_2^2)
} \, 
\log  \left( \frac{\Sigma_{12} - p^2 -  \lambda^{1/2} (p^2,M_1^2,M_2^2)}{
\Sigma_{12} - p^2 +  \lambda^{1/2} (p^2,M_1^2,M_2^2)} \right)
\Bigg\} 
\nn \\
\bar{J}^{(0,0,1)} (p^2,M_1^2,M_2^2) & = & \frac{1}{32 \pi^2} \,  \Bigg\{
- \displaystyle\frac{2}{\Delta_{12}} - 
\displaystyle\frac{ \Delta_{12}^2 + 2 M_1^2 p^2 
}{p^2 \,  \Delta_{12}^2} 
\log \frac{M_2^2}{M_1^2} 
\nn \\
&+&
\displaystyle\frac{ p^2 + \Delta_{12}}{
 p^2  \, \lambda^{1/2} (p^2,M_1^2,M_2^2)
} \, 
\log  \left( \frac{\Sigma_{12} - p^2 -  \lambda^{1/2} (p^2,M_1^2,M_2^2)}{
\Sigma_{12} - p^2 +  \lambda^{1/2} (p^2,M_1^2,M_2^2)} \right)
\Bigg\} 
\nn \\
\bar{J}^{(0,0,1)} (p^2,M^2,M^2) & = &
\frac{1}{32 \pi^2} \,  \Bigg\{
\displaystyle\frac{1}{M^2} + 
\displaystyle\frac{2}{p^2 \, \sigma } \, 
\log  \left(\frac{ \sigma + 1}{\sigma - 1} \right) 
\Bigg\} \ . 
\eea

We recall here that we expand all our amplitudes around the neutral  
pion and kaon masses $M_\pi$ and $M_K$ to define the isospin limit
(see Ref.~\cite{grs03} for a more general discussion of the
splitting between strong and electromagnetic contributions). This
applies also to the $\eta$ mass given in Eq.~(\ref{eq:treemass}). 
Therefore, in all (loop) amplitudes where $M^2_\eta$ appears
explicitly it actually stands for $(4 M^2_K - M^2_\pi)/3$ instead of
the physical value in (\ref{eq:treemass}). This concerns all loop 
functions in Apps.~B and C.

\subsubsection{$\Delta I= 1/2$ amplitudes}

In this section we list the one-loop corrections to the $\Delta I=
1/2$ amplitude.

\bea
\Delta_L \cA_{1/2}^{(27)} &=& 
- \frac{M_{\pi}^2}{2\,F_{\pi}^2} \,\bar{J}(M_K^2,M_{\eta}^2,M_{\eta}^2)
+ 
\frac{\left( 2\,M_K^2 - M_{\pi}^2 \right)}{2\,F_{\pi}^2}  
\,\bar{J}(M_K^2,M_{\pi}^2,M_{\pi}^2) 
\nn \\ 
&+&
  \frac{M_K^4}{3\,F_{\pi}^2\,M_{\pi}^2} 
\,\bar{J}(M_{\pi}^2,M_K^2,M_{\eta}^2) 
-
  \frac{M_K^2\,\left( M_K^2 - 4\,M_{\pi}^2 \right)}
{4\,F_{\pi}^2\,M_{\pi}^2} \,
\bar{J}(M_{\pi}^2,M_K^2,M_{\pi}^2) 
\nn \\
&-& 
  \frac{3\,\left( 12\,M_K^4 - 11\,M_K^2\,M_{\pi}^2 + 
3\,M_{\pi}^4 \right)}{8\,F_{\pi}^2\,
\left( -M_K^2 + M_{\pi}^2 \right) }  \,T(M_{\eta}^2)
+ 
  \frac{\left( 6\,M_K^4 - 11\,M_K^2\,M_{\pi}^2 \right)}
{4\,F_{\pi}^2\,\left( 
M_K^2 - M_{\pi}^2 
   \right) }  \,T(M_K^2) 
\nn \\
&+&
\frac{\left( 8\,M_K^4 - 35\,M_K^2\,M_{\pi}^2 + 25\,M_{\pi}^4 \right)}
{8\,F_{\pi}^2\,
\left( M_K^2 - M_{\pi}^2 \right) }  \,T(M_{\pi}^2) 
+ \frac{-M_K^2 + M_{\pi}^2}{16\,F_{\pi}^2\,{\pi }^2} 
\\
\Delta_L \cA_{1/2}^{(8)} &=& 
  \frac{M_{\pi}^2}{18\,F_{\pi}^2} \,\bar{J}(M_K^2,M_{\eta}^2,M_{\eta}^2)
+ 
  \frac{\left( 2\,M_K^2 - M_{\pi}^2 \right)}{2\,F_{\pi}^2} 
 \,\bar{J}(M_K^2,M_{\pi}^2,M_{\pi}^2)
\nn \\
&-& 
  \frac{M_K^4}{12\,F_{\pi}^2\,M_{\pi}^2} 
\,\bar{J}(M_{\pi}^2,M_K^2,M_{\eta}^2)
- 
  \frac{\left( M_K^4 - 4\,M_K^2\,M_{\pi}^2 \right)}
{4\,F_{\pi}^2\,M_{\pi}^2} 
 \,\bar{J}(M_{\pi}^2,M_K^2,M_{\pi}^2)
\nn \\
&-& 
  \frac{\left( 36\,M_K^4 - 73\,M_K^2\,M_{\pi}^2 + 19\,M_{\pi}^4 \right)
  }{72\,F_{\pi}^2\,\left( M_K^2 - M_{\pi}^2 \right) }\,T(M_{\eta}^2)  
+ 
\frac{M_K^2}{4\,F_{\pi}^2}\,T(M_K^2)
\nn \\
&+& 
  \frac{\left( 8\,M_K^4 - 35\,M_K^2\,M_{\pi}^2 + 25\,M_{\pi}^4 \right)}
   {8\,F_{\pi}^2\,\left( M_K^2 - M_{\pi}^2 \right) }  \,T(M_{\pi}^2)
+ \frac{-9\,M_K^2 + 4\,M_{\pi}^2}{144\,F_{\pi}^2\,{\pi }^2} 
 \\
\Delta_L \cA_{1/2}^{(\epsilon)} &=& 
+ 
  \frac{5\,M_{\pi}^2}{6\,F_{\pi}^2} \,\bar{J}(M_K^2,M_{\eta}^2,M_{\eta}^2)
+ 
  \frac{3\,M_K^2}{2\,F_{\pi}^2} \,\bar{J}(M_K^2,M_K^2,M_K^2)
\nn \\
&+& 
  \frac{\left( 5\,M_K^2 - 6\,M_{\pi}^2 \right)}{3\,F_{\pi}^2}  
\,\bar{J}(M_K^2,M_{\pi}^2,M_{\eta}^2)
+ 
  \frac{\left( 2\,M_K^2 - M_{\pi}^2 \right)}{2\,F_{\pi}^2}  
\,\bar{J}(M_K^2,M_{\pi}^2,M_{\pi}^2)
\nn \\
&-& 
  \frac{\left( 3\,M_K^4 - 4\,M_K^2\,M_{\pi}^2 \right)}
{12\,F_{\pi}^2\,M_{\pi}^2} 
 \,\bar{J}(M_{\pi}^2,M_K^2,M_{\eta}^2)
- 
\frac{\left( M_K^4 - 2\,M_K^2\,M_{\pi}^2 \right)}{F_{\pi}^2\,M_{\pi}^2} 
 \,\bar{J}(M_{\pi}^2,M_{\pi}^2,M_K^2)
\nn \\
&-&
\frac{\left( 22\,M_K^4 - 71\,M_K^2\,M_{\pi}^2 + 43\,M_{\pi}^4 \right)
}{12\,F_{\pi}^2\,\left( M_K^2 - M_{\pi}^2 \right) }  \, T(M_{\eta}^2)
+ 
  \frac{5\,M_K^2\,M_{\pi}^2}{4\,F_{\pi}^2
\,\left( M_K^2 - M_{\pi}^2 \right) } 
\,T(M_K^2) \nl
&+& 
  \frac{\left( 4\,M_K^4 - 11\,M_K^2\,M_{\pi}^2 \right) 
}{4\,F_{\pi}^2\,\left( M_K^2 - M_{\pi}^2 \right) } \,T(M_{\pi}^2)
+ 
\frac{-8\,M_K^4 - 22\,M_K^2\,M_{\pi}^2 + 7\,M_{\pi}^4}
   {32\,F_{\pi}^2\,\left( 4\,M_K^2 - M_{\pi}^2 \right) \,{\pi }^2} 
\nn \\
&+& 
  \frac{4\,M_{\pi}^2\,\left( M_K^2 - M_{\pi}^2 \right) 
   }{9\,F_{\pi}^2}  \,\bar{J}^{(0,0,1)}(M_K^2,M_{\eta}^2,M_{\eta}^2) 
\nn \\
&-&
 \frac{M_K^4\,\left( M_K^2 - M_{\pi}^2 \right)
 }{3\,F_{\pi}^2\,M_{\pi}^2}  
\,  \left[   \bar{J}^{(0,0,1)}(M_{\pi}^2,M_K^2,M_{\eta}^2) 
  + \bar{J}^{(0,1,0)}(M_{\pi}^2,M_K^2,M_{\eta}^2) \right]
\\
\Delta_L \cA_{1/2}^{(Z)} &=& 
- 
  \frac{3\,M_K^2}{8\,F_{\pi}^2} \,\bar{J}(M_K^2,M_K^2,M_K^2)
- 
  \frac{\left( 2\,M_K^2 - 3\,M_{\pi}^2 \right)}{2\,F_{\pi}^2}
  \,\bar{J}(M_K^2,M_{\pi}^2,M_{\pi}^2)
\nn \\
&-& 
  \frac{\left( 2\,M_K^6 + 5\,M_K^4\,M_{\pi}^2 - 4\,M_K^2\,M_{\pi}^4 \right) 
}{24\,F_{\pi}^2\,M_{\pi}^4}  \,\bar{J}(M_{\pi}^2,M_K^2,M_{\eta}^2)
\nn \\
&-& 
\frac{\left( M_K^6 - 4\,M_K^2\,M_{\pi}^4 \right)}{4\,F_{\pi}^2\,M_{\pi}^4}  
\,\bar{J}(M_{\pi}^2,M_K^2,M_{\pi}^2)
\nn \\
&+& 
\frac{M_{\pi}^2\,\left( 4\,M_K^2 - M_{\pi}^2 \right)}{
8\,F_{\pi}^2\,\left( M_K^2 - M_{\pi}^2 \right) }  \,T(M_{\eta}^2)
- 
  \frac{3\,\left( 7\,M_K^4 - 7\,M_K^2\,M_{\pi}^2 + M_{\pi}^4 \right)}
   {8\,F_{\pi}^2\,\left( M_K^2 - M_{\pi}^2 \right) }  \,T(M_K^2)
\nn \\
&+& 
  \frac{\left( 7\,M_K^2 - 18\,M_{\pi}^2 \right)}{8\,F_{\pi}^2}  \,T(M_{\pi}^2)
+ 
\frac{8\,M_K^4 -
 13\,M_K^2\,M_{\pi}^2 + 2\,M_{\pi}^4}{128\,F_{\pi}^2\,M_{\pi}^2\,{\pi }^2} 
\nn \\
&-& 
  \frac{\left( 2\,M_K^4 - 3\,M_K^2\,M_{\pi}^2 + M_{\pi}^4 \right)
}{F_{\pi}^2}  \, \bar{J}^{(0,0,1)} (M_K^2,M_{\pi}^2,M_{\pi}^2)
\nn \\
&+& 
  \frac{\left( M_K^6 - 5 M_K^4\,M_{\pi}^2 + 4 M_K^2\,M_{\pi}^4 \right)
 }{4\,F_{\pi}^2\,M_{\pi}^2}   \left[ 
\bar{J}^{(0,0,1)}(M_{\pi}^2,M_K^2,M_{\pi}^2) + 
\bar{J}^{(1,0,0)}(M_{\pi}^2,M_K^2,M_{\pi}^2) \right]
\nn \\
& + & 
  \frac{\left( M_K^6 - M_K^4\,M_{\pi}^2 \right) }{12\,F_{\pi}^2\,M_{\pi}^2} 
\, \left[ \bar{J}^{(0,1,0)}(M_{\pi}^2,M_K^2,M_{\eta}^2)
+ \bar{J}^{(1,0,0)}(M_{\pi}^2,M_K^2,M_{\eta}^2) \right]
\\
\Delta_L \cA_{1/2}^{(g)} &=& 
- 
  \frac{3\,M_K^2}{8\,F_{\pi}^2} \,\bar{J}(M_K^2,M_K^2,M_K^2)
+ 
  \frac{\left( 2\,M_K^2 - M_{\pi}^2 \right)}{2\,F_{\pi}^2}
  \,\bar{J}(M_K^2,M_{\pi}^2,M_{\pi}^2)
\nn \\
&-& 
  \frac{M_K^4}{8\,F_{\pi}^2\,M_{\pi}^2} \,\bar{J}(M_{\pi}^2,M_K^2,M_{\eta}^2)
- 
  \frac{\left( M_K^4 - 4\,M_K^2\,M_{\pi}^2 \right)}{4\,F_{\pi}^2\,M_{\pi}^2}  
\,\bar{J}(M_{\pi}^2,M_K^2,M_{\pi}^2)
\nn \\
&-& 
  \frac{\left( 8\,M_K^4 - 6\,M_K^2\,M_{\pi}^2 + M_{\pi}^4 \right)}{
8\,F_{\pi}^2\,\left( M_K^2 - M_{\pi}^2 \right)  } \,T(M_{\eta}^2)
+ 
  \frac{\left( 2\,M_K^4 + 7\,M_K^2\,M_{\pi}^2 \right)}{
8\,F_{\pi}^2\,\left( M_K^2 - M_{\pi}^2 \right) }  \,T(M_K^2)
\nn \\
&+& 
  \frac{\left( 8\,M_K^4 - 35\,M_K^2\,M_{\pi}^2 + 21\,M_{\pi}^4 \right)}{
   8\,F_{\pi}^2\,\left( M_K^2 - M_{\pi}^2 \right) }  \,T(M_{\pi}^2)
+ 
\frac{-5\,M_K^2 + 4\,M_{\pi}^2}{128\,F_{\pi}^2\,{\pi }^2}~.
\eea

\subsubsection{$\Delta I= 3/2$ amplitudes}

In this section we list the one-loop 
corrections to the $\Delta I= 3/2$ amplitude. 

\bea
\Delta_L \cA_{3/2}^{(27)} &=& 
- 
  \frac{\left( M_K^2 - 2\,M_{\pi}^2 \right)}{2\,F_{\pi}^2}  
\,\bar{J}(M_K^2,M_{\pi}^2,M_{\pi}^2)
- 
  \frac{M_K^4}{24\,F_{\pi}^2\,M_{\pi}^2} 
\,\bar{J}(M_{\pi}^2,M_K^2,M_{\eta}^2)
\nn \\
&-& 
  \frac{M_K^2\,\left( 5\,M_K^2 - 8\,M_{\pi}^2 \right) }
{8\,F_{\pi}^2\,M_{\pi}^2} 
\,\bar{J}(M_{\pi}^2,M_K^2,M_{\pi}^2)
\nn \\
&+& 
  \frac{M_{\pi}^2\,\left( 4\,M_K^2 - M_{\pi}^2 \right)
 }{8\,F_{\pi}^2\,\left( M_K^2 - M_{\pi}^2 \right) } 
 \,T(M_{\eta}^2)
+ 
  \frac{\left( 3\,M_K^4 + M_K^2\,M_{\pi}^2 \right) 
}{4\,F_{\pi}^2\,\left( M_K^2 - M_{\pi}^2 \right) }  
\,T(M_K^2)
\nn \\
&-& 
  \frac{\left( 4\,M_K^4 - 22\,M_K^2\,M_{\pi}^2 + 29\,M_{\pi}^4 \right) 
}{8\,F_{\pi}^2\,\left( M_K^2 - M_{\pi}^2 \right) }
\,T(M_{\pi}^2)
+ 
\frac{M_K^2 - 2\,M_{\pi}^2}{32\,F_{\pi}^2\,{\pi }^2} 
\\
\Delta_L \cA_{3/2}^{(\epsilon)} &=& 
+ 
  \frac{M_K^2}{6\,F_{\pi}^2}  \,\bar{J}(M_K^2,M_{\pi}^2,M_{\eta}^2)
- 
  \frac{\left( M_K^2 - 2\,M_{\pi}^2 \right)}{2\,F_{\pi}^2} 
 \,\bar{J}(M_K^2,M_{\pi}^2,M_{\pi}^2)
\nn \\
&-& 
  \frac{\left( 21 M_K^4 - 8 M_K^2\,M_{\pi}^2 \right) 
 }{24\,F_{\pi}^2\,M_{\pi}^2}  \,\bar{J}(M_{\pi}^2,M_K^2,M_{\eta}^2)
- 
\frac{\left( 11 M_K^4 - 16 M_K^2\,M_{\pi}^2 \right) 
}{8\,F_{\pi}^2\,M_{\pi}^2} \, \bar{J}(M_{\pi}^2,M_{\pi}^2,M_K^2)
\nn \\
&-& 
  \frac{\left( 44\,M_K^4 - 47\,M_K^2\,M_{\pi}^2 + 9\,M_{\pi}^4 \right) 
}{24\,F_{\pi}^2\,\left( M_K^2 - M_{\pi}^2 \right) }  
\,T(M_{\eta}^2)
- 
  \frac{\left( 12\,M_K^4 - 17\,M_K^2\,M_{\pi}^2 \right) 
}{4\,F_{\pi}^2\,\left( M_K^2 - M_{\pi}^2 \right) }  
\,T(M_K^2)
\nn \\
&-& 
  \frac{\left( 4\,M_K^4 - 11\,M_K^2\,M_{\pi}^2 + 15\,M_{\pi}^4 \right) 
}{8\,F_{\pi}^2\,\left( M_K^2 - M_{\pi}^2 \right) }  
\,T(M_{\pi}^2)
+
\frac{2\,M_K^2 - M_{\pi}^2}{16\,F_{\pi}^2\,{\pi }^2} 
\nn \\
&+& 
  \frac{M_K^4\,\left( M_K^2 - M_{\pi}^2 \right) }{6\,F_{\pi}^2\,M_{\pi}^2}
\,\bar{J}^{(0,1,0)}(M_{\pi}^2,M_K^2,M_{\eta}^2)
\\
\Delta_L \cA_{3/2}^{(Z)} &=& 
- 
  \frac{\left( 13\,M_K^2 - 18\,M_{\pi}^2 \right)}{10\,F_{\pi}^2} 
 \,\bar{J}(M_K^2,M_{\pi}^2,M_{\pi}^2) 
\nn \\
&-& 
  \frac{\left( 10\,M_K^6 + 13\,M_K^4\,M_{\pi}^2 - 
32\,M_K^2\,M_{\pi}^4 + 24\,M_{\pi}^6 \right) 
}{120\,F_{\pi}^2\,M_{\pi}^4}  \, \bar{J}(M_{\pi}^2,M_K^2,M_{\eta}^2)
\nn \\
&-& 
  \frac{\left( 10\,M_K^6 + 3\,M_K^4\,M_{\pi}^2 - 28\,M_K^2\,M_{\pi}^4 \right)
}{40\,F_{\pi}^2\,M_{\pi}^4}  \, \bar{J}(M_{\pi}^2,M_K^2,M_{\pi}^2)
\nn \\
&+& 
  \frac{\left( 48\,M_K^4 - 40\,M_K^2\,M_{\pi}^2 + 7\,M_{\pi}^4 \right) 
}{40\,F_{\pi}^2\,\left( M_K^2 - M_{\pi}^2 \right) } 
\,T(M_{\eta}^2)
- 
  \frac{3\,\left( 21\,M_K^4 - 20\,M_K^2\,M_{\pi}^2 \right)
}{20\,F_{\pi}^2\,\left( M_K^2 - M_{\pi}^2 \right) }  
 \,T(M_K^2)
\nn \\
&-& 
  \frac{\left( 58\,M_K^4 - 22\,M_K^2\,M_{\pi}^2 - 27\,M_{\pi}^4 \right) 
}{40\,F_{\pi}^2\,\left( M_K^2 - M_{\pi}^2 \right) }  
\,T(M_{\pi}^2)
+
\frac{-M_K^4 + 14\,M_K^2\,M_{\pi}^2 - 
10\,M_{\pi}^4}{80\,F_{\pi}^2\,M_{\pi}^2\,{\pi }^2} 
\nn \\
&+& 
  \frac{2\,\left( M_K^4 - 3\,M_K^2\,M_{\pi}^2 + 2\,M_{\pi}^4 \right) }
{5\,F_{\pi}^2} 
 \,  \bar{J}^{(0,0,1)}(M_K^2,M_{\pi}^2,M_{\pi}^2)
\nn \\
&+& 
\frac{\left( 5 M_K^6 - 13 M_K^4\,M_{\pi}^2 + 8 M_K^2\,M_{\pi}^4 \right)
}{20\,F_{\pi}^2\,M_{\pi}^2}   
\left[  \bar{J}^{(1,0,0)}(M_{\pi}^2,M_K^2,M_{\pi}^2) + 
\bar{J}^{(0,0,1)}(M_{\pi}^2,M_K^2,M_{\pi}^2) \right]
\nn \\
&+& 
  \frac{\left( M_K^6 - M_K^4\,M_{\pi}^2 \right) }{12\,F_{\pi}^2\,M_{\pi}^2} 
\, \left[  \bar{J}^{(0,1,0)}(M_{\pi}^2,M_K^2,M_{\eta}^2)
+  \bar{J}^{(1,0,0)}(M_{\pi}^2,M_K^2,M_{\eta}^2) \right]
\\
\Delta_L \cA_{3/2}^{(g)} &=& 
- 
  \frac{\left( M_K^2 - 2\,M_{\pi}^2 \right)}{2\,F_{\pi}^2}  
\,\bar{J}(M_K^2,M_{\pi}^2,M_{\pi}^2)
- 
  \frac{M_K^4}{8\,F_{\pi}^2\,M_{\pi}^2}  
\,\bar{J}(M_{\pi}^2,M_K^2,M_{\eta}^2)
\nn \\
&-& 
  \frac{\left( 5\,M_K^4 - 8\,M_K^2\,M_{\pi}^2 \right) }
{8\,F_{\pi}^2\,M_{\pi}^2} 
 \,\bar{J}(M_{\pi}^2,M_K^2,M_{\pi}^2) 
\nn \\
&-& 
  \frac{\left( 8\,M_K^4 - 6\,M_K^2\,M_{\pi}^2 + M_{\pi}^4 \right) 
}{8\,F_{\pi}^2\,\left( M_K^2 - M_{\pi}^2 \right) }  
\,T(M_{\eta}^2)
- 
  \frac{\left( 2\,M_K^4 - 5\,M_K^2\,M_{\pi}^2 \right)
 }{4\,F_{\pi}^2\,\left( M_K^2 - M_{\pi}^2 \right) } 
 \,T(M_K^2)
\nn \\
&-& 
  \frac{\left( 4\,M_K^4 + 2\,M_K^2\,M_{\pi}^2 - 3\,M_{\pi}^4 \right) 
}{8\,F_{\pi}^2\,\left( M_K^2 - M_{\pi}^2 \right) } 
 \,T(M_{\pi}^2)
+ 
\frac{M_K^2 - 2\,M_{\pi}^2}{32\,F_{\pi}^2\,{\pi }^2} ~.
\eea

\subsubsection{$\Delta I= 5/2$ amplitudes}

In this section we report the one-loop $\Delta I= 5/2$ amplitude
generated by insertions of $e^2 p^0$ vertices from ${\cal L}_{\rm elm}$.

\bea
\Delta_L \cA_{5/2}^{(Z)} &=& 
- 
  \frac{8\,\left( M_K^2 - M_{\pi}^2 \right) }{5\,F_{\pi}^2} 
\,\bar{J}(M_K^2,M_{\pi}^2,M_{\pi}^2)
- 
\frac{2\,\left( M_K^4 - M_K^2\,M_{\pi}^2 \right)
 }{5\,F_{\pi}^2\,M_{\pi}^2}   \,\bar{J}(M_{\pi}^2,M_K^2,M_{\pi}^2)
\nn \\
&-& 
  \frac{2\,\left( M_K^4 + M_K^2\,M_{\pi}^2 - 2\,M_{\pi}^4 \right) 
}{15\,F_{\pi}^2\,M_{\pi}^2}  \,\bar{J}(M_{\pi}^2,M_K^2,M_{\eta}^2)
\nn \\
&-& 
\frac{2\,\left( 4\,M_K^2 - M_{\pi}^2 \right) 
}{5\,F_{\pi}^2}  \,T(M_{\eta}^2)
- 
  \frac{4\,M_K^4 }{5\,F_{\pi}^2\,\left( M_K^2 - M_{\pi}^2 \right) } 
\,T(M_K^2)
\nn \\
&-& 
  \frac{2\,\left( 6\,M_K^4 - 19\,M_K^2\,M_{\pi}^2 + 11\,M_{\pi}^4 \right)
}{5\,F_{\pi}^2\,\left( M_K^2 - M_{\pi}^2 \right) } 
 \,T(M_{\pi}^2)
+
\frac{-M_K^4 + 9\,M_K^2\,M_{\pi}^2 - 
10\,M_{\pi}^4}{40\,F_{\pi}^2\,M_{\pi}^2\,{\pi }^2} 
\nn \\
&+& 
  \frac{4\,\left( M_K^4 - 3\,M_K^2\,M_{\pi}^2 + 2\,M_{\pi}^4 \right)
 }{5\,F_{\pi}^2}  \, \bar{J}^{(0,0,1)}(M_K^2,M_{\pi}^2,M_{\pi}^2)
\nn \\
&-& 
 \frac{4\,\left( M_K^4 - M_K^2\,M_{\pi}^2 \right) }{5\,F_{\pi}^2} 
\, \left[  \bar{J}^{(0,0,1)}(M_{\pi}^2,M_K^2,M_{\pi}^2)
+   \bar{J}^{(1,0,0)}(M_{\pi}^2,M_K^2,M_{\pi}^2) \right].
\eea

\subsubsection{Divergent parts}

For completeness, we list here the divergent parts of the 
mesonic loop amplitudes. We have checked explicitly that 
they get absorbed by the independently known 
renormalization of NLO chiral couplings.  

\bea 
\left[ \Delta_L \cA_{1/2}^{(27)} \right]_{\rm div} &=& 
\displaystyle\frac{ - 28 \, M_K^2  + 17   \,  M_\pi^2  }{2 \,   F_{\pi}^2} 
\, \Lambda (\nu_\chi) 
\nn \\
\left[ \Delta_L \cA_{1/2}^{(8)} \right]_{\rm div} &=& 
\displaystyle\frac{-27 \,  M_K^2 + 103 \,     M_\pi^2  }{ 18 \,  F_{\pi}^2} 
\, \Lambda (\nu_\chi) 
\nn \\
\left[ \Delta_L \cA_{1/2}^{(\epsilon)} \right]_{\rm div} &=& 
\displaystyle\frac{ 10 \,  M_K^2   - 43 \,   M_\pi^2  }{ 6 \,  F_{\pi}^2} 
\, \Lambda (\nu_\chi) 
\nn \\
\left[ \Delta_L \cA_{1/2}^{(Z)} \right]_{\rm div} &=& 
\displaystyle\frac{ 7 \, (M_K^2 +  M_\pi^2)  }{2 \,    F_{\pi}^2} 
\, \Lambda (\nu_\chi) 
\nn \\
\left[ \Delta_L \cA_{1/2}^{(g)} \right]_{\rm div} &=& 
\displaystyle\frac{-   M_K^2  + 10 \,    M_\pi^2  }{ 2 \,   F_{\pi}^2} 
\, \Lambda (\nu_\chi) 
\nn \\
\left[ \Delta_L \cA_{3/2}^{(27)} \right]_{\rm div} &=& 
\displaystyle\frac{ - M_K^2  - 15 \,    M_\pi^2  }{2  \,  F_{\pi}^2} 
\, \Lambda (\nu_\chi) 
\nn \\
\left[ \Delta_L \cA_{3/2}^{(\epsilon)} \right]_{\rm div} &=& 
\displaystyle\frac{ 64  \, M_K^2   - 27 \,   M_\pi^2  }{ 6\,   F_{\pi}^2} 
\, \Lambda (\nu_\chi) 
\nn \\
\left[ \Delta_L \cA_{3/2}^{(Z)} \right]_{\rm div} &=& 
\displaystyle\frac{ 17 \,  (4 \, M_K^2 +  M_\pi^2)  }{10 \,    F_{\pi}^2} 
\, \Lambda (\nu_\chi) 
\nn \\
\left[ \Delta_L \cA_{3/2}^{(g)} \right]_{\rm div} &=& 
\displaystyle\frac{8 \,  M_K^2  +    M_\pi^2  }{ 2 \,  F_{\pi}^2} 
\, \Lambda (\nu_\chi) 
\nn \\
\left[ \Delta_L \cA_{5/2}^{(Z)} \right]_{\rm div} &=& 
\displaystyle\frac{ 48 \,  (M_K^2 -  M_\pi^2)  }{5 \,   F_{\pi}^2} 
\, \Lambda (\nu_\chi)~. 
\eea

\setcounter{equation}{0}
\addtocounter{zaehler}{1}
\section{Alternative convention for LO LECs}
\label{app:convention2}

In the effective chiral Lagrangians of Sec.~\ref{sec:ECL}, the meson
decay constant in the chiral limit $F$ is the only dimensionful
parameter in addition to the Fermi coupling constant $G_F$. This is
the original convention of Cronin \cite{cronin67} for the nonleptonic
weak Lagrangian of lowest order and it is used throughout this
work. It has definite advantages, e.g., for the renormalization of the
various Lagrangians.

However, this convention has a certain aesthetic drawback in that the
$K \to 2 \pi$ amplitudes (the $K \to 3 \pi$ amplitudes as well, for
that matter) depend at NLO on the strong LECs $L_4$ and $L_5$ even in
the isospin limit. These LECs account for the renormalization of $F$
to $F_\pi$ and $F_K$ at NLO. The associated uncertainties propagate
into the uncertainties of the LO LECs $G_8$, \dots Since $F_\pi$ and
$F_K$ are much better known than $F$, it may be useful for
phenomenological purposes to redefine the LO LECs so that they are
then free of the uncertainties in $L_4^r, L_5^r$.

A first step consists in generalizing the convention first used in
Ref.~\cite{kmw90}, albeit with a different notation:
\begin{eqnarray} 
\bar G_8 = G_8 F^4 / F_\pi^4 ~,\qquad & \qquad \bar G_{27} = G_{27} 
F^4 / F_\pi^4 \nl 
\bar g_{\rm ewk} = g_{\rm ewk} F^2 / F_\pi^2 ~,\qquad & \qquad 
\bar Z = Z F^2 / F_\pi^2~.
\label{eq:conv2}
\end{eqnarray} 
At lowest order, the barred quantities are identical to the original
unbarred ones because we always set $F=F_\pi$ at lowest order. Writing
the NLO amplitudes (\ref{eq:structure1}) in terms of the barred LECs of
lowest order, the strong LEC $L_4^r$ disappears completely from all
$K \to 2 \pi$ amplitudes. To get rid of $L_5^r$ as well (at least in
the isospin limit), one can introduce a scale factor $F_\pi/F_K$ 
\cite{kmw91}. The amplitudes of 
Eq.~(\ref{eq:structure1}) then take the following form:
\begin{eqnarray} 
\cA_n &=& \bar G_{27} \, F_\pi \, \bigg( M_K^2 - M_\pi^2 \bigg) \,  
\bar \cA_{n}^{(27)} 
 \label{eq:structure3}  \\
&+& 
\bar G_{8} \, F_\pi \,  \Bigg\{ \left( M_K^2 - M_\pi^2 \right)  
\bigg[ \bar \cA_{n}^{(8)} +  \varepsilon^{(2)} \, \bar \cA_{n}^
{(\varepsilon)} \bigg] -   e^2 \, F_\pi^2 \,  \bigg[
  \cA_{n}^{(\gamma)} 
+ \bar Z \,  \bar \cA_{n}^{(Z)} + 
 \bar g_{\rm ewk} \,  \bar \cA_{n}^{(g)} \bigg]  \Bigg\}~, \nn
\end{eqnarray} 
where
\begin{equation} 
\bar \cA_{n}^{(X)} = \left\{ \begin{array}{lll}
a_n^{(X)} \, \displaystyle\frac{F_\pi}{F_K} \, \left[ 1  + 
\Delta_{L} \bar \cA_{n}^{(X)} + \Delta_{C} \bar \cA_{n}^{(X)}
\right]  & \ \ \  \mbox{if} \ \ \  &  a_n^{(X)} \neq 0~, \\
  &   &     \label{eq:structure4}  \\
\qquad \quad \quad \ \Delta_{L} \cA_{n}^{(X)} + \Delta_{C} \cA_{n}^{(X)} &
\ \ \ \mbox{if} \ \ \ & 
 a_n^{(X)} = 0~.
\end{array}
\right.
\end{equation} 
The change in notation only affects those amplitudes that are
non-zero at lowest order. 

The amplitudes $\bar \cA_{n}^{(X)}$ are related to the original
$\cA_{n}^{(X)}$ as follows
(only amplitudes for $n=$ 1/2 or 3/2 are affected):
\begin{eqnarray} 
{\rm for}~~X=27,8,\varepsilon~: && \nl
\Delta_{C} \bar \cA_{n}^{(X)} &=& \Delta_{C}
\cA_{n}^{(X)}|_{\widetilde{\Delta}_C=L_4^r=0} +
\displaystyle\frac{24(M_K^2-M_\pi^2)}{F_\pi^2} L_5^r(\nu_\chi) 
\delta_{n,1/2} \delta_{X,\varepsilon} \nl
\Delta_{L} \bar \cA_{n}^{(X)} &=& \Delta_{L} \cA_{n}^{(X)} + \Delta_K
+ 3 \Delta_\pi - \displaystyle\frac{3}{2} (E_K + 3 E_\pi)
\delta_{n,1/2} \delta_{X,\varepsilon} ~;
\label{eq:X278eps} \\
{\rm for}~~X=Z,g~: && \nl 
\Delta_{C} \bar \cA_{n}^{(X)} &=& \Delta_{C}
\cA_{n}^{(X)}|_{\widetilde{\Delta}_C^{({\rm ew})}=0} \nl
\Delta_{L} \bar \cA_{n}^{(X)} &=& \Delta_{L} \cA_{n}^{(X)} + \Delta_K
+ 5 \Delta_\pi ~. \label{eq:Zg}
\end{eqnarray} 

 From the definitions of $F_\pi$ and $F_{K^\pm}$ in 
Ref.~\cite{Knecht:1999ag} one obtains [$T(M^2)$ is defined in 
(\ref{eq:TM}) and $M^2_\eta$ stands for $(4 M^2_K - M^2_\pi)/3$ as in
  all loop amplitudes]
\begin{eqnarray} 
\Delta_\pi &=& \displaystyle\frac{M_\pi^2}{F_\pi^2} T(M_\pi^2)+
 \displaystyle\frac{M_K^2}{2 F_\pi^2} T(M_K^2) \nl
\Delta_K &=& \displaystyle\frac{3 M_\pi^2}{8 F_\pi^2} T(M_\pi^2)+
\displaystyle\frac{3 M_K^2}{4 F_\pi^2} T(M_K^2) +
\displaystyle\frac{(4 M_K^2 - M_\pi^2)}{8 F_\pi^2} T(M_\eta^2) \nl
E_\pi &=& - ~\displaystyle\frac{(M_K^2 - M_\pi^2)}{ F_\pi^2} T(M_K^2)
+ \displaystyle\frac{(M_K^2 - M_\pi^2)}{(4\pi)^2 F_\pi^2} \nl
E_K &=&  \displaystyle\frac{3 M_\pi^2}{4 F_\pi^2} T(M_\pi^2)
- \displaystyle\frac{2(M_K^2 - M_\pi^2)}{ F_\pi^2} T(M_K^2) -
\displaystyle\frac{(8 M_K^2 - 5 M_\pi^2)}{4 F_\pi^2} T(M_\eta^2) \nl
&& + ~\displaystyle\frac{3(M_K^2 - M_\pi^2)}{(4\pi)^2 F_\pi^2}~. 
\label{eq:DE}
\end{eqnarray} 

As can be seen from Eqs.~(\ref{eq:structure3},\dots,\ref{eq:DE}), $L_4^r$
has disappeared completely from the amplitudes $\cA_n$ whereas $L_5^r$
occurs only in the isospin violating amplitude $\Delta_{C} \bar \cA_{1/2}^
{(\varepsilon)}$. In spite of its conceptual advantages, we have not
used this alternative convention in this paper because
$L_4, L_5$ reappear anyway through the large-$N_c$ relations for the
NLO LECs $N_i, D_i, Z_i$. Moreover, the large-$N_c$ relations for
$g_8$, $g_{27}$ and $g_{\rm ewk}$ would also be affected.
Finally, consistent with the expansion to leading
order in $1/N_c$, $L_4^r$ and $L_5^r$ are set equal to their
large-$N_c$ limits as discussed in Sec.~\ref{sec:LECp4}.

\setcounter{equation}{0}
\addtocounter{zaehler}{1}
\section{Details on the optical theorem analysis}
\label{app:optical}

In this appendix we report the explicit form of functions needed 
when studying the unitarity condition in the presence of isospin 
breaking. Let us start with the IR divergent factors:
\bea 
B_{\pi \pi} &=&   B_{+-} (M_\gamma) \qquad \mbox{(see Eq.(\ref{eq:appD1}))}
   \\
C_{\pi \pi} &=&   16 \pi \bigg[ (u - 2 M_\pi^2) \, G_{+ - \gamma} (u) \, - \,  
(t - 2 M_\pi^2) \, G_{+ - \gamma} (t) \bigg] \ .  
\eea
The definition of the variables $t,u$ and the function 
$G_{+ - \gamma} (x)$ can be found in Refs.~\cite{ku98,kn02}.

In order to define the remaining ingredients, we need to fix 
the notation.  The four-momenta are denoted as follows:  
$$ K (P) \longrightarrow \pi^+ (q_+) \, \pi^- (q_-) \, \gamma (k)  
\longrightarrow \pi^+ (p_+) \, \pi^- (p_-) \ . $$ 
The differential phase space is then given by 
\beq
d \Phi_{+ - \gamma} = \frac{d^3 \, q_+}{(2 \pi)^3 \, 2 q_+^0} \, 
\frac{d^3 \, q_-}{(2 \pi)^3 \, 2 q_-^0} \, 
\frac{d^3 \, k}{(2 \pi)^3 \, 2 k^0} \, 
(2 \pi)^4 \, \delta^{(4)} (q_+ + q_- + k  - p_+ - p_-) \ . 
\eeq
Then, after performing the sum over photon polarizations,
the radiative amplitudes in leading Low approximation generate
the following factors:
\bea
f_{1}^{\rm rad} &=& - \displaystyle\frac{q_+^2}{(q_+ \cdot k + 
\frac{M_\gamma^2}{2})^2} - 
\displaystyle\frac{q_-^2}{(q_- \cdot k + \frac{M_\gamma^2}{2})^2} + 
\displaystyle\frac{2 q_+ \cdot q_-}{(q_+ \cdot k + \frac{M_\gamma^2}{2})
(q_- \cdot k + \frac{M_\gamma^2}{2})} \\
f_{2}^{\rm rad} &=& 
\displaystyle\frac{ p_+ \cdot q_+}{(p_+ \cdot k - \frac{M_\gamma^2}{2})
(q_+ \cdot k + \frac{M_\gamma^2}{2})} +  
\displaystyle\frac{ p_- \cdot q_-}{(p_- \cdot k - \frac{M_\gamma^2}{2})
(q_- \cdot k + \frac{M_\gamma^2}{2})}  \nonumber \\
 &-&
\displaystyle\frac{ p_+ \cdot q_-}{(p_+ \cdot k - \frac{M_\gamma^2}{2})
(q_- \cdot k + \frac{M_\gamma^2}{2})} - 
\displaystyle\frac{ p_- \cdot q_+}{(p_- \cdot k - \frac{M_\gamma^2}{2})
(q_+ \cdot k + \frac{M_\gamma^2}{2})} \ . 
\eea

\end{document}